\documentclass{article}

\usepackage{packages}
\usepackage{pdfpages}
\usepackage{titling}
\usepackage{setspace}
\begin{document}

\begin{singlespace}

\begin{titlepage}
  \begin{center}
       \vspace*{3cm} 

       \textbf{\huge{\bf{Predicting laboratory earthquakes with machine learning}}}

       \vspace{3cm} 
       
       {\LARGE{S.D. van Klaveren, I. Vasconcelos and A.R. Niemeijer}}

       \vspace{8cm}

       \vfill

     \begin{multicols}{2}
     \raggedright
             \includegraphics[width=0.4\textwidth, left]{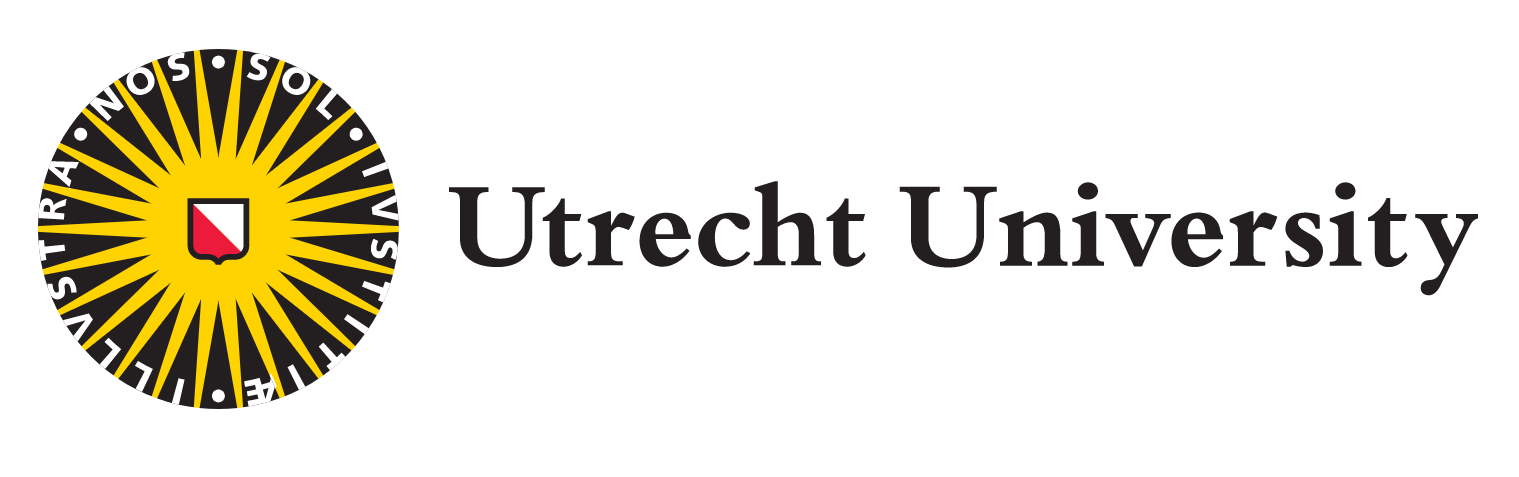}
             \vfill\null
             \columnbreak
    \raggedleft
       \normalfont{Department of Earth Sciences\\
       Utrecht University\\
       The Netherlands\\
       8 November 2020 \\ }

    \vfill
    \end{multicols}
   \end{center}
\end{titlepage}

\tableofcontents

\vspace{3cm}

\end{singlespace}

\newpage
\addcontentsline{toc}{section}{Abstract}
\section*{Abstract}
The successful prediction of earthquakes is one of the holy grails in Earth Sciences. Traditional predictions use statistical information on recurrence intervals, but those predictions are not accurate enough. In a recent paper, a machine learning approach was proposed and applied to data of laboratory earthquakes. The machine learning algorithm utilizes continuous measurements of radiated energy through acoustic emissions and the authors were able to successfully predict the timing of laboratory earthquakes. Here, we reproduced their model which was applied to a gouge layer of glass beads and applied it to a data set obtained using a gouge layer of salt. In this salt experiment different load point velocities were set, leading to variable recurrence times. The machine learning technique we use is called random forest and uses the acoustic emissions during the interseismic period. The random forest model succeeds in making a relatively reliable prediction for both materials, also long before the earthquake. Apparently there is information in the data on the timing of the next earthquake throughout the experiment. For glass beads energy is gradually and increasingly released whereas for salt energy is only released during precursor activity, therefore the important features used in the prediction are different. We interpret the difference in results to be due to the different micromechanics of slip. The research shows that a machine learning approach can reveal the presence of information in the data on the timing of unstable slip events (earthquakes). Further research is needed to identify the responsible micromechanical processes which might be then be used to extrapolate to natural conditions. 

\newpage

\section{Introduction}

Research on earthquakes predictions is of high importance since it could improve warning systems before catastrophic failure, which could help prevent significant loss of life and financial damage. However, predicting earthquakes is difficult and predictions of both timing and magnitude are usually not very accurate \citep{rouet17}. 
\\ \\
Plate tectonics move the Earth's lithosphere, causing an increased state of stress in rocks along weak zones which leads to elastic strain. The elastically stored energy builds up and when the stress finally exceeds the strength of the rock, the rock slips along weak zones. An earthquake occurs when this energy is rapidly released. Reid's elastic rebound theory describes this cyclic process of elastic loading and failure. Reid concluded, based on an earthquake in California in 1906, that the earthquake must have involved an elastic rebound of previously stored elastic stress \citep{reid1908california}. When the source of stress is continuous, the process repeats itself and is referred to as the earthquake cycle as it starts to accumulate stress again in the interseismic period until the next slip event. The earthquake cycle is a stick-slip process, with stick during the interseismic and slip during the seismic event. The mechanical properties of fault zones that accommodate deformation are very complex and geographic areas involved are very large, which further increase the challenge of quantifying the built-up stress and the fault strength it has to overcome before rupture. Besides stick-slip events there are also other slip mechanisms. \citet{leeman2016laboratory} reported a spectrum of tectonic fault slip behaviour, from stable sliding (aseismic slip) at one end of the spectrum to unstable sliding (stick-slip) at the other end. At the threshold from stable to unstable sliding they observed a spectrum of slow slip behavior and found that transient fault slip behaviours arise from the same frictional dynamics as regular earthquakes. The type of fault slip is determined by the interplay of fault frictional properties, effective normal stress and the elastic stiffness of the surrounding material \citep{leeman2016laboratory}.
\\ \\
Different approaches on earthquake prediction are proposed in literature. For example based on their recurrence intervals \citep{zechar2012} or based on a combination of measured slip rates and timing of the most recent earthquake \citep{shimazaki1980time}. The model of \citet{shimazaki1980time} states that an earthquake occurs when the fault recovers the stress relieved in the most recent earthquake. Stress is difficult to measure in situ, however, the Earth's crust is assumed to be linearly elastic, so strain is linearly proportional to stress. The subsurface is also assumed to be a smooth continuum, so that surface strain observations can be used in inferring stress and strain states at depth. Taking advantage of this, geodetic measurements of the Earth's surface may be used to estimate both the strain released in an earthquake and the interseismic strain rate \citep{murray2002testing}. This model is not suitable for every geological setting as found by \citet{murray2002testing}. They applied the model to the Parkfield segment of the San Andreas fault: the earthquake prediction model fails.
\\ \\
If stress is released only in earthquakes with characteristic size and tectonic loading and fault strength are relatively constant, periodic or at least semi-periodic earthquake sequences might be expected \citep{zechar2012}. A model based on these simple conditions with characteristic, repeating earthquakes was previously believed to describe the occurrence of a sequence of earthquakes of magnitude 5-7 at the Parkfield section of the San Andreas fault, California \citep{bakun1985}. Six similar earthquakes occurred and were recorded between 1857 and 1966, suggesting a recurrence interval of 21.9 $\pm$ 3.1 years. The next earthquake was therefore expected between 1988 and 1993. The Parkfield High-Resolution Seismic Network was monitoring the area to capture the seismic evolution before and after the predicted earthquake. The predicted earthquake ultimately took place in 2004, 38 years later. This shows that prediction based on recurrence intervals is not accurate and reliable enough. To date, there is no model that accurately predicts earthquakes in terms of timing, magnitude and location.

\subsection{Previous work}
Recently it has been suggested that machine learning (ML) could be used for earthquake predictions. \citet{rouet17} successfully predicted laboratory earthquakes with a reasonable $r{^2}$ score. They sheared a fault gouge-like sample of glass beads and recorded the acoustic emissions continuously during the experiment. During the laboratory experiment slip events occur which appear to be similar to earthquakes in nature \citep{brace1966stick}. Those slip events are therefore a good analogue for earthquakes. Leading up to this failure event, the system releases several bursts of energy. These are precursors, a manifestation of critical stress conditions preceding shear failure. Precursors take place when the material is still modestly dilating, yet while the macroscopic frictional strength is no longer increasing \citep{johnson13}. The rate of impulsive precursors accelerates as failure approaches \citep{johnson13}, suggesting that the timing of the upcoming laboratory earthquake can be predicted \citep{rouet17}. This pattern of increase in seismic activity before failure form the foundation of the ML model. The promising part of this research is that once the predictive time to failure model is build, the timing of the slip event can be predicted long before and close to the actual time of the failure event.

\subsection{This work}
This research aims to reproduce the prediction model of \citet{rouet17}, based on the data from their experiment. Once the ML model is built, we use the same algorithm for another data set from a different experiment. The main differences between the two data sets are the experimental setup and the material used. \citet{rouet17} used glass beads, micrometre-scale spheres representing fault gouge material and they submitted the sample to double direct shear. The shear velocity is constant during the experiment. After obtaining the model based on glass beads data, the same algorithm is applied to a data set obtained using a gouge layer of salt. \citet{korkolis} conducted an experiment on salt during his PhD research in which he sheared a fine grained salt sample in a ring shear set up, testing five different load point velocities. Besides comparing the performance of the machine learning model on both data sets and comparing the effect of using different materials, the effect of this change in load point velocities on the model is investigated in this research.
\\ \\
The machine learning model makes use of the scikit learn packages in Python \citep{pedregosa2011scikit}. The machine learning technique used is called \textit{random forest}. A random forest model is a ensemble learning method for classification and regression problems and is used when a pattern is hidden in a data set. In this research our random forest model aims to predict time to failure. Based on the acoustic emission data, it can predict how long it will take before the next slip event will occur. The outcome of the model, i.e. the predictability performance for both data sets is discussed extensively in this research. The research questions are: is the random forest model able to predict time to failure? And what is the effect of changing load point velocity on the prediction model? Moreover, when interpreting the results, we try to answer the following question: do the feature patterns of the estimated RF model provide insight into the process or mechanics of precursors for different physical experiments? This research works with data sets of two different experiments, one on glass beads and one on salt. We speculate that the differences are mostly caused by the difference in micromechanical characteristics of both materials. 
\\ \\
This new approach of earthquake prediction considers the characteristics of the acoustic signal in between two earthquakes. During the laboratory earthquake experiment, the applied stress is controlled and the acoustic emissions coming from the sheared sample are continuously recorded. This model uses seismic data which is known to arise from the experiment, in contrast to a seismometer on Earth that records signals from all directions, which may or may not be associated with a specific fault system. To observe small changes in the acoustic signal one must take measures to ensure that the signal originated from a specific fault, e.g. by placing sensors in proximity to the fault and by appropriate data processing. Therefore, earthquake prediction on real faults is substantially more challenging. In regions of induced seismicity, instrumentation is often located close to the fault. Unlike natural seismicity, induced seismicity does not follow a cycle because the loading rate is variable due to human activity. Our experiments are made to simulate natural earthquakes so induced seismicity data would likely not be suitable for our ML model. ML-driven experimental prediction of induced seismicity requires further investigation and is not in the scope of this project. However, the appearance of seismic activity before the main event, called precursors, seems to tell something about the state of stress. A similar ML model might be able to give an indication of the stress level on the faults by recording an increase in activity.
\\ \\
The observations of precursors prior to slip may help constrain periods of increased seismic hazard. An earthquake precursor can be observed by anomalies in for example electric and magnetic fields, gas emissions, groundwater level changes, temperature changes, surface deformations and seismicity \citep{scholz1973earthquake} \citep{cicerone2009systematic}. \citet{cicerone2009systematic} found that the frequency and amplitude of precursors increase towards failure. In this research we focus on seismicity as a precursor, specified by a peak in acoustic signal prior to the main event. \citet{ferdowsi2013microslips} also found that large slip events are preceded by a sequence of small slip events, called microslips, whose occurrence accelerates exponentially before the large slip event. These precursors, or at least the increase in acoustic emissions towards failure, are the foundation of the prediction model. Precursors are observed in different types of faults: thrust, strike-slip and normal faults. In extensional environments normal faulting is controlled by gravity, since the maximum stress axis is parallel to the lithostatic load \citep{doglioni2015normal}. Those earthquakes are sometimes called graviquakes. Reid's elastic rebound theory is likely true for contractional and strike-slip tectonic settings, but in tensional environments the influence of gravity seems to dominate. \citet{doglioni2015normal} observed precursors in normal faults and they interpret that they plausibly originate from instability of the hangingwall. In this research we focus on strike-slip faults, therefore we assume that the principles of the elastic rebound theory hold in our experiments.
\\ \\
Although precursors have been observed for a signiﬁcant number of earthquakes, many earthquakes apparently exhibit no precursor activity \citep{johnson13}. It shows that not every earthquake is the same as a consequence of an interplay of physical processes and variable boundary conditions of pressure, temperature, fault rock type and loading rate, which are difficult to unravel. Besides improving prediction models, this kind of research helps in revealing microphysics that operates in these materials under these conditions. Because of its high social and economical impact, it is of high importance to investigate earthquakes on large scales but also on controlled microscales to improve our understanding of fault mechanisms, eventually towards enabling early earthquake hazard warning. 
\vfill\null
\raggedcolumns

\newpage

\section{Experimental set up}

\subsection{Glass-bead experiment}
\begin{figure}[ht]
\centering 
\includegraphics[height=2.9in]{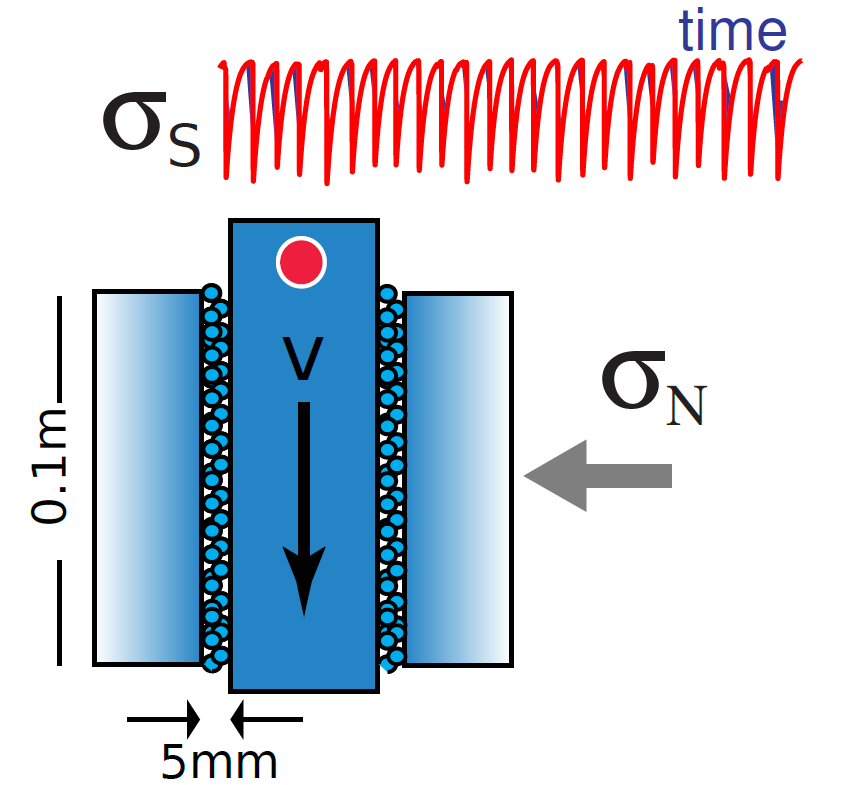}
\caption{Double shear set up. A slider block with fault gouge layers on two sides is loaded by a constant stress $\sigma_{N}$. The slider is driven downward at constant displacement rate $v$, inducing slide-slip behavior (as seen in the shear stress $\sigma_{S}$). An accelerometer continuously records the acoustic emission at a sampling rate of 330 kHz. Figure from \citet{rouet17}.}
\label{setup_glassbeads}
\end{figure}

The friction experiment by \citet{rouet17} consists of a two-fault configuration that contains simulated fault gouge material: glass beads with a very small diameter (105-149 $\mu m$). They submit the sample to double direct shear, where they place two 5 mm-thick fault gouge layers between the three blocks, which are hold in place by a fixed normal load (figure \ref{setup_glassbeads}) \citep{rouet17}. The central block induces shear with a constant displacement rate of 5 $\mu$m/s. An accelerometer, positioned on the central block, $\sim$2.5 cm from the sample, records the acoustic emission (AE) coming from the shearing layers, with a sampling rate of 330kHz. The applied load and shearing rate are controlled and monitored. The imposed shear, gouge layer thickness and friction are monitored as well during the experiment \citep{rouet17}.

\subsection{Salt experiment with different load point velocities}
Dr. Evangelos Korkolis simulated laboratory earthquakes in a rotary ring shear experiment in the high temperature and pressure lab at Utrecht University. A dry salt (halite, NaCl) sample is sheared and the acoustic emissions (AE) are recorded as well as the shear stress through the experimental run time. In the pre-compaction stage the salt sample is sheared for a one hour and 45 minutes. After the pre-compaction phase the sensors start recording. Four sensors (ch9, ch11, ch13 and ch15) are positioned 90 degrees apart on the top stationary piston at a distance of 5 mm from the fault and record the acoustic emissions.
\\ \\
At room temperature, salt produces stick-slips while releasing acoustic emissions at relative low rotation velocities. Five different loading velocities are set in this experiment, from high loading velocities to low loading velocities, all velocities being in the stick-slip regime \citep{voisin2007}. Five blocks are distinguished in which the loading velocity is constant. The total rotation is $\sim70 ^{\circ}$, i.e. 51.9 mm displacement determined at the center radius of the sample. Table \ref{slidingvel} shows the different load point velocities for every block. The rotation rate corresponds to the rate at the mean radius of the sample. 
\vspace{0.3cm}
\begin{table*}[ht]
\centering
\caption{Load point velocities for the five different blocks. The last column indicates the data time spans of the experiment that we will use to make the prediction model, we select only three seconds per block.}
\begin{tabular}{l|l|l|l|}
\small
         & \begin{tabular}[c]{@{}l@{}}Load point \\velocity (deg/s)\end{tabular}   & \begin{tabular}[c]{@{}l@{}}Load point \\velocity (mm/s)\end{tabular} & \begin{tabular}[c]{@{}l@{}}Selected time\\span ($\pm$ 1.5 s)\end{tabular}
         \\ \hline
         
         Block 1 & $\sim$ 0.15            &  $\sim$ 0.2   & 110            \\ \hline
         
Block 2                 & $\sim$ 0.075           & $\sim$ 0.1    & 280          \\ \hline
Block 3                 & $\sim$ 0.06            & $\sim$ 0.08   & 450          \\ \hline
Block 4                  & $\sim$ 0.03            & $\sim$ 0.04 & 620           \\ \hline
Block 5                 & $\sim$ 0.015           & $\sim$ 0.02  &790         \\ \hline
\end{tabular}
\label{slidingvel}
\end{table*}

\begin{figure}[ht]
    \centering
    \includegraphics[width=0.8\textwidth]{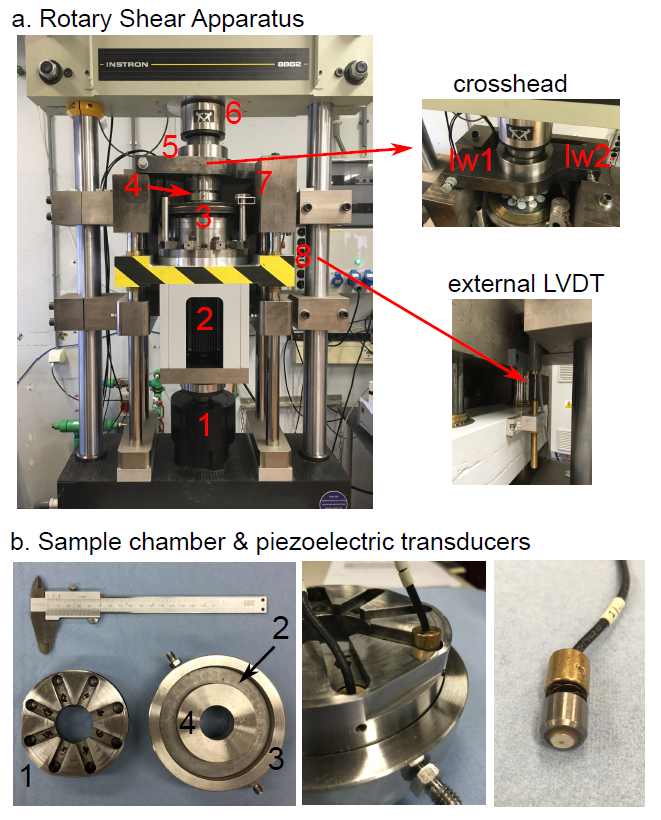}
    \caption{From PhD thesis E. Korkolis: a) View of the Rotary Shear Apparatus. Only relevant numbers are described. 4: Sample chamber. 5: Crosshead, equipped with two load cells for measuring torque. 7: Angular potentiometer. 8: External LVDT. b) Sample chamber and piezoelectric transducers. (b, left) 1: Top piston. 2. Bottom piston. 3: Outer ring. 4: Inner ring. (b, middle) close-up view of the assembled sample chamber. One piezoelectric transducer is properly installed (left), whereas a second one is partially inserted into its slot. (b, right) A piezoelectric transducer. When installed, the piezoelectric elements lie approximately 5 mm away from the sample.}
    \label{apparatus}
\end{figure}

\noindent The sampling frequency of the mechanical data (shear stress) is 10 kHz and is 1 MHz for the AE data. The effective normal stress for these experiments is 8 MPa. The four AE transducers are installed inside the top piston, in a cross configuration, shown in figure \ref{apparatus}. The shear stress is calculated by converting the time series of the load, recorded by the two load cells installed in the crosshead, to torque on the sample and dividing by the surface area of the piston ring \citep{korkolis}. 
\\ \\
The rotary shear configuration is able to capture the evolution of AE characteristics over large shear displacements. Larger than common experimental designs as, for example the double direct shear described earlier, used in the study of \citet{rouet17}, that can only shear to limited extent. The thickness of the sample at any point during the experiment is measured by the time series of the two linear variable differential transducers (LVDT) and the measurement of the sample chamber's thickness after the experiment. The relative angular displacement of the two piston rings is calculated using the time series of two angular potentiometers (LPD), expressed in degrees referenced to the mean radius of the sample \citep{korkolis}.

\clearpage

\section{Theoretical background}

\subsection{Material properties}
\begin{figure}[ht]
    \centering
    \includegraphics[width=0.8\textwidth]{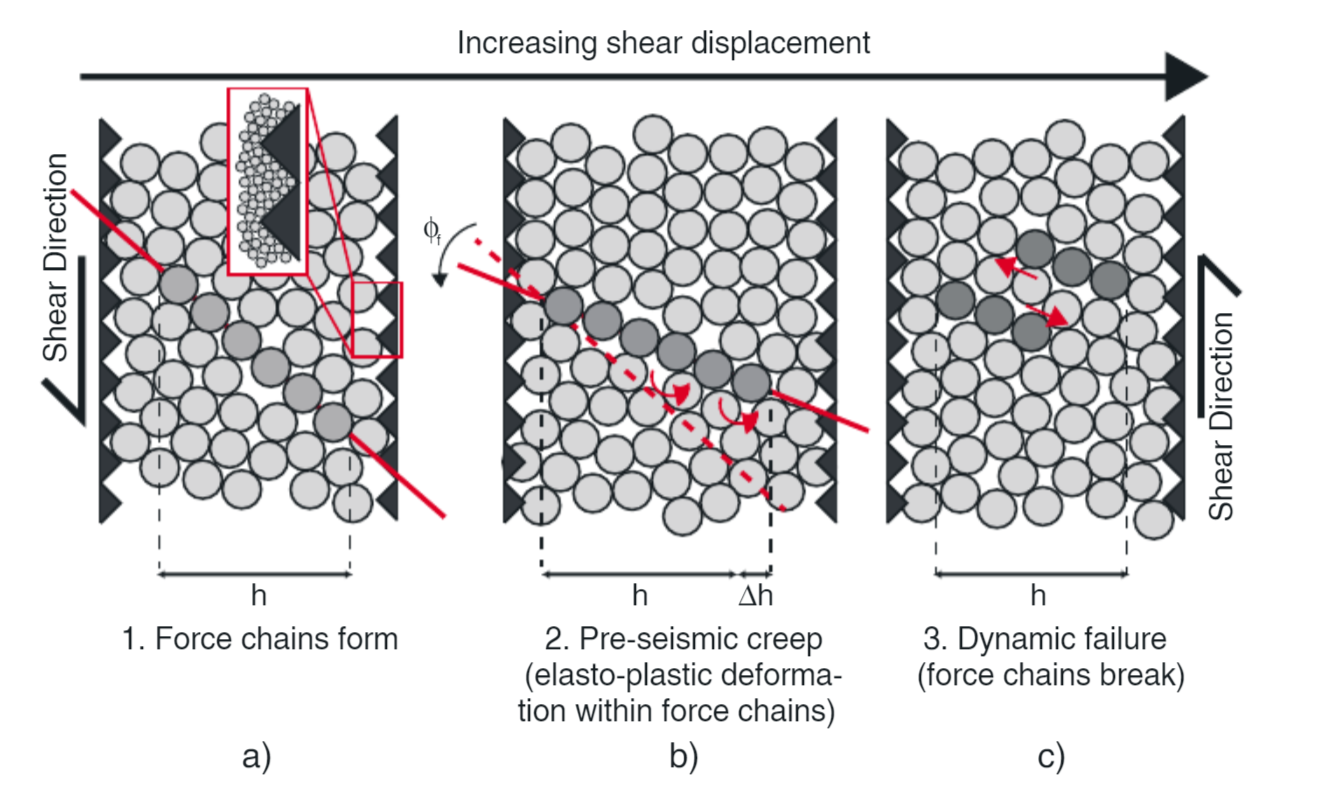}  
    \caption{Cartoons of shearing glass beads. The showed grsins are larger than the actual grain size. The inset in a) shows the true scale relative to the surface roughness. The developed force chains are shown in darker grey. (b) While shear is continued, force chains deform elastoplastically due to interparticle slip and rolling, causing layer dilation. (c) Once the maximum strength is reached, grains slide and the force chains break. The cycle is repeated for each stick-slip event throughout the experiment. Figure from \citet{scuderi2014}}.
    \label{forcechain}
\end{figure}

Figure \ref{forcechain} shows the deformation mechanism that operates in the glass beads experiment. Deformation takes place largely in the gouge since the steel blocks are extremely stiff ($\sim160 GPa$). When stress is applied the glass beads form force chains. During the interseismic period, shear stress increases and the force chains deform. When approaching failure, it exhibits characteristics of a critical stress regime, including many small shear failures that emit impulsive AEs \citep{johnson13}, the hypothesis is that those AEs originate from rotational friction of single beads. During the failure event, the shearing block rapidly displaces and friction and shear stress drop. The force chain breaks and the layer simultaneously compact. 

\begin{figure}[ht]
    \centering
    \includegraphics[width=0.7\columnwidth]{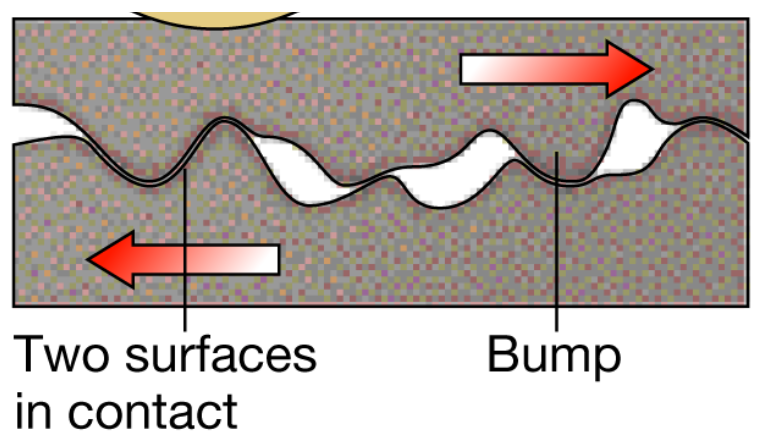}
    \caption{Cartoon showing the roughness of the contact interface of a fault. The amount of stress that can be build up is dependent of the contact area of the two surfaces and the strength of the material.}
    \label{frictionsalt}
\end{figure}

\noindent The deformation mechanisms in salt are different. Because of its physical properties, salt, an analogue for natural faults, allows for frictional processes as well as plastic deformation and pressure solution creep to operate on lab timescales \citep{voisin2007}. The salt gouge starts as a granular aggregate. The pre-shear creates a densified gouge layer leading to localization of shear on the interface. There is still compaction and/or material loss, indicating that some of the deformation might still occur within the bulk layer, however most deformation will now occur on the slip interface. Friction is among other things dependent on the normal load and the roughness of the slip interface described by contact asperities \citep{dieterich1992fault} \citep{dieterich1994direct}. Friction is higher for rough interfaces than for smooth interfaces because the more contact asperities there are or the bigger the size, the higher the frictional strength (figure \ref{frictionsalt}). Frictional strength is governed by the resistance to shear on the interface that separates two sliding bodies \citep{ben2010slip}. The overall strength of the interface is determined by the real contact area and the shear strength of the contacts. For motion to occur, the asperities must fracture, undergo internal damage or deformation which costs significant energy \citep{ben2010slip}. Energy can build up in the form of elastic energy until the strength of the material, i.e. the frictional strength of the contacts, is reached and the two bodies slip with respect to each other. The roughness of the contact interface and the surface morphology turns out to be crucial for generating slips events rather than stable sliding \citep{voisin2007}.

\subsection{Rate and state friction theory}
\begin{figure}[ht]
    \centering
    \includegraphics[width=0.75\columnwidth]{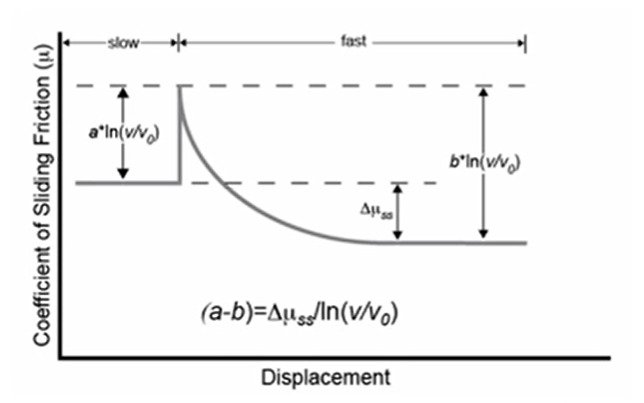}
    \caption{Steady state coefficient of friction with displacement. The coefficient of sliding friction $\mu$ will increase when $a>b$, the level of friction will decrease when $b>a$.}
    \label{frictionss}
\end{figure}

\begin{figure}[ht]
    \centering
    \includegraphics[width=0.4\columnwidth]{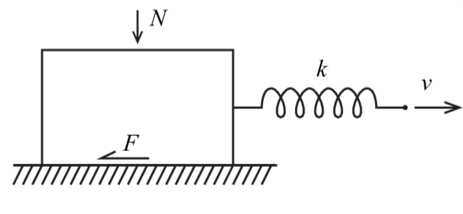}
    \caption{Spring slider experiment as analogue to stick slip behavior in rocks. The slider and the surface have a certain friction F, resisting movement of the slider. The string is pulled with constant velocity v and the spring will extend, loading elastically. Until at one point the level of friction is exceeded and the slider slides to the right. The stored elastic energy converts to kinetic energy and the spring will move back to its original extension.}
    \label{springslider}
\end{figure}

The rate and state theory (RSF) describes friction and is important in earthquake and fault mechanics. Several parameters play a role in the state of friction and some of them are time dependent. Figure \ref{frictionsalt} visualizes friction. The RSF provides relations between the measured coefficient of friction ($\mu$), rate of deformation ($V$), and “state”, the evolution of the frictional interface with time and displacement ($\theta$). The rate and state equation takes into account the following principles: \\
1. Friction is to a first order in slip velocity $V$ a constant: $\mu_0$. \\
2. Increase in contact area is time dependent due to for example pressure solution creep which increases the surface contact. \\
3. Slip breaks the surfaces in contact, causing a decrease in contact area / increase in dilantancy. \\
4. Friction is slip rate dependent. \textsc{\char13}Static\textsc{\char13} friction is higher than \textsc{\char13}dynamic\textsc{\char13} friction because contacts are older (larger, principle 2), this implies that contact size decreases as slip velocity increases.
\\ \\
The most widely used form of the RSF equation is \citep{dieterich1979modeling} \citep{ruina1983slip}:
\begin{equation}
    \mu = \mu_{0} + a \, ln\big( \frac{V}{V_{0}} \big) + b \, ln \big( \frac{V_{0} \theta }{D_{c}} \big)
\end{equation}
with $\mu_{0}$ being the reference coefficient of friction measured at sliding velocity $V_{0}$. The parameter \textit{a} is a constant, describing the magnitude of the instantaneous direct effect and parameter \textit{b} is a constant describing the magnitude of the time-dependent evolution effect. Together they are thought to represent material properties \citep{van2018comparison}. $D_{c}$ controls the slip distance over which the evolution towards the new steady-state takes place. The evolution of the state parameter $\theta$ is described by the aging law \citep{dieterich1979modeling}:
\begin{equation}
    \frac{d\theta}{dt}=1-\frac{\theta V}{D_{c}}
    \label{aginglaw}
\end{equation}
If only interested in change of friction with respect to time, it is enough to look at the (a-b) parameter. At steady state ($\frac{d\theta}{dt}=0$), the evolution parameter (equation \ref{aginglaw}) reduces to $D_{c}/V$. The steady-state coefficient of friction becomes:
\begin{equation}
    \mu_{ss}(V) = \mu_{0} + (a-b)ln \big( \frac{V}{V_{0}} \big)
\end{equation}
The parameter (a-b) describes the velocity-dependence of $\mu$ at steady-state. Figure \ref{frictionss} visualizes the effect of the parameters in the RSF equation. Positive values ($a>b$) result in velocity strengthening (i.e. frictional resistance increases with sliding velocity) and negative values result in velocity weakening behaviour (i.e. frictional resistance decreases with sliding velocity). For stick-slip behaviour a material is characterised by negative values ($a<b$) \citep{ruina1983slip}. It is therefore assumed that faults with seismic activity exhibit a negative (a-b) \citep{van2018comparison}.
\\ \\
A way of using the RSF equations to calculate slip velocity is by comparing the experiment to a spring-slider (figure \ref{springslider}), where a block with a spring is pulled with constant velocity. A spring-slider is a model for unstable slip. The block is in this case the salt or glass beads sample, having a certain friction. The spring is the apparatus (including all non active parts of the sample), the spring will be pulled elastically during time when the block is not moving. The stress increases linearly with time (assuming the spring is linear elastic and velocity is constant). When the pulling velocity is lower, it takes longer to build the same amount stress so it takes longer before stress exceeds strength and slip occurs. The faster the sliding velocity, the more stick-slips are generated in the same amount of time. Because of the unbalance of forces the fault can accelerate, leading to stick slip behavior.

\noindent According to the rate-and-state theory stick-slips are only possible when the net effect is a decrease in friction with increasing velocity $(b>a)$, because only then a fault can accelerate (figure \ref{frictionss}). The friction must drop faster than the elastic unloading to excess energy for acceleration. During the interseismic period slip velocity is zero and the fault gets stronger during that time. In that case one would expect that, in the stick slip regime, stress drops decrease with increasing velocity because there is less time between events during which the fault can gain strength. The strength of the material will be lower so stress will exceed strength on a lower level resulting is smaller stress drops. 

\noindent For an experimental fault obeying rate- and state friction laws, the occurrence of stick-slip instabilities is controlled by the interaction between the system elastic stiffness, k, and the fault frictional properties, which can be written in terms of a critical rheologic stiffness $k_{c}$ \citep{scuderi2017}. The system is stable when the stiffness of the system is higher than the stiffness of the fault. The stability criterion: 
\begin{equation}
    k_{c} = -\frac{\sigma_{n}(a-b)}{D_{c}}
    \label{stability}
\end{equation}
where $\sigma _{n}$ is the effective normal stress. The ratio $K = k/k_{c}$ describes fault stability: sliding is unstable when $k_{c}>k$ so when $K<1$. These researches are conducted in the stick slip regime so $K<1$ and show velocity weakening behavior. This theoretical framework is kept in mind when we speculate about the potential microphysics that operate in the samples.

\newpage

\section{Data description}
Before going into depth about the machine learning model, it is important to have a detailed look at the data. The two important data sets are the shear stress data and the acoustic emission (AE) data. From the shear stress data we infer the time to failure curve. The AE data and the time to failure curve form the basis of the model: from the AE data the model makes time to failure predictions and it compares those predictions to the actual time to failure curve to test the performance of the model. In addition to shear stress data and AE data, displacement and sample thickness are continuously measured.

\subsection{Acoustic seismic data}
The acoustic signal consists of signals originated from the sample, background noise and resonance of the apparatus. The first one being the most interesting feature and the latter one being the most undesirable. When a slip event occurs we observe large acoustic-signal amplitudes (figure \ref{data_280}a and \ref{data_620}a). The amplitude of the AE data gives information about the amount of energy release. Before the failure event, we observe peaks with a smaller amplitude, so-called precursors.
\\ \\
For the salt experiment, four sensors are recording the acoustic emissions. All four sensors record the signal simultaneously, so the signal recorded is very similar. In the result section we point out the differences between the signals from different sensors, but these differences are only visible when zoomed at relatively very small timescales. For now we describe the data from sensor ch9. From every load point velocity block we take three-seconds intervals to build and test the prediction model. Below we see the acoustic data from two blocks, around 280 and 620 seconds (figure \ref{data_280}a and \ref{data_620}a) as well as a zoomed in plot (figure \ref{data_280}b and \ref{data_620}b). The other times, from block 1, 3 and 5 (110s, 450s and 790s) are shown in appendix \ref{appendix1}.

\subsection{Shear stress data}
\begin{figure*}[ht!]
    \centering
    \includegraphics[width=0.85\textwidth,scale=0.1]{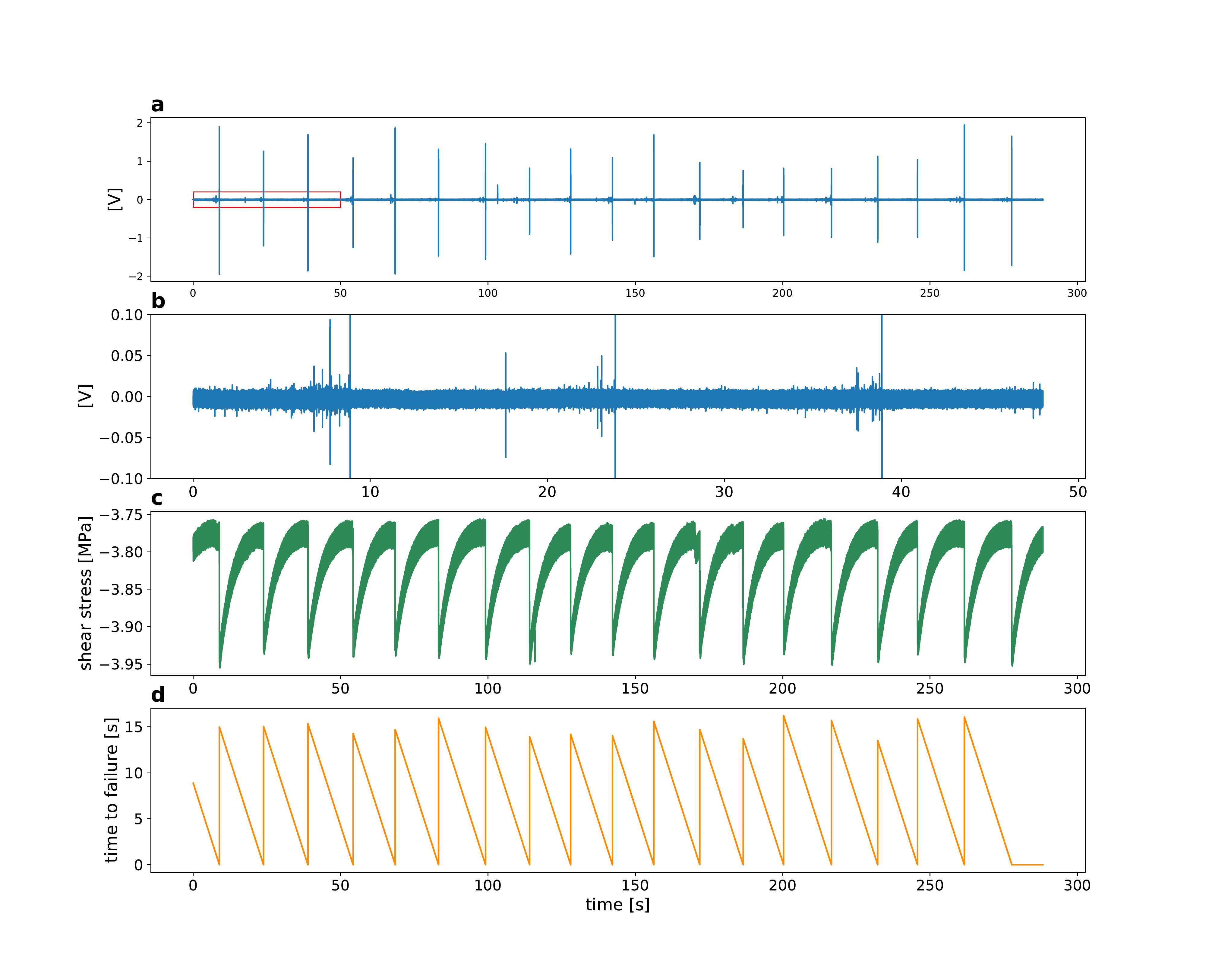}
    \caption{Continuously recorded data from the glass beads experiment. a) Acoustic data. b) Zoom in acoustic data, red box in panel above. c) Shear stress data. d) Time to failure curve. }
    \label{acoustic_glassbeads}
\end{figure*}

\begin{figure*}[ht!]
    \centering
    \includegraphics[width=0.82\textwidth,scale=0.1]{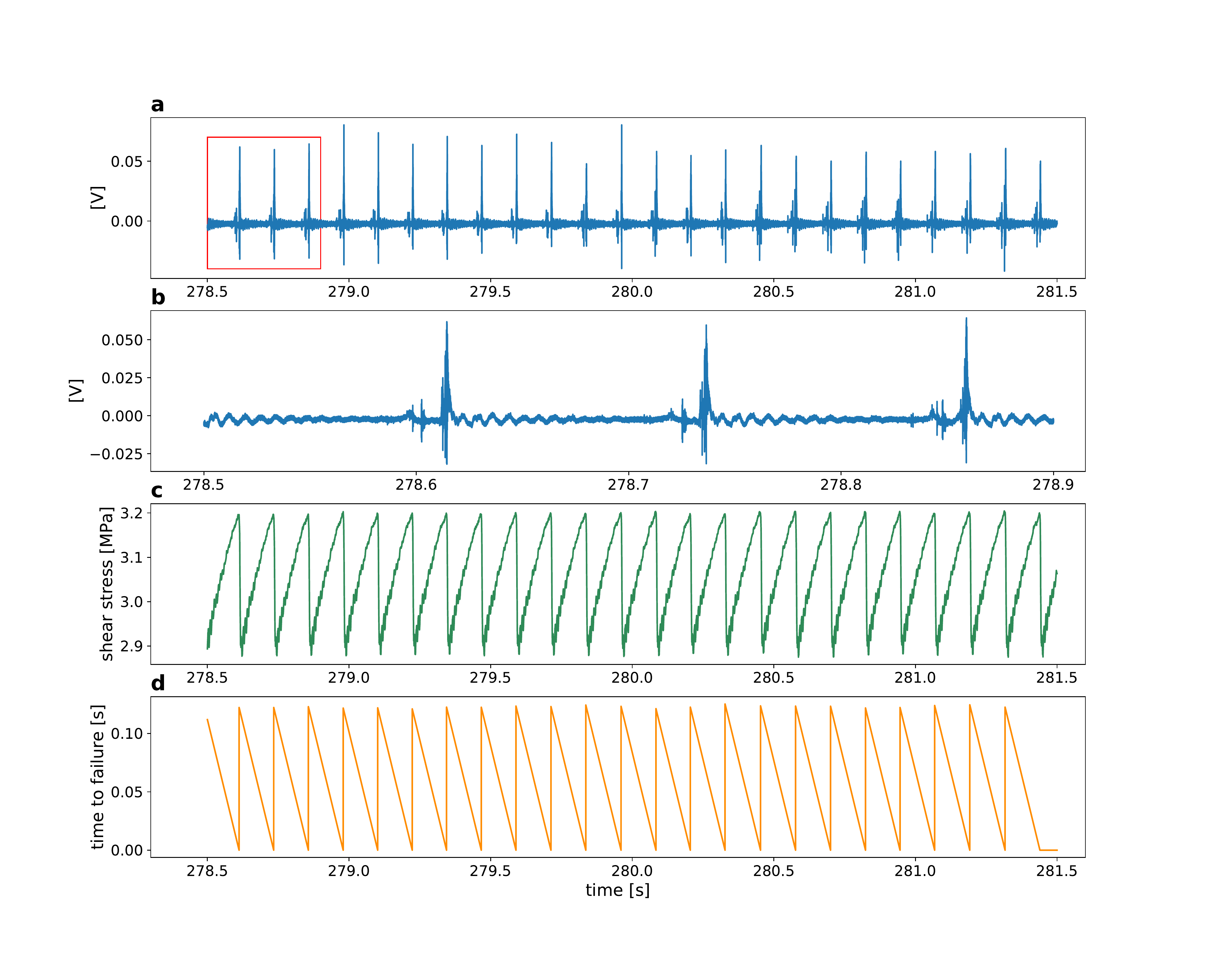}
    \caption{Continuously recorded data from the salt experiment. Sensor ch9, block 2, 280$\pm$1.5 seconds. a) Acoustic data. b) Zoom in acoustic data, red box in panel above. c) Shear stress data. d) Time to failure curve. }
    \label{data_280}
\end{figure*}

\begin{figure*}[ht!]
    \centering
    \includegraphics[width=0.82\textwidth,scale=0.1]{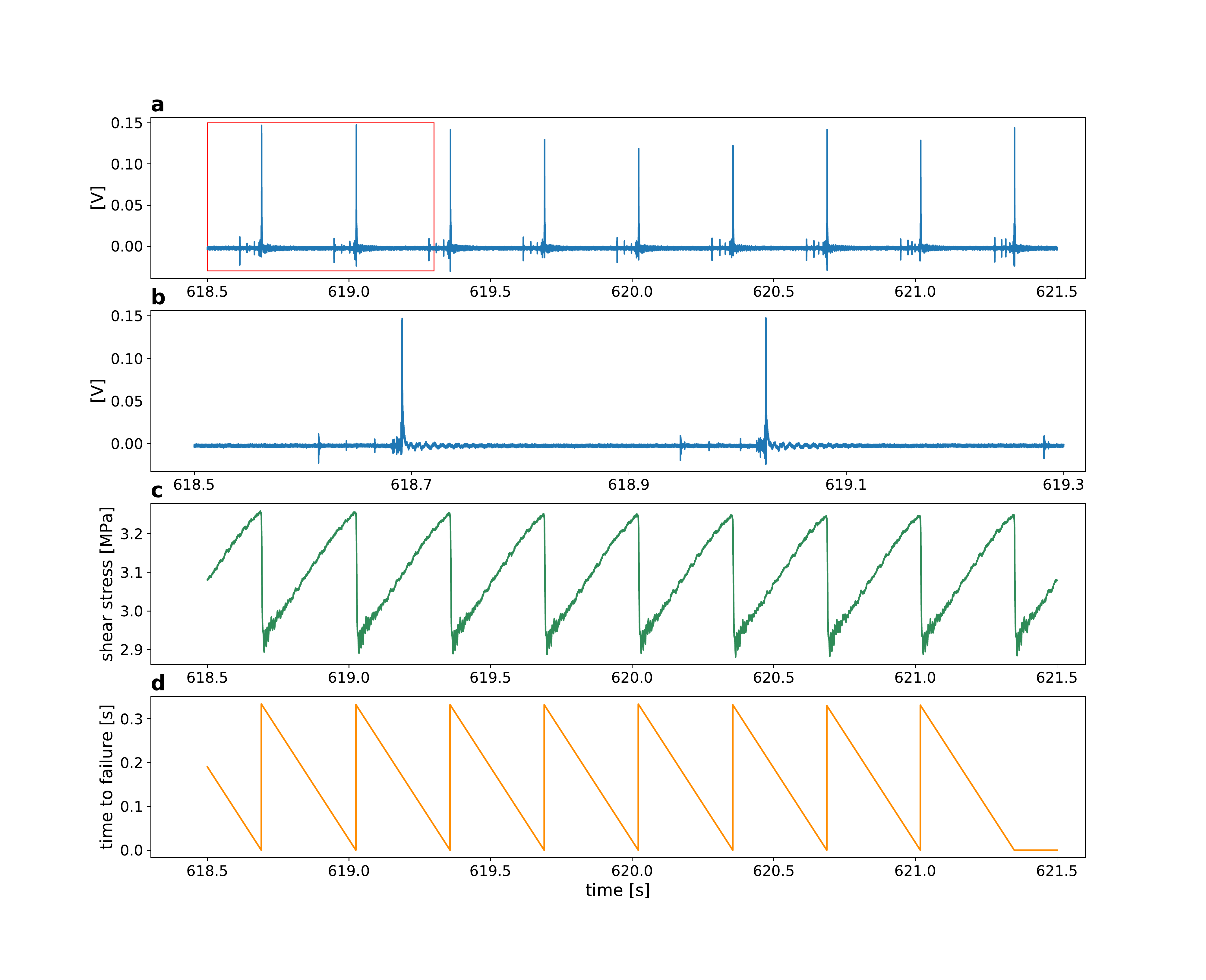}
    \caption{Continuously recorded data from the salt experiment. Sensor ch9, block 4, 620$\pm$1.5 seconds. a) Acoustic data. b) Zoom in acoustic data, red box in panel above. c) Shear stress data. d) Time to failure curve. }
    \label{data_620}
\end{figure*}

Figure \ref{acoustic_glassbeads}c (glass beads) and figure \ref{data_280}c and \ref{data_620}c (salt) show the shear stress data. The abrupt stress drops indicate slip events. Note that the shear stress build-up for glass beads is non-linear and the shear stress build up for salt is linear. This is related to the gouge behaviour, which is related to the different material properties. The glass beads do not break so porosity is high, the salt probably breaks, mainly in the pre-compaction phase, so that porosity is low. This causes deformation to become localized. The linear trend for salt indicates that salt behaves elastically. Glass beads deform non-linearly, indicating that small amounts of macroscopic slip occur before failure (aseismic creep). The shear stress measurements are a measure of the macroscopic shear stress state in the sample and are therefore a spatial average.
\\ \\
For the different load point velocities tested in the salt experiment, we observe that the higher the sliding velocity (280s), the more stick-slips are generated, in line with the rate-and-state theory. Within a three-seconds time window, block 2 (280s) counts 24 stick-slip events whereas block 4 (620s) counts 9 stick-slip events (figure \ref{data_280}a and \ref{data_620}a). 
\\ \\
The time to failure curves (figure \ref{acoustic_glassbeads}d, \ref{data_280}d and \ref{data_620}d) are based on the shear stress data as in the panel above. The duration of the interseismic period is the time between slip events, which are the large stress drops in the shear stress data. The linear decreasing trend corresponds to the time remaining before the next labquake will occur. To obtain such a curve the timing of the prior event and the event itself need to be known. 

\clearpage

\begin{figure}[ht]
    \centering
    \includegraphics[width=0.75\columnwidth]{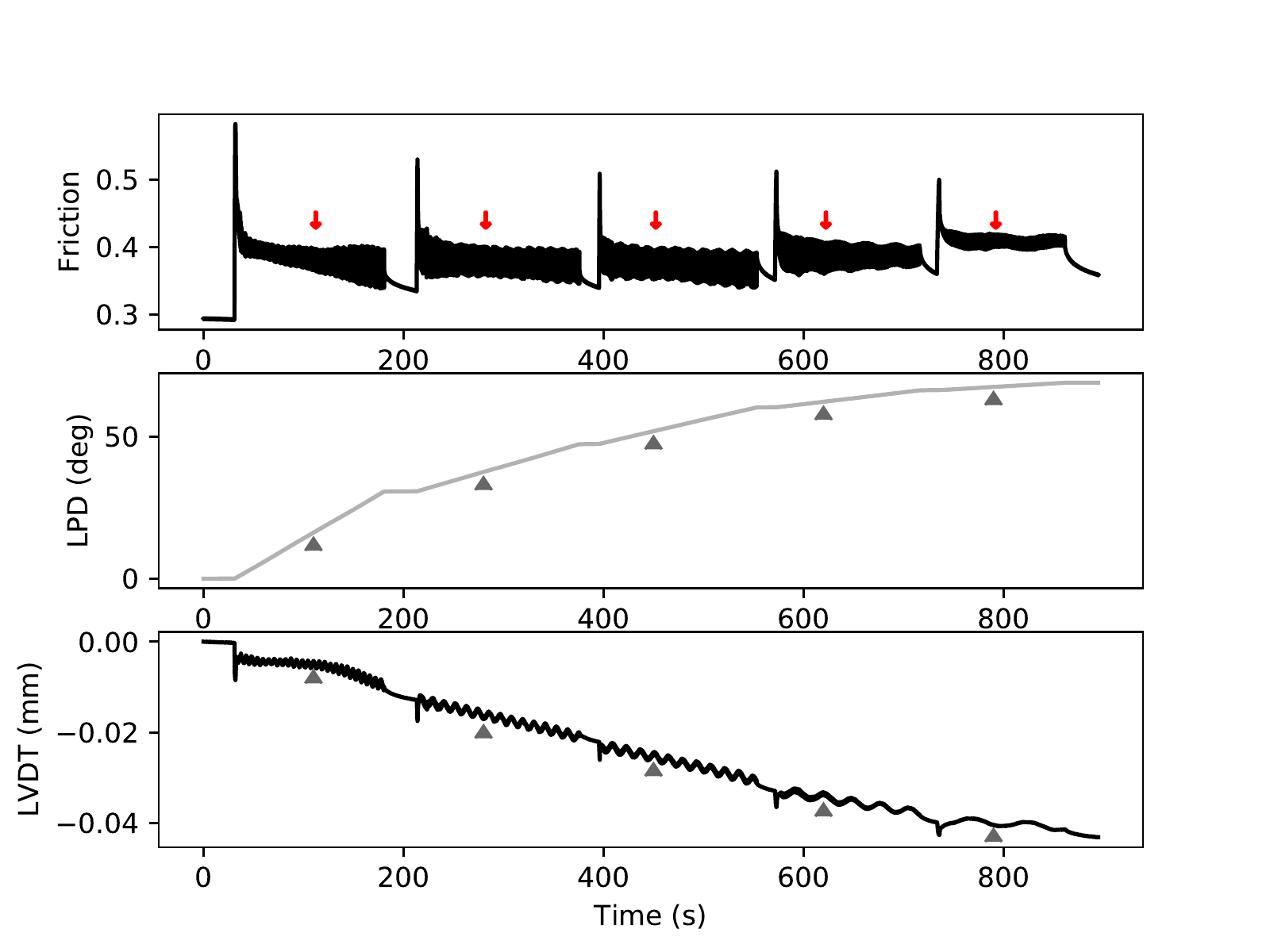}
    \caption{Upper plot: friction during the salt experiment. In the first tens of seconds after changing the load point velocity, the friction peaks, after which it enters a periodic stick slip phase. Middle plot: total amount of rotation in degrees during the experiment. Lower plot: reduction in thickness of the sample in mm during the experiment. The red arrows and grey triangles indicate the times used in the ML model: 110, 280, 450, 620, 790 seconds ($\pm$1.5 s). }
    \label{evan_viewer}
\end{figure}

\subsection{Friction, rotation and thickness}
Figure \ref{evan_viewer} shows extra recorded data during the salt experiment. The apparent coefficient of friction is the ratio of shear stress to normal stress, shown in the upper panel. As the normal stress is fixed at 8 MPa, variations in friction reflect variations in shear stress (actually figure \ref{data_280}c and \ref{data_620}c are a zoom in of this upper plot of figure \ref{evan_viewer}). The five blocks with different loading velocities can be distinguished. After changing the load point velocity the friction peaks after which it enters a periodic stick slip phase. The size of the stress drops does not change much for block 2, 3 and 4 but for (the beginning of) block 1 and for block 5 the stress drops are much smaller. 
\\ \\
The second panel (figure \ref{evan_viewer}) shows the relative angular displacement expressed in degrees. The total amount of rotation decreases relatively per block since the angular shear rate decreases. The bottom panel shows the thickness of the sample. Lower values indicate a thinner sample, so over time the sample compacts (or material is lost). We observe oscillations within a block, which seem dependent on displacement. There are more oscillations when load point velocity is high and the sample thus obtained more displacement. These oscillations are plausibly related to the influence of the extremely stiff teeth of bottom and top of the piston where the sample is placed and rotated in. The spacing between these teeth is $\sim$0.9 mm, with a slight difference between inner radius and outer radius. When looking at the displacement found by comparing the load point velocity with the period of an oscillation, we find values in the order of the teeth spacing ($\sim$ 1.2 mm). 

\subsection{Resonance}
For the salt experiment, the AE and shear stress signals show oscillations after the slip event, as we can observe in figure \ref{data_280} and \ref{data_620}, panel a, b and c. In the glass bead data we do not observe these oscillations. The three options from where this oscillating signal in salt can be generated from are: from sensor properties, from the sample itself or from apparatus resonances. If it were only caused by sensor properties the oscillation would be seen in either the AE data or the shear stress data or would at least appear at different frequencies. If it were coming from the sample it would be seen in AE and/or shear stress data but also in the LVDT data, where it is not present. The oscillations in the AE data and in the shear stress data both have a period of $\sim$7 ms, making the most plausible source resonance of the apparatus: the shaking of the total system.
\\ \\
Resonance is not desired, because the ML model will train itself on that feature in the data, which is not representative of what is happening inside the sample. However, in the result section we show that even when not taking into the account the resonance data, the model still performs well. Therefore we use the total data set, including resonance. Especially for high load point velocities, with many stick-slip events, the time span of the resonance is so large that no data would be left if it was cut out. One possibility to reduce the influence of resonance would be to filter out the resonance, that is however beyond the scope of this project.

\newpage

\section{Model description}

This research uses a machine learning technique called random forest. A random forest model is an average over a set of decision trees. Each decision tree predicts the time remaining before the next failure using a sequence of decisions based on statistical features derived from the time windows. Every split (decision) made is based on splitting the data set in two where the resulting two groups differ as much as possible. Splits are made until the data set left describes a time to failure prediction. The input data of the machine learning model is the AE data and the output is a prediction of time to failure. How long before the stick-slip event takes place? To do so the first step is splitting the full data set in two: a training set and a testing set. The model is trained on the training set where no information from the testing data is present. The performance of the model can be tested afterwards, because from the shear stress data it can be inferred when stick-slip events actually happen. This actual time to failure curve is then compared to the time to failure prediction curve, i.e. outcome of the model to see how well the prediction fits the direct time-to-failure observations. This is in short what the model does, the following sections provide more details on the process and principles from beginning to end.

\subsection{Sliding window}
First, the acoustic data are split in windows to represent a sliding window capturing the evolution of the AE data through time. The width of a window is dependent on the time between two failure events and is based on the glass beads experiment where about 5-6 adjacent window lengths fit within one interseismic period (between two slip events). For example for block 2 the interseismic time is about 0.12 seconds so the window width taken is 0.02 seconds. The chosen window widths are documented in table \ref{windowwidth}.

\begin{table}[ht]
\small
\centering
\caption{Interseismic period and window width for the five different blocks}
\begin{tabular}{l|l|l|l|}
        & \begin{tabular}[c]{@{}l@{}}Time\\($\pm$ 1.5 s)\end{tabular} & \begin{tabular}[c]{@{}l@{}}Interseismic\\period (s)\end{tabular}  & \begin{tabular}[c]{@{}l@{}}Window \\width (s)\end{tabular} \\ \hline
        
Glass beads & - & $\sim$8 & 1.8 \\ \hline
Salt block 1      & 110                  & $\sim$0.05            & 0.01            \\ \hline
Salt block 2 & 280                  & $\sim$0.12           &  0.02            \\ \hline
Salt block 3 & 450                  & $\sim$0.15            & 0.03           \\ \hline
Salt block 4 & 620                  & $\sim$0.32            & 0.06           \\ \hline
Salt block 5 & 790                  & $\sim$0.38           & 0.08           \\ \hline
\end{tabular}
\label{windowwidth}
\end{table}

\noindent This window slides through time. Every next window shifts with 10\% to the right (through time) so that the next window overlaps with 90\% of the previous window. The time sample corresponding to each individual set of data features lies at the middle of the window. For example, from the AE data from block 2, 0-0.02s form the first window, which represents the feature set at 0.01s, the second window is 0.002-0.022s, the third is 0.004-0.024s. However, windows that contain a failure event are not taken into account (figure \ref{acoustic_glassbeads}, \ref{data_280} and \ref{data_620}). The aim is to show what happens leading up to failure, not at failure itself. Note that with this approach, the time to failure predictions never reach zero due to the discretization in time imposed by the moving window approach and due to eliminating windows with failure in it. This problem vanishes with smaller windows, at the cost of increased computation time \citep{rouet17}. 

\subsection{Features}
From every window we determine 72 features. The features are a combination of signal properties and statistical measures and jointly characterise describe the physical state of the acoustic signal during that time window. The type of features can be divided in two categories: signal distribution and energy \& precursors. 

\subsubsection{Signal distribution and energy}
To capture the evolution of the energy content of the signal through time, several higher order moments are calculated as features. For every time window the mean and higher moments of the signal are determined; variance, skewness, kurtosis, all features normalized and not normalized. The equations used to determine these 6 features are described below, nn stands for not normalized.
\begin{equation}
    mean = \frac{\sum x_{i}}{N}
\end{equation}

\begin{equation}
    variance = \frac{\sum (x_{i}-mean)^{2}}{N}
    \label{variance}
\end{equation}

\begin{equation}
    skewness = \frac{1}{N}\sum \left( \frac{(x_{i}-mean)^{3}}{(\sqrt{variance})^{3}} \right)
\end{equation}

\begin{equation}
    kurtosis = \frac{1}{N}\sum \left( \frac{(x_{i}-mean)^{4}}{(\sqrt{variance})^{4}} \right)
\end{equation}

\begin{equation}
    skewness-nn = \frac{1}{N}\sum \left((x_{i}-mean)^{3} \right)
\end{equation}

\begin{equation}
    kurtosis-nn = \frac{1}{N}\sum \left( (x_{i}-mean)^{4} \right)
\end{equation}

\newpage
\noindent The mean is the average value. The higher moments (variance, skewness and kurtosis) are features based on the distribution curve of the values within the window, which describes the probability of a value to be in the data set. Variance measures how far the values are spread out from the average value. It is the average distance from the mean found in a window. High variance corresponds to a large spread of values. Skewness is a measure of the asymmetry of the probability distribution curve. High skewness indicates a more asymmetric distribution curve. Kurtosis describes the shape of the tails of the distribution curve. Data sets with high kurtosis tend to have heavy tails/outliers. Data sets with low kurtosis tend to have light tails or lack of outliers.

\subsubsection{Precursors}
The sample emits bursts of acoustic signals when the system enters a critical state leading up to failure. To quantitatively monitor this precursory activity, different percentiles and thresholds are determined during the considered time window. The 1st to 9th and 91th to 99th percentiles, by increments of 1\%, are calculated from the signal distribution curve as an indication of the amount of extreme amplitudes (18 features). The minimum and maximum value of every window are also determined (2 features).
\\ \\
The threshold measures the fraction of time that the acoustic signal spends over a threshold value $f_{0}$ within each window, given by: $\frac{1}{T} \int \mathcal{H} (f(t)-f_{0})dt$, with $\mathcal{H}(x)$ the Heaviside step function ($\mathcal{H}(x) = 0$ is $x<0$ and $\mathcal{H}(x)=1$ otherwise). The thresholds used differ for every sensor since we observe differences in noise levels. These small differences are probably related to the coupling between the sensors and the piston. The thresholds we use are shown in table \ref{thresholds} (10 features). For glass beads a strain threshold is used, which is linearly related to the AE signal. The strain thresholds used are $f_{0} = 10^{-9}, 5 \cdot 10^{-9}, 10^{-8}, 5 \cdot 10^{-8}, 10^{-7}$ and their opposite for negative strains.
\begin{table*}[ht]
\centering
\small
\caption{Threshold values used for the threshold features}
\begin{tabular}{l|lllll|lllll}
Sensor & \multicolumn{5}{l|}{Negative thresholds}        & \multicolumn{5}{l}{Positive thresholds}    \\ \hline
ch9    & -0.007  & -0.0065 & -0.006  & -0.0055 & -0.005  & 0       & 0.0005 & 0.001  & 0.0015 & 0.002  \\ \hline
ch11   & -0.0075 & -0.007  & -0.0065 & -0.006  & -0.0055 & -0.0005 & 0      & 0.0005 & 0.001  & 0.0015 \\ \hline
ch13   & -0.006  & -0.0055 & -0.005  & -0.0045 & -0.004  & 0.001   & 0.0015 & 0.002  & 0.0025 & 0.003  \\ \hline
ch15   & -0.0032 & -0.0027 & -0.0022 & -0.0017 & -0.0012 & 0.0038  & 0.0043 & 0.0048 & 0.0053 & 0.0058
\label{thresholds}
\end{tabular}
\end{table*}
\\ \\
Moreover, to obtain more features describing the physical state of the window, the window is split in two. We determine all the features described above on the first half of the data in the window, as well as for the second half of the window. We refer to those features as \textsc{\char13}before\textsc{\char13} and \textsc{\char13}after\textsc{\char13} heretoforth as well as within in the code we use for our data processing. All 72 features (6+18+10+2=36, 36x2) are normalized, between -0.5 and 0.5, to the maximum value of that feature. When looking at the feature evolution (result section: figure \ref{features_glassbeads}, \ref{features_salt280} and \ref{features_salt620}) most features increase towards failure. Therefore a specific value of a feature can be related to the amount of time left before failure. As said earlier, we do not use windows that contain failure, because at the moment failure takes place, all the statistical features can vary within several orders of magnitude compared to their inter-failure counterparts. After all, we want to find what happens leading up to failure. The combination of 72 features describes one window, so one data point in time. The decision tree decides which features are most important and thus contribute most to the model.

\subsection{Decision Trees}
\begin{figure}[ht]
    \centering
    \includegraphics[width=0.85\columnwidth]{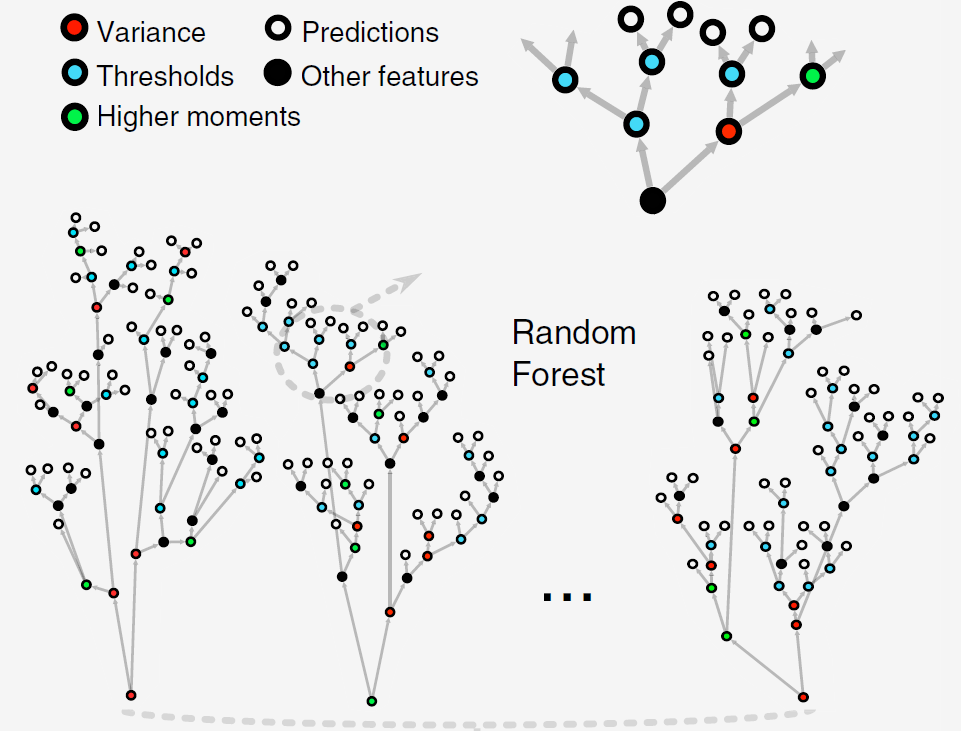}
    \caption{Visualization of decision trees and a random forest. The random forest model predicts the time remaining before the next slip event by averaging the predictions of many (in our model: 1000) decision trees. Each tree makes its prediction (white leaf node), following a series of decisions (coloured nodes) based on features of the acoustic signal of the considered window. Figure from \citep{rouet17}.}
    \label{decisiontree}
\end{figure}

The building blocks of a random forest model are decision trees. A single tree is a predictor on its own. It is build purely on the training data set, i.e. the first half of the data. To build a decision tree we take a subset of the training data set. A decision tree can be thought of as a series of yes/no questions asked about the data, eventually leading to a predicted value, visualized in figure \ref{decisiontree}. For example, is kurtosis higher than 0? All data points (windows) that answer yes go to one node, when the answer is no the data points go to the other node. Thus, a tree represents a map from inputs (acoustic emission features) to output labels (time remaining before failure prediction) \citep{rouet17}. 
\\ \\
A split is made by dividing the data set in two subsets which are as different as possible from each other but at the same time with minimal difference within the subset. A split is made only by one feature, the one that causes the largest reduction in variance. The function to measure the quality of a split is called MSE (the mean squared error), which is equal to variance reduction (equation \ref{variance}). The variance of both created subsets is calculated and the weighted average of both is the variance of the split, which is minimized \citep{Prratek_2018}. We are dealing with a regression problem, where the target (time to failure) is a discrete, real-valued function and not a label or catagory. The leaf/end node will therefore be a time to failure value with a probability range.

\subsubsection{Bias and variance}
\begin{figure}
    \centering
    \includegraphics[width=0.8\textwidth]{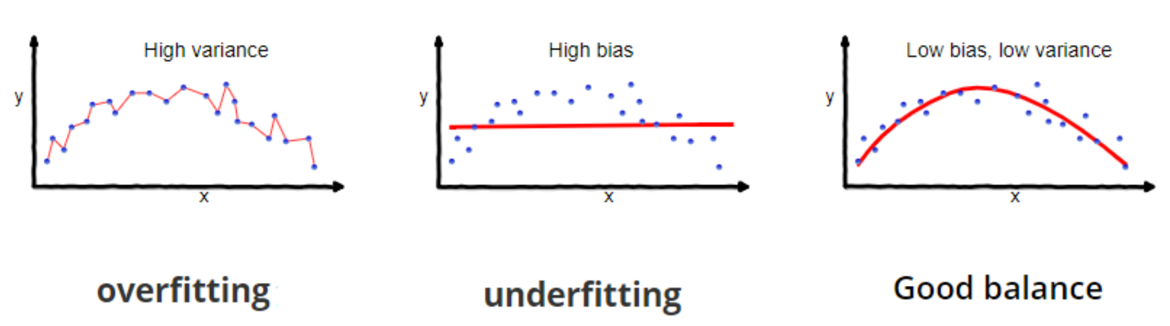}
    \caption{Bias-variance trade off. High variance leads to error on the test data, high bias leads to error on both training and test data. The key is to find a balance between bias and variance.}
    \label{biasvariance}
\end{figure}

A tree is based on the actual time to failures and the maximum depth (number of splits) is not necessarily limited. If as many splits as possible can be made, all data points will fall in the correct leaf, because in that case a leaf would consist of only one sample. The machine learning model however needs to be more general because it needs to be applicable to data it has never seen before, data that can be slightly different. If the depth of the tree is not set to a maximum level, overfitting occurs since the model memorizes the training data by fitting the data perfectly. The model will learn not only the actual relationships in the training data, but also any noise that is present, while ignoring the potential existence of a model null space in the prediction model. If a new tree with a slightly different training data set is produced, this tree will vary considerably from the other, when the maximum depth is not limited. 
\\ \\
On the other hand, if the depth of the tree is limited too much, it makes assumptions about the training data which might not be true, thus inducing regularisation-driven bias. The model is biased if it under or overpredicts the target variable (time to failure) systematically. An essential concept in machine learning is the bias-variance trade off (figure \ref{biasvariance}). It is the balance between creating a model that is so flexible it memorizes the training data versus an inflexible model that cannot learn the training data. As an alternative to limiting the depth of the tree, which reduces variance (good) and increases bias (bad), we combine many decision trees into a single ensemble model known as the random forest \citep{koehrsen_2018}.

\subsection{Random Forest}
Our random forest is an ensemble model of many decision trees (figure \ref{decisiontree}), in our model a thousand trees are combined. Rather than simply averaging the prediction of trees (which would be a normal forest), this model uses two key concepts that makes it random and it ensures that the behaviour of each individual tree reduces correlation between the trees in the model: \\ \\
1. Random sampling of training data points when building trees: bagging. \\
2. Random subsets of features considered when splitting nodes.

\subsubsection{Bagging (bootstrap aggregating)}
\begin{figure}[ht]
    \centering
    \includegraphics[width=0.8\columnwidth]{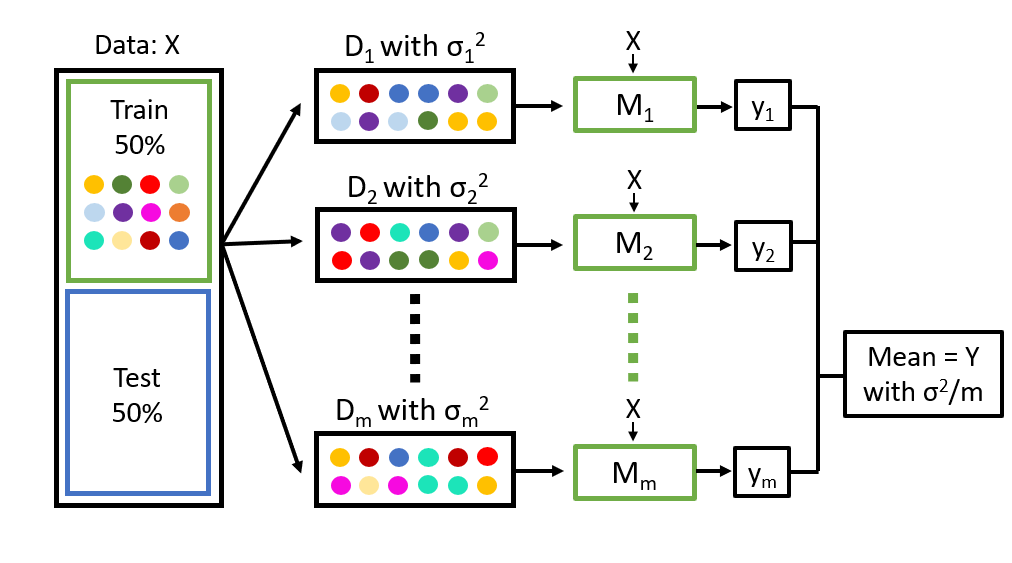}
    \caption{Schematic overview of a random forest with bootstrapping indicated with the colored dots. Each data set D contains a slightly different data set than the original data set X, with a corresponding new variance $\sigma$. A decision tree is build on this new data set resulting in model M. Data X is put in all the models resulting in slightly different predictions. The average is taken over all the different predictions to obtain prediction Y which has as result a lower variance than each individual tree. Letter m indicates the number of trees.}
    \label{bagging}
\end{figure}

The random forest trains each tree on a slightly different data set, a random sample of the data points. The samples are drawn with replacement from the original training set, referred to as bootstrapping, invented by \citet{breiman2001random}. With replacement we mean that we form the new data set by selecting random data points one by one from the total original data set. When selecting that data point, it is not discarded from the original set. So when selecting a new random data point to fill the new data set it can choose from the total original data set again. In that way the new data set will not have all data points included and at the same time some data points will be present multiple times. This is visualized in figure \ref{bagging}. 
\\ \\
The idea is that by training each tree on different samples, although each tree might have high variance with respect to a particular set of the training data, overall, the entire forest will have lower variance but not at the cost of increasing the bias \citep{koehrsen_2018} \citep{breiman2001random}. Each tree runs independently and the outputs are aggregated at the end, without preference to any model. A large number of uncorrelated errors average out to zero, so bagging reduces the variance of the prediction \citep{Prratek_2018}. The key assumption here is thus that both the variance and bias of individual trees are uncorrelated due to the randomness of the bagging process. This improves both the stability and the accuracy of the model. Such an approach is called an ensemble method: together the trees make more accurate predictions than an individual tree.

\subsubsection{Random feature selection}
The random forest not only trains each tree on a slightly different set of the training data (bagging) but also makes the splits in each tree by considering a limited number of the available features. The number of features that can be the cause of split at each node is limited to some percentage of the total. This ensures that the ensemble model does not rely too heavily on any individual feature, and makes fair use of all potentially predictive features \citep{chakure_2019}.

\subsection{Hyperparameter tuning}
There are model parameters that are set by the user before making the model, called hyperparameters. For example the number of trees: 1000 in this research. For some parameters the model uses a grid search, where the model tests different values for the parameters and the value that turns out to make the best prediction is chosen. These parameters are: the maximum features to consider to make a split, the minimum amount of data points on a leaf (end node) and the minimum amount of data points on a node to make a split. The model decides which values are most optimal by testing a given set of options.
\\ \\
The grid search is based on a 3-fold cross-validation. The training set (first half of AE data) is split into three smaller sets. The model trains itself on two of the three folds. The resulting model is validated on the remaining part of the data, the third fold (i.e., it is used as a test set to compute a performance measure). This is repeated three times, taking different folds as training data. At the very end of training, the performance on each of the folds is averaged to come up with final validation metrics for the model. The best hyperparameters are chosen to build the random forest model.

\subsection{Feature reduction}
To make the model perform even better recursive feature elimination (RFE) is applied. RFE is a feature selection method that fits a model and removes the weakest features until the specified number of features is reached. Features are ranked by their importance, and by recursively eliminating a small number of features per loop, RFE attempts to eliminate dependencies and correlated features that may exist in the model. Having irrelevant features in the data can decrease the accuracy of the model.
\\ \\
The ranking criterion used by the RFE algorithm is the permutation importance measure \citep{breiman2001random}. It directly measures the impact of each feature on the accuracy of the model. The general idea is to permute (shuffle around) the values of each feature one by one and measure how much the permutation decreases the accuracy of the model. For unimportant features, the permutation should have little to no effect on model accuracy, while permuting important variables should significantly decrease it. The most useful features are then selected recursively until a desired number of features is reached. RFE has been shown to be among the best methods for feature selection with random forest \citep{gregorutti2017correlation}.

\newpage

\section{Results}

\subsection{Model results}
This section shows the RF model results and describes the most important features for both glass beads and salt. In the case of salt we present the results of two blocks: 2 and 4. The prediction results of the other three blocks are found in appendix \ref{appendix2}.

\subsubsection{Glass beads experiment}
\begin{figure}[ht!]
    \centering
    \includegraphics[page=13,width=0.8\textwidth]{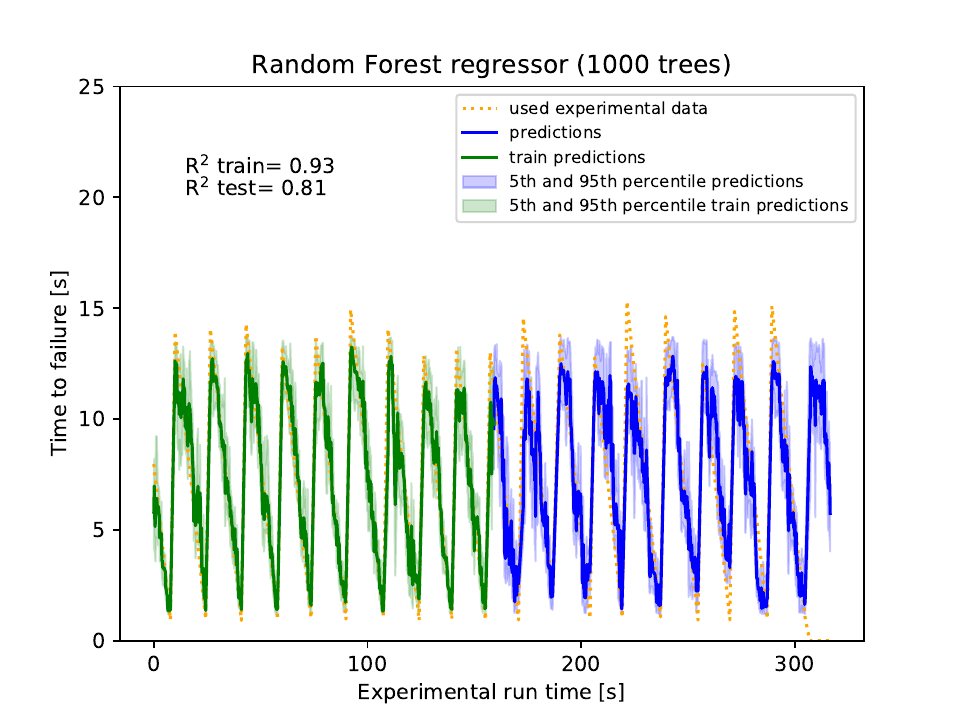}
    \caption{Random forest model for the glass beads experiment. The performance of the training data set (green) is 0.93 and the performance of the testing data set (blue) is 0.81 as indicated with the $r^{2}$ score. The shaded green and blue colors show the 5th and 95th percentile of the predictions corresponding to uncertainties in the predictions. The orange dashed line is the used time to failure data inferred from the shear stress data. It excludes the time windows that contain the failure event.}
    \label{Glassbeads}
\end{figure}

\begin{figure}[ht!]
\centering
\subfigure[Most important features according to the random forest model (figure \ref{Glassbeads}). Feature reduction is applied till 24 features were left. The relative importance between these feature is shown in the graph, the black bar indicates one standard deviation.]{
\label{most_imp_glassbeads}
\includegraphics[page=14,height=2.7in]{figures/RFE_randomforest_regressor_double_remove_outliers.pdf}}
\qquad
\subfigure[The four most important features. The upper panel shows the AE signal, the lower four panels the features. Since windows with failure are not taken into account, there are data gaps between stick-slips. Note the zoomed y-axis for the kurtosis nn plots to show the trend in the data.]{
\label{features_glassbeads}
\includegraphics[height=3.1in]{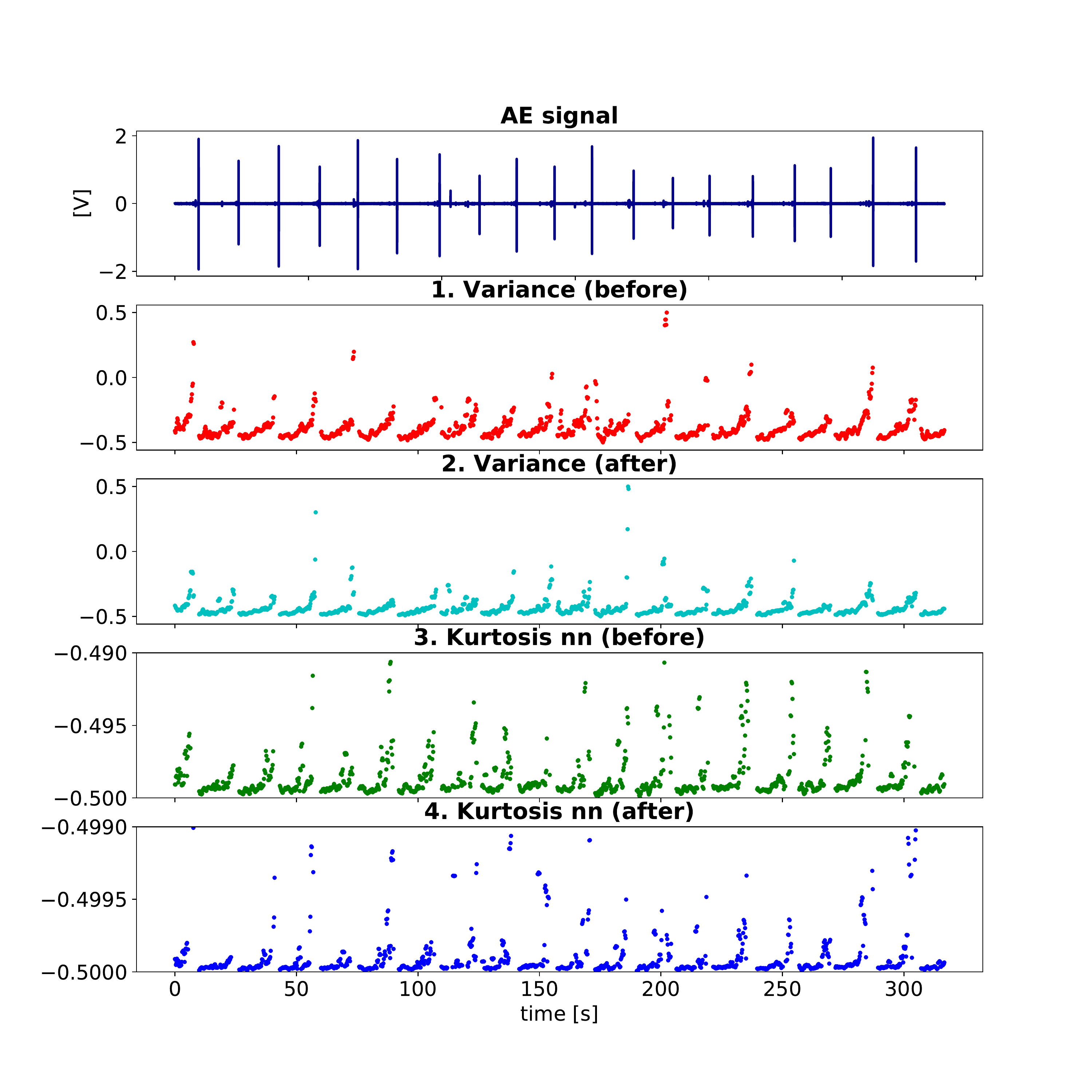}}
\caption{Visualization of the features that are used in the random forest model of the glass beads experiment and their relative importance.}
\end{figure}

Figure \ref{Glassbeads} shows the result of the prediction model. The green line shows the training data, the blue line the testing data. The dashed orange line represents the actual time to failure curve inferred from the shear stress data. The predictions are quite accurate, with a little better $r^{2}$ score for the training data than for the testing data. The shaded colours show the prediction range between 5 and 95 percentile, a 95\% certainty measure that the actual value falls within this range. Note that the predictions never reach zero since the windows that contain failure are not included in the model. The peak height of the time to failure curve varies between $\sim$13 and 16 seconds within the experiment. The prediction model of \citet{rouet17} reached $r^{2}$ values of 0.91 for training and 0.89 for testing. Comparing those values to the values we found with our model, we find that they are quite similar. We rather well achieved to reproduce their result and we can convincingly set our prediction model as a benchmark to compare the salt experiment data against.
\\ \\
Following the recursive feature elimination approach described above, the model eliminates the least usable features recursively until 24 features are left (figure \ref*{most_imp_glassbeads}). Their relative importance is ranked, with variance (before and after) as the most important features. There are still many features with very low importance. Figure \ref{features_glassbeads} shows the four most important features. For all four, we observe a gradual increase towards failure. For the \textit{kurtosis not normalized} plots we zoom in on the y-axis to observe a similar trend as the trend seen for variance. Close to failure the values for kurtosis are so high that, after normalizing the values between [-0.5,0.5], the trend would not be visible otherwise.

\subsubsection{Salt experiment}
\begin{figure}[ht]
    \centering
    \includegraphics[page=16,width=0.8\textwidth]{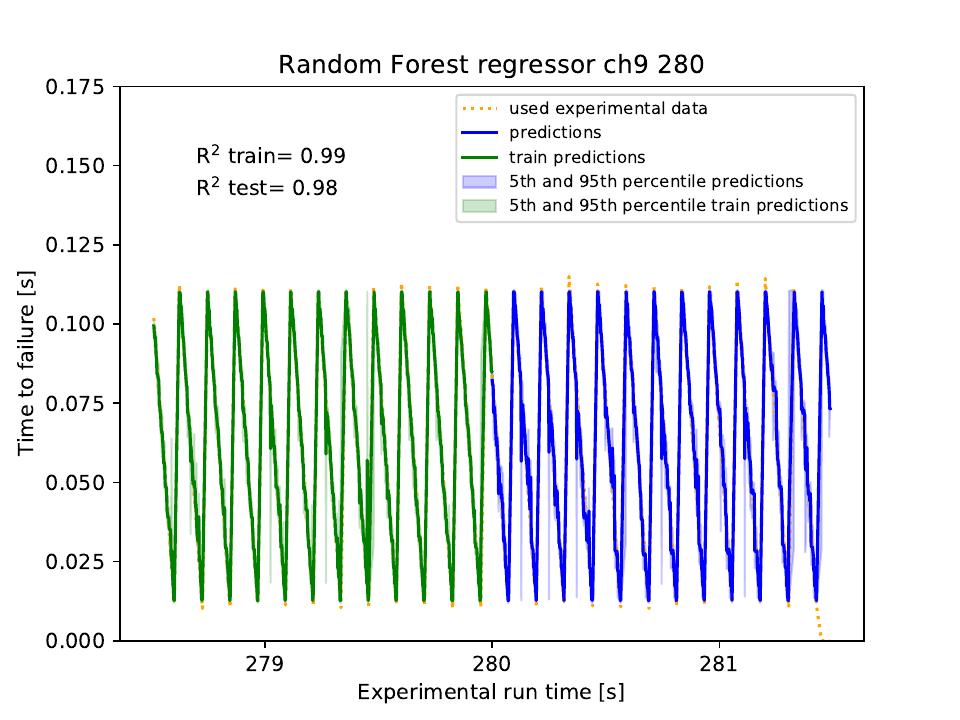}
    \caption{Random forest model for the salt experiment: sensor 9 from 278.5 - 281.5 seconds. The performance of the training data set (green) is as good as the performance of the testing data set (blue) as indicated with the $r^{2}$ score. The shaded green and blue colors show the 5th and 95th percentile of the predictions corresponding to uncertainties in the predictions. The orange dashed line is the used time to failure data, inferred from the shear stress data, excluding the time windows that contain the failure event.}
    \label{ch9_280}
\end{figure}

This section presents the results for block 2 and 4 of the salt experiment. Figure \ref{ch9_280} shows the outcome of the model for block 2. The actual time to failure curve in dashed orange is almost not visible, because it is hidden behind the prediction curve. These predictions are highly accurate with a very high $r^{2}$ score. The peaks are all of almost equal height at a time of $\sim$0.11 seconds.

\clearpage

\begin{figure}[ht!]
\centering
\subfigure[Most important features according to the random forest model (figure \ref{ch9_280}). Feature reduction is applied till 32 features were left. The relative importance between these feature is shown in the graph, with the black bar indicating one standard deviation.]{
\label{most_imp_salt280}
\includegraphics[page=18,height=2.9in]{figures/E_RFE_randomforest_ch9_280.pdf}}
\qquad
\subfigure[The four most important features. The upper panel shows the AE signal, the lower four panels the features with gaps when failure occurs, since those windows are not taken into account.]{
\label{features_salt280}
\includegraphics[height=2.9in]{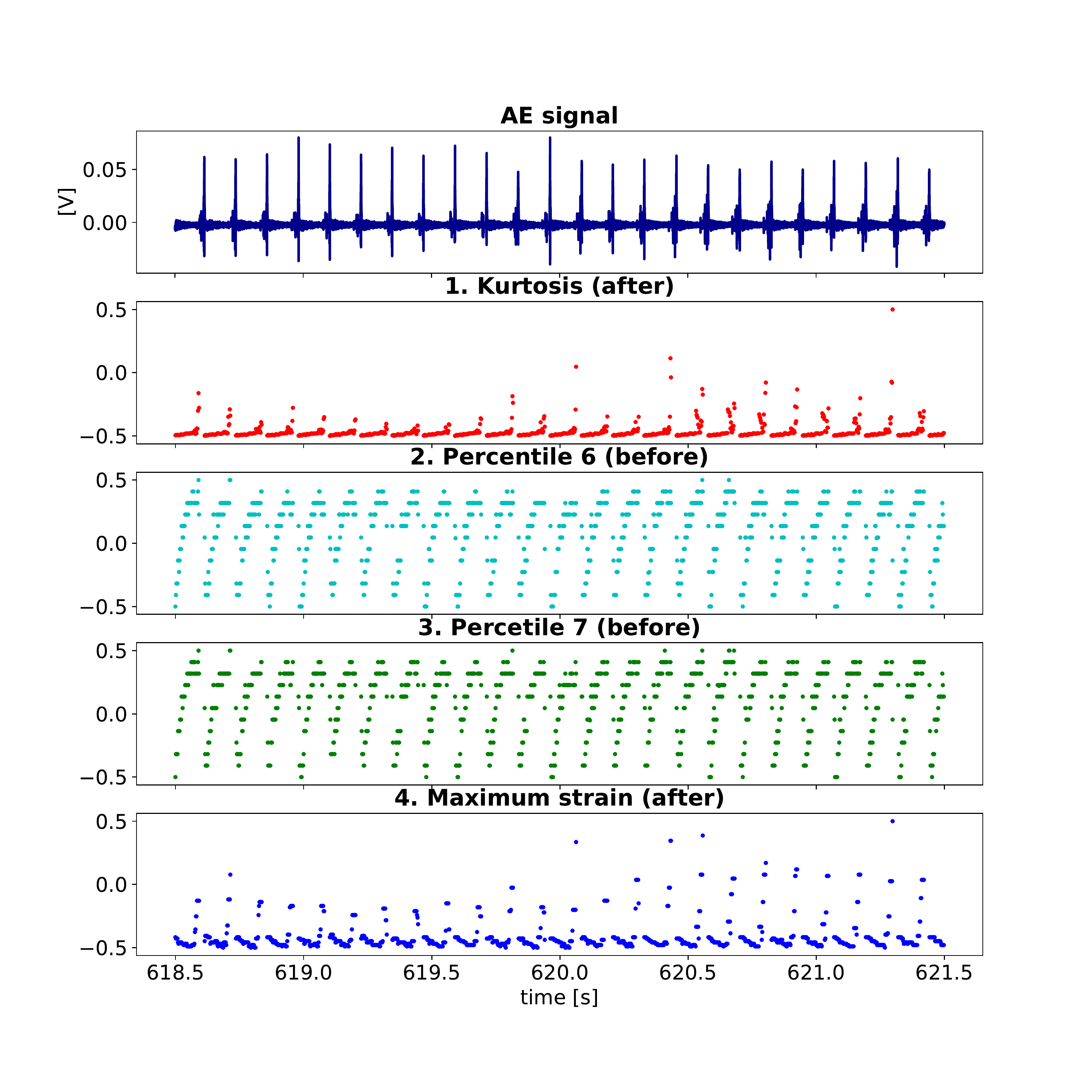}}
\caption{Visualization of the features that are used in the random forest model of block 2 and their relative importance.}
\end{figure}

\begin{figure}[ht!]
    \centering
    \includegraphics[page=16,width=0.8\textwidth]{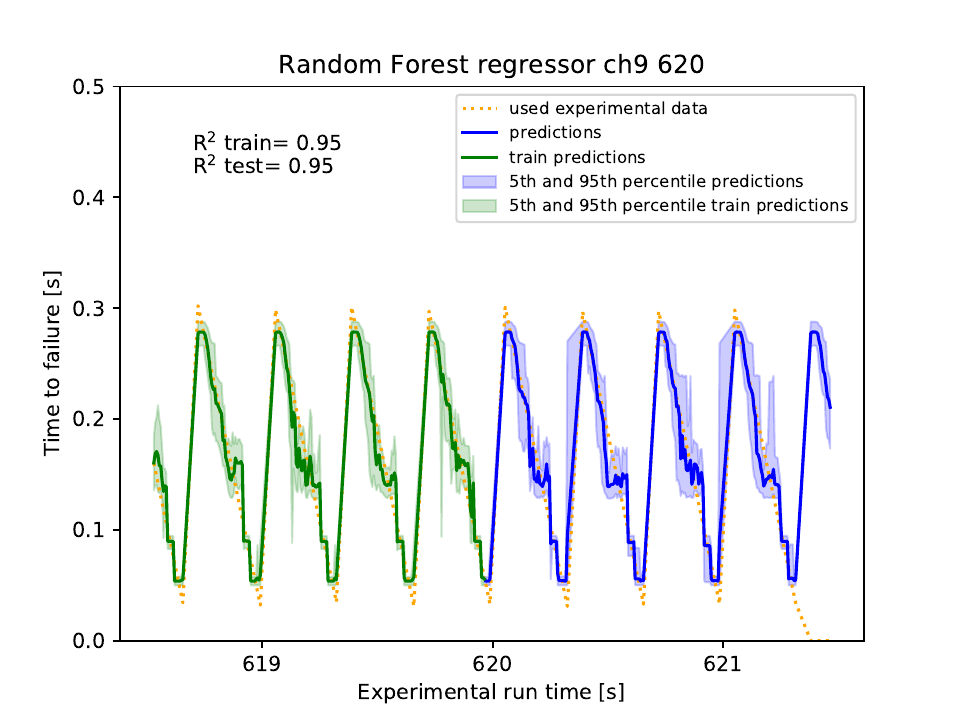}
    \caption{Random forest model for the salt experiment: sensor 9 from 618.5 - 621.5 seconds. The performance of the training data set (green) is as good as the performance of the testing data set (blue) as indicated with the $r^{2}$ score. The shaded green and blue colors show the 5th and 95th percentile of the predictions corresponding to uncertainties in the predictions. The orange dashed line is the used time to failure data, inferred from the shear stress data, excluding the time windows that contain the failure event.}
    \label{ch9_620}
\end{figure}

\begin{figure}[ht!]
\subfigure[Most important features according to the random forest model (figure \ref{ch9_620}). Feature reduction is applied till 32 features were left. The relative importance between these feature is shown in the graph, with the black bar indicating one standard deviation.]{
\label{most_imp_salt620}
\includegraphics[page=18,height=2.8in]{figures/E_RFE_randomforest_ch9_620.pdf}}
\qquad
\subfigure[The four most important features. The upper panel shows the AE signal, the lower four panels the features with gaps when failure occurs since those windows are not taken into account.]{
\label{features_salt620}
\includegraphics[height=2.8in]{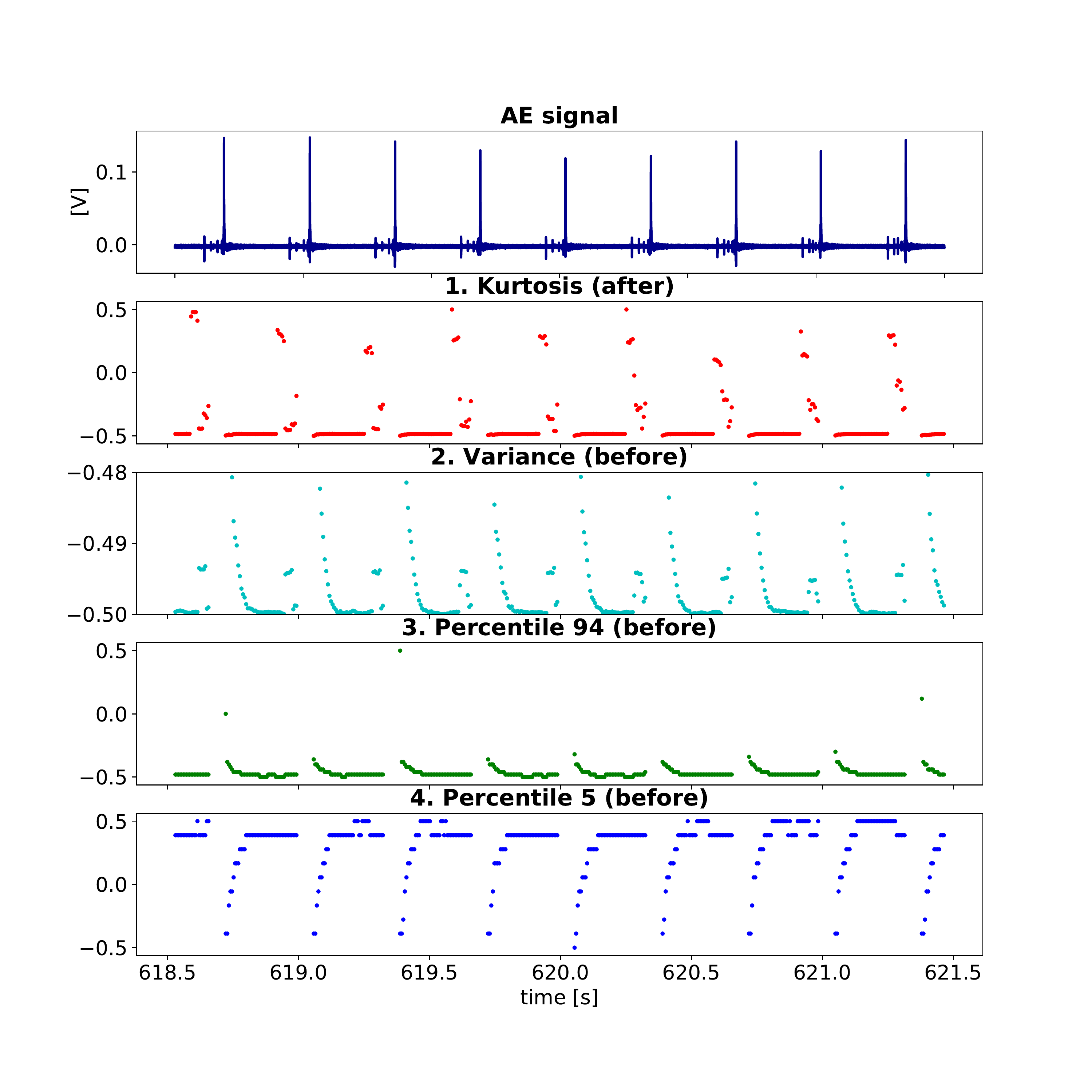}}
\caption{Visualization of the features that are used in the random forest model of block 4 and their relative importance.}
\end{figure}

Figure \ref{most_imp_salt280} shows the most important features, with kurtosis clearly as the most important feature. Percentile 6 and 7 seem also important. For this experiment the feature variance is not a very important parameter according to the random forest model. Figure \ref{most_imp_salt280} shows the four most important features. Kurtosis increases towards failure. Percentile 6 and 7 are also increasing in the interseismic period but have a convex shaped curve. The maximum strain decreases slightly in the beginning of the interseismic period and increases again towards the end of the interseismic period.
\\ \\
Figure \ref{ch9_620} shows the result of block 4. The prediction curve for every stick-slip looks very similar. In the middle of the time to failure 
curve, the curve flattens a little, with higher uncertainties. The peaks within this block are again of similar height. The difference between block 2 and 4 is the maximum height which is related to the amount of slip events within this time span of three seconds (figure \ref{ch9_280} and \ref{ch9_620}). The most importance features are kurtosis, variance and percentile 94 and 5 (figure \ref{most_imp_salt620} and \ref{features_salt620}). Other features are of less importance but contribute more to the outcome than the less important features for the model of block 2 and the glass beads experiment. 
\\ \\
Kurtosis is the most important feature with an increase just before failure. Variance decreases steeply in the beginning and rises a little before the slip event. Both the percentiles are changing in the first part of the interseismic period and do not change much close to failure (figure \ref{most_imp_salt620}).

\newpage
\subsection{Without resonance}
\begin{figure}[ht!]
    \centering
    \includegraphics[page=16,width=0.78\textwidth]{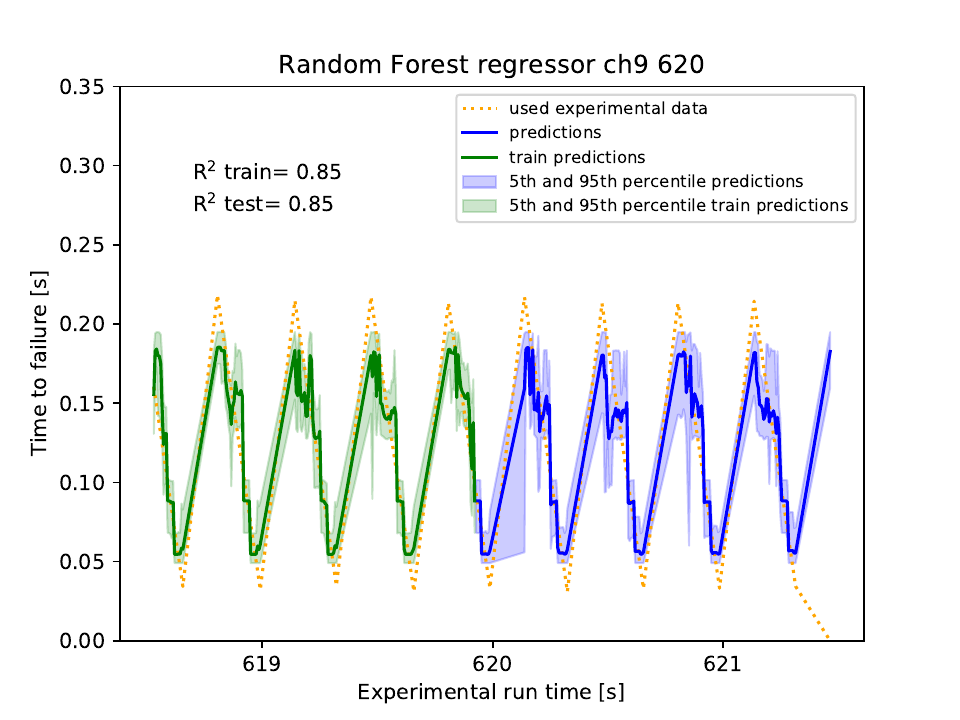}
    \caption{Random forest model for the salt experiment: sensor 9 from 618.5 - 621.5 seconds, without resonance. The performance of the training data set (green) is as good as the performance of the testing data set (blue) as indicated with the $r^{2}$ score. The shaded green and blue colors show the 5th and 95th percentile of the predictions corresponding to uncertainties in the predictions. The orange dashed line is the used time to failure data, inferred from the shear stress data, excluding the time windows that contain the failure event.}
    \label{ch9_620_wr6}
\end{figure}

One of the main differences between the glass beads experiment and the salt experiment is the method of measuring the AE data. For glass beads an accelerometer is used, for salt four sensors. \citet{rouet17} point out that they improve the signal to noise ratio by selecting the peak of the system resonances. The glass beads data does not show visible resonance after a slip event.
\\ \\
As observed in figure \ref{most_imp_salt280} and \ref{most_imp_salt620}, features do show variation just after the slip event. It would make sense if these variations help the model in predicting long times before failure, since resonance always appears when the slip event just happened and its amplitude decreases with time. We do not want to include resonance in the prediction model. In this case it provides information about a next slip event, but in nature there is no machine constantly resonating (even though aftershocks are observed for some earthquakes). The information has to come from the sample itself, so we test the influence of resonance on the model.
\\ \\ 
To check whether the model still performs well without the resonance, the resonance is cut out the data for block 4. Figure \ref{ch9_620_wr6} shows the model result. The performance is slightly worse, probably due to the fact that there is less data, however a performance of 0.85 is still very accurate. The model is able to predict time to failure also without the resonance data. It seems that the confidence level of the prediction systematically decreases as we approach failure. This shows that there is more information close to failure. Bare in mind that the upward slope does not contain any information, since those times represent windows that contain failure. Those upward slopes are basically data gaps. 

\begin{figure}[ht!]
\subfigure[The most important features when removing the windows that contain the resonance of the system, according to the random forest model (figure \ref{ch9_620_wr6}). Feature reduction is applied till 32 features were left. The relative importance between these feature is shown in the graph, with the black bar indicating one standard deviation.]{
\label{most_imp_salt620_wr6}
\includegraphics[page=18,height=2.9in]{figures/E_RFE_randomforest_ch9_620_wr6_2.pdf}}
\qquad
\subfigure[The four most important features of the random forest model without system resonance. The upper panel shows the AE signal, the lower four panels the features with gaps when failure occurs since those windows are not taken into account.]{
\label{features_salt620_wr6}
\includegraphics[height=2.9in]{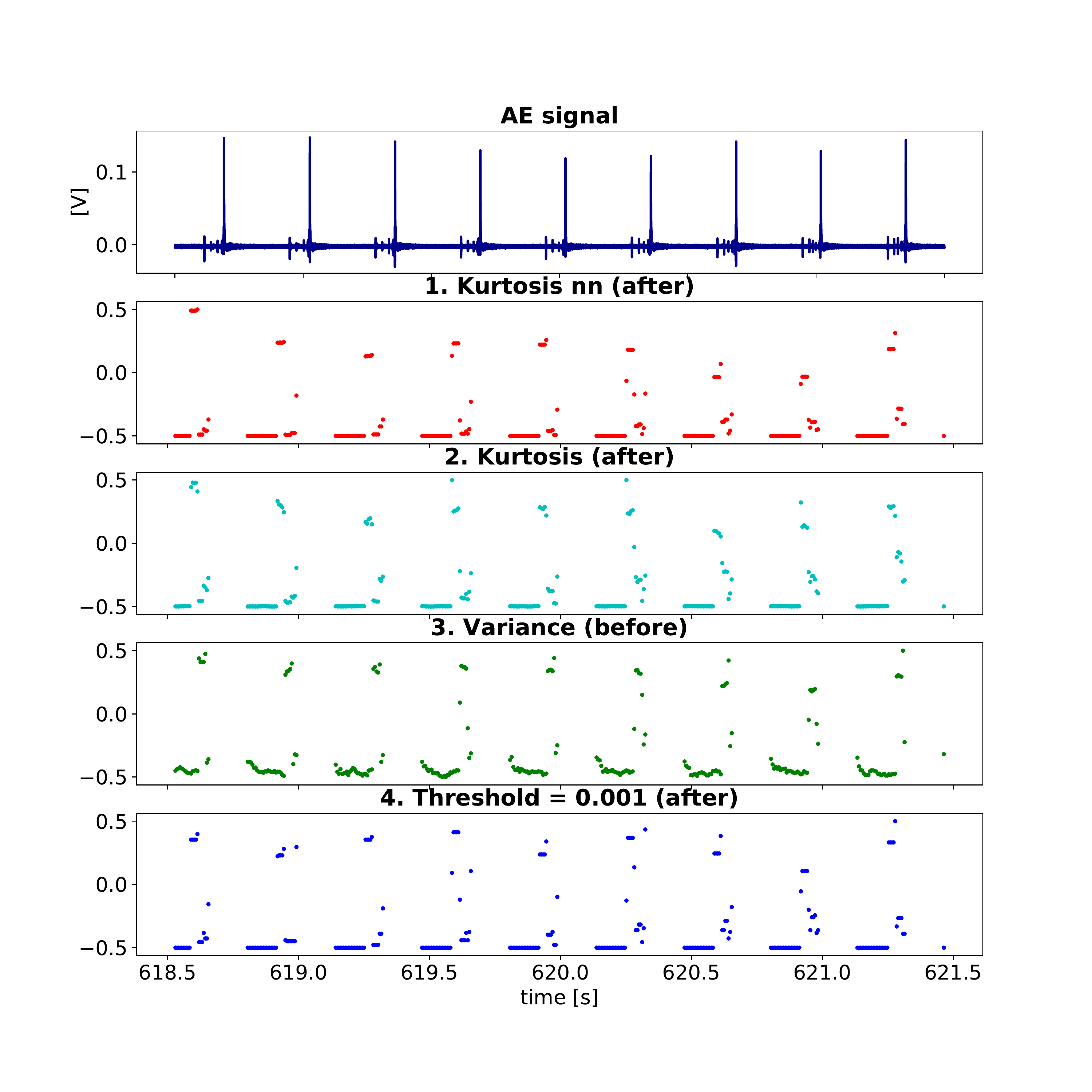}}
\caption{Visualization of the features that are used in the random forest model without the data that contains resonance and the relative importance of the features.}
\end{figure}

\clearpage
\noindent It is interesting to compare what features are most important in this case, since resonance will not affect the features. Therefore these features are the most important ones concerning what is happening within the sample. Figure \ref{most_imp_salt620_wr6} shows that all features increase when precursors are observed and do not change much in the beginning, in contrast to most features in the model with resonance. The four most important features did change: the percentiles are not important in the model without resonance. \textit{Kurtosis} and \textit{variance} are important for both models.

\subsection{Mock data set}
\begin{figure}[ht]
\subfigure[The data from one block is split in two. Half is used for training, the other half is used for testing. The half size data sets of the different blocks are placed next to each other. The $r^{2}$ scores are again very high, the model is capable of predicting the time to failures.]{
\label{super}
\includegraphics[page=1,width=.5\textwidth]{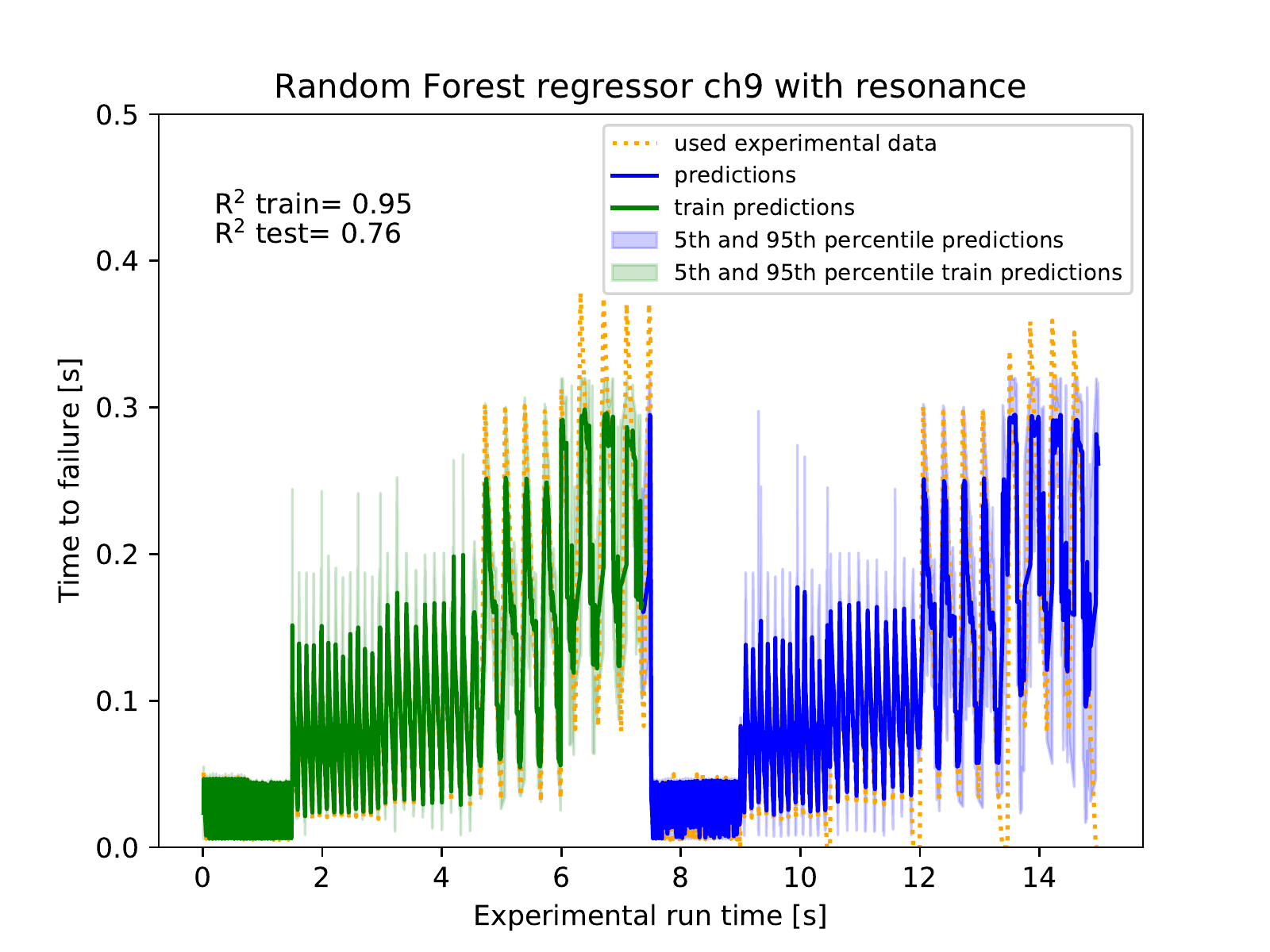}}
\qquad
\subfigure[Same result as figure  \ref{super} but before ordering the indices. This graph shows that the model treat one window at a time so only the data in that window is considered when making the prediction. The prediction is not based on neighbouring windows.]{
\label{superrandom}
\includegraphics[page=1,width=.5\textwidth]{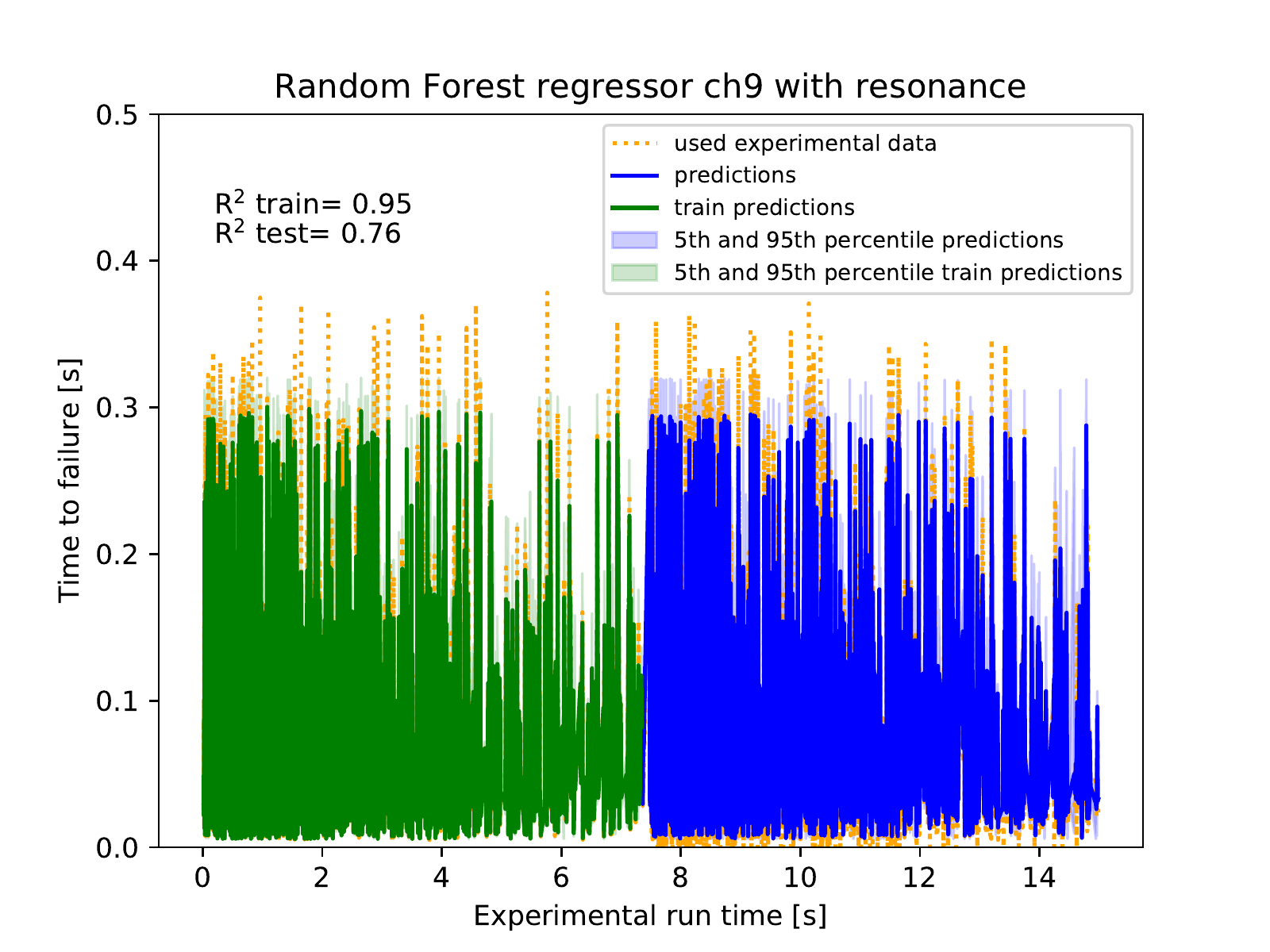}}
\caption{Model of mock dataset containing salt data from all five blocks, sensor 9.}
\end{figure}

To test whether it would make a difference to have data from all five blocks with different sliding velocities in the training set, we build a mock data set containing information from all load velocities. Every block participates for 1.5 seconds. The test set has a similar structure with 1.5 seconds from every block. The predictions are again very accurate (figure \ref{super}) and the model is flexible enough to predict time to failure from every block. To clarify, placing the blocks in a different order does not make a difference in model performance since all the data from both training and testing sets is randomized before making the decision trees in the RF model (figure \ref{superrandom}). After building the model, the original order is re-obtained (figure \ref{super}). The prediction of one data point is based on one window and no information of neighbouring windows is taken into account when building the trees. Every window is evaluated independently. Figure \ref{super} shows the data how we presented it to build the model, figure \ref{superrandom} shows how the model treats all the information. It always randomize the windows before building and testing the model. The prediction model works, confirming that the method of building a prediction model works also with very variable data. Moreover we show with this exercise that the random forest model evaluates single data points and that the order of the data does not matter.

\subsection{Different load point velocities}
\begin{wrapfigure}{R}{2.7in}
\includegraphics[width=2.7in]{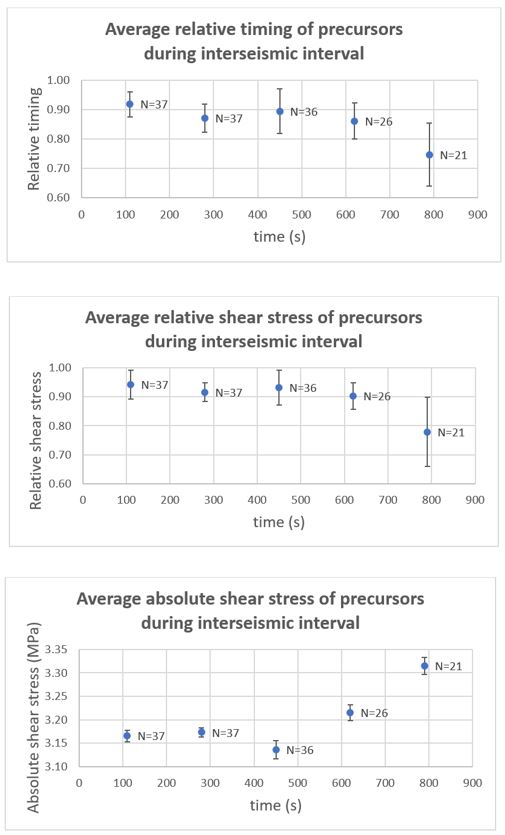}
\caption{Data on the relative timing, relative shear stress level and absolute shear stress level of 157 precursors from the five different blocks.}
\label{precursors}
\end{wrapfigure}

For the salt experiment we checked whether there is a specific critical shear stress threshold at which precursors occur. From figure \ref{acoustic_glassbeads}, \ref{ch9_280}, \ref{ch9_620} and appendix \ref{appendix1} the exact shear stress values corresponding to the precursor times are noted (appendix \ref{appendix4}) and visualized in figure \ref{precursors}. There is no specific threshold found and also no correlation in the absolute shear stress data of precursors. Additionally, we checked what the relative timing of a precursor is during the interseismic interval: a relationship is found. The lower the load point velocity, the relatively earlier a precursor is established. For block 1 the first precursors appeared at $\sim85\%$, for block 5 at $\sim60\%$ of both the time and shear stress interval (appendix \ref{appendix4}). Figure \ref{precursors} shows the average taken of all precursors that are evaluated.

\subsection{Difference between sensors}
\begin{figure}[ht]
\subfigure[The acoustic emissions recorded by four different sensors for block 4 (618.5-621.5s). The amplitudes for every sensor are different and some peaks are not symmetric around the y-axis. The scales are the same for all subplots.]{
\label{differentsensors}
\includegraphics[width=.5\textwidth]{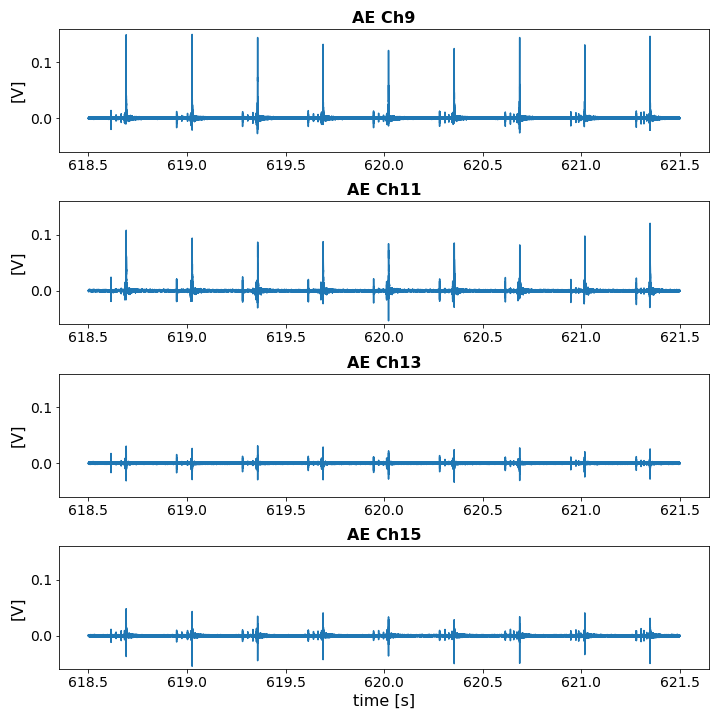}}
\qquad
\subfigure[A precursor recorded by the four sensors. The signal is recorded first by sensor ch11.]{
\label{zoomprecursors}
\includegraphics[width=.5\textwidth]{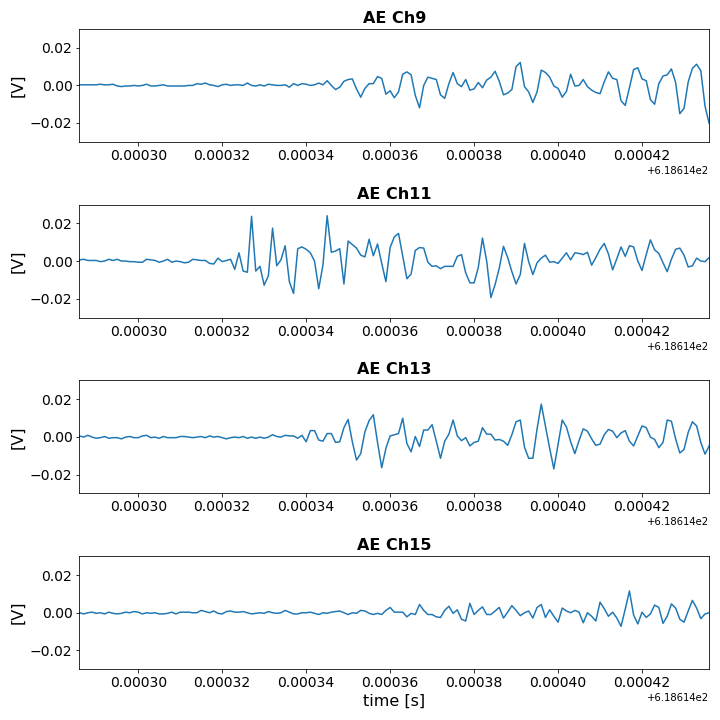}}
\caption{Acoustic emission data from the four sensors (block 4).}
\end{figure}

The four sensors that record the data differ slightly from each other due to the fact that waves have a different angle of incidence, depending on the location of the source of the signal. In this case, these incidence angle differences can lead to noticeable differences in which signals are recorded due to the use of piezoelectric transducers, which are mostly sensitive to signals at near-vertical incidence only. These differences are shown in figure \ref{differentsensors}. For some sensors the amplitude of the slip events are higher than for others and some peaks are more reflected to the positive side of the y-axis. The timing of all the amplitudes seems to correspond between all four sensors. However, when we zoom in on a precursor we observe a difference in first arrival timing, on very small time scales.

\subsubsection{First arrival precursors}
\begin{figure}[ht!]
\centering
\includegraphics[height=2in]{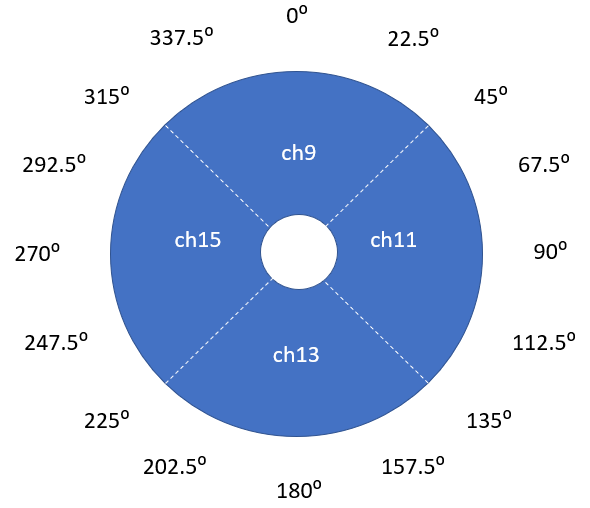}
\caption{A map view of the top piston with 16 distinguished angles representing the location classes.}
\label{classes}
\end{figure}

To illustrate spatial variations in recorded signals, figure \ref{zoomprecursors} shows the difference in AE data from the four different sensors for a random precursor. The first arrival is recorded by sensor ch11, followed by ch9 and ch13 and lastly by ch15. From this we can infer where the signal may have originated. The nucleation of rupture has to start somewhere, so the signal will be recorded in one place first, and last by the opposite sensor. If all sensors record the signal with the same amplitude, it would represent a sample scale slip. However, we observe a decrease in overall amplitude, meaning the energy was released closest to the first arrival: ch11 and not everywhere on the slip plane. We now try to visualize where the precursor signals plausibly originated, based on the first arrival of the precursor. To do so we create a map view of the top piston (figure \ref{classes}) and distinguish 16 different sectors with a width of $22.5\degree$. 

\begin{figure*}[ht!]
    \centering
    \includegraphics[width=\textwidth]{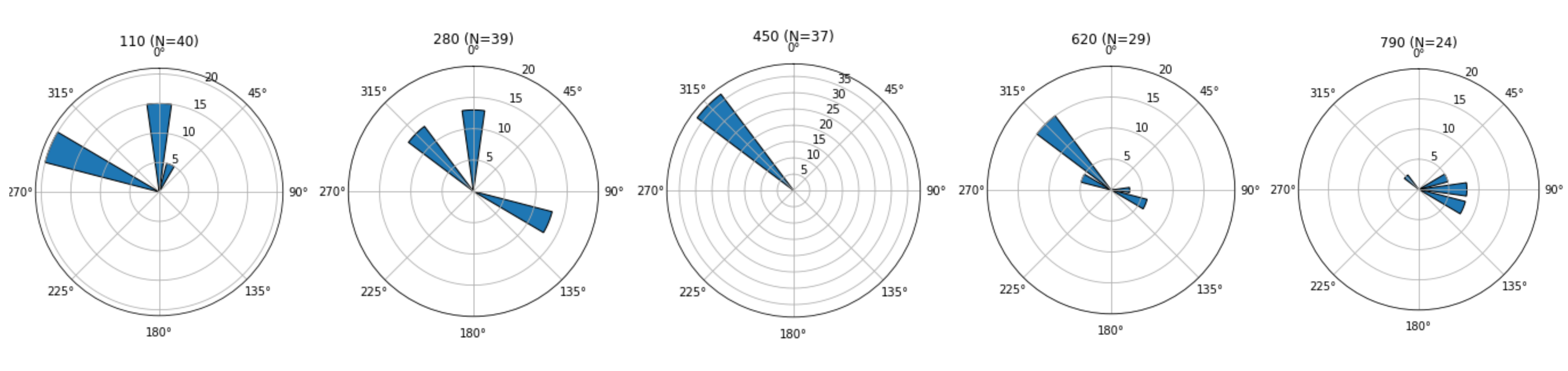}
    \caption{A circle diagram of the inferred asperity location based on the first arrival of the precursors, for the five different blocks. N indicates the number of precursors evaluated. Appendix \ref{appendix3} shows which precursors are evaluated. The locations are divided in 12 angular locations. The figure shows the number of precursors that are originated in that angle range. Note, the radial scale of the third block (450) is different than the others.}
    \label{circle}
\end{figure*}

\noindent Taking figure \ref{zoomprecursors} as an example, the location of the energy release is closest to ch11. Ch13 shows a slightly earlier arrival than ch9 and also a slightly higher amplitude. From this we can infer that the location of the energy release is located a little more towards ch13: at $112.5\degree$ (figure \ref{classes}).
\\ \\
Almost all precursors of the five blocks are evaluated likewise, to check if there is an evolution in the data which corresponds to the amount of rotation obtained by rotary shear. Since the first blocks have much more precursors than the later blocks not all of them are evaluated, figure \ref{precursors} shows the number of evaluated precursors (\textit{N}) per block. Figure \ref{circle} shows the result of the location of the energy release of the precursors for the five different blocks. The used precursors can be found in appendix \ref{appendix3}. The circle diagram represents a location between [0\degree,360\degree]. For the first 4 blocks the main location is the fourth quadrant between [270\degree,360\degree], i.e. between sensor ch15 and ch9. For block 3 the only inferred location is 315\degree, note the different radial scale for this block. Block 5 has most locations between [45\degree,135\degree]. According to figure \ref{evan_viewer}, the total rotation is $\sim70\degree$, where in the first blocks a higher load point velocity is set with a decrease in velocity for the following blocks. We do not observe a clear trend in the distribution of the inferred locations of the energy releases. Take for example the difference between 280s and 450s (block 2 and 3). A rotation of $\sim20\degree$ (figure \ref{evan_viewer}) is expected but not observed. Another interesting observation is the lack of locations between [112.5\degree,292.5\degree].

\subsection{Data quartz gouge}
\begin{figure}[ht!]
\subfigure[Full experiment]{
\label{wendata}
\includegraphics[page=1,width=.5\textwidth]{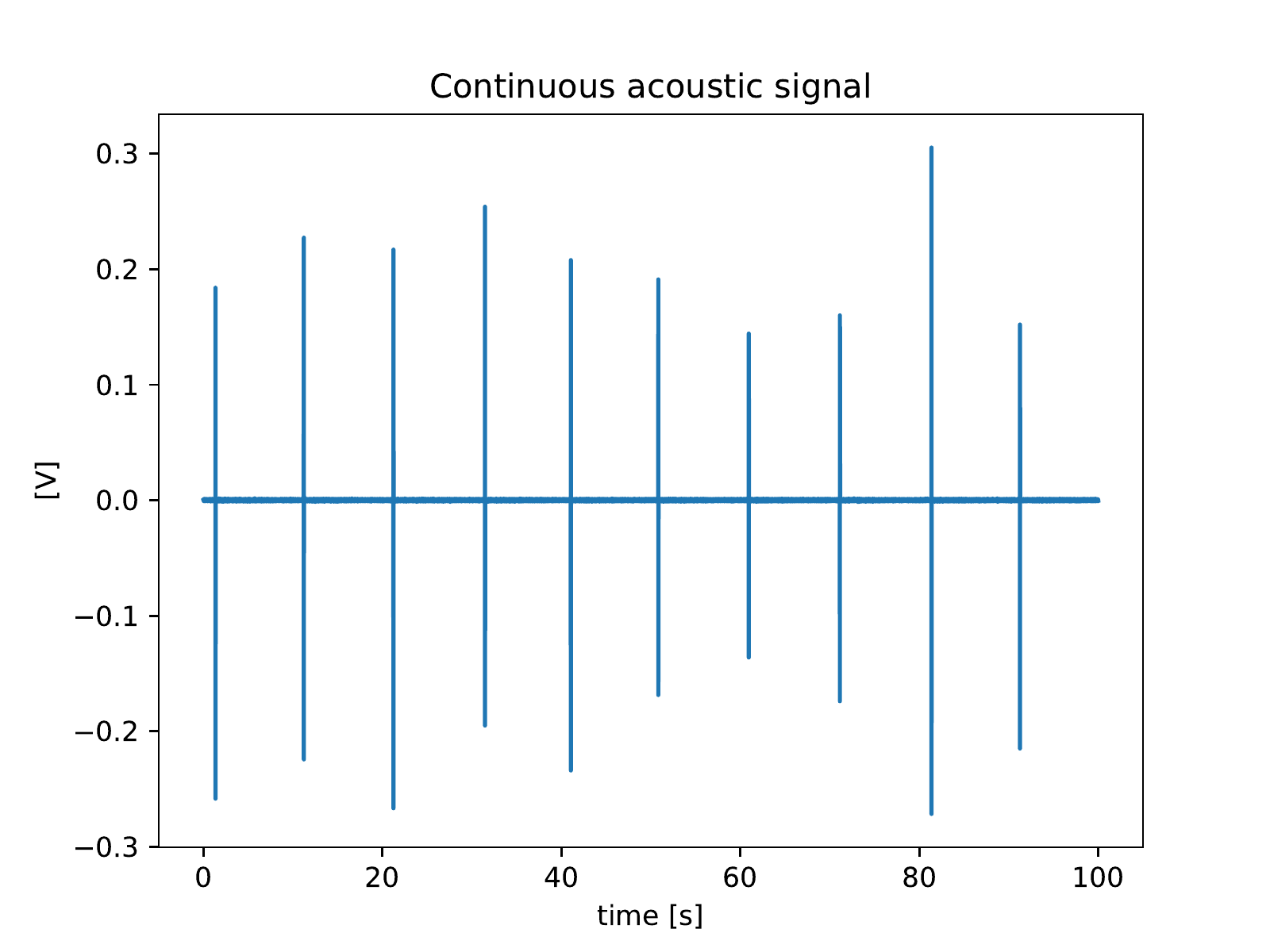}}
\qquad
\subfigure[Zoom in of a)]{
\label{zoomwen}
\includegraphics[page=2,width=.5\textwidth]{figures/Wen_acoustic_data.pdf}}
\caption{Acoustic data from an experiment on quartz gouge on a ring shear by dr. Wen Zhou.}
\end{figure}

In addition to the salt experiment, Dr. Wen Zhou conducted a similar stick slip experiments on the ring shear, not with salt gouge but with quartz gouge and he continuously recorded the acoustic emissions coming from the sample. The sensors were placed on top of the piston and not inside the piston as done in the salt experiment. Figure \ref{wendata} shows the acoustic data. We do observe high peaks during slip events but we do not observe any precursors. The magnitude of the slip events is about 100 times the size of the noise. Figure \ref{zoomwen} shows that the noise level is of similar magnitude before and after the slip event. The lack of observable precursors and overall signal variety in the interseismic period makes this data not suitable for building a RF model. 

\clearpage 

\section{Discussion}
First of all, the model is able to predict slip events very well, all $R^{2}$ scores are high. The $R^{2}$ scores for the salt experiments are slightly higher than for the glass beads experiment. This might have to do with periodicity. The maximum time to failure peaks for the salt experiments are very regular, all peaks are about the same height, whereas the peaks for the glass beads are less regular. In general, if there are fewer correlations between features and time to failure within the data set, the prediction will be less accurate. It could also have to do with the resonance that is present in the salt experiment, which helps the model predict in the early stage of the interseismic period. Nevertheless, the random forest model for the glass bead experiment still does an excellent job in predicting failure time, showing that the approach can be generalized to more aperiodic fault cycles.
\\ \\
The difference in model performance between block 2 and block 4 can be explained by the amount of data points used. The more data points, the more information to train the model on. In general, more data leads to better predictions. Since the window width is different for both, 0.02s (block 2) and 0.06s (block 4) (table \ref{windowwidth}), block 2 has about three times more data points than block 4. Besides that, the amount of slip events differ within the blocks. Since the windows with a slip event are not taken into account we end up with 1259 data points for block 2 and 400 data points for block 4. The performance of the model of block 2 might therefore be slightly better than block 4. For the glass bead experiment 1390 data points are used.
\\ \\
It is important to note and perhaps the only disadvantage of using a ML model: when the training set has not seen data similar to the testing set it will never be able to predict it well. The model is trained on window examples in the training set, thus when a completely different window, from for example a totally different experiment, is shown to the model, it does not know what to predict. The training set has to contain similar type of information as the information that should to be predicted. If we go back to the Parkfield example, a random forest model would not have predicted an earthquake exactly in 2004, since such a long interval was never present in the training data. Besides, there are almost no precursors observed in the Parkfield study so it is questionable whether a RF model would even be applicable in that case.

\subsection{Features}
Kurtosis is for all experiments a very important feature. Kurtosis describes the tailness of the distribution curve. When many extreme events are recorded in the window, during for example precursor events, the tails of the curve are fuller which means kurtosis is high. It therefore tells us something about precursor activity.
\\ \\
Variance is the most important feature for the glass beads experiment. The amplitude of the AE data increases towards the slip event, indicating the more energy is released the closer we get to failure. For the glass beads experiment it shows a gradual evolution, for both variance and kurtosis. The values gradually increase, with a steep increase close to the slip event. For the salt experiment there is no evolution in increase in variance or kurtosis, it is just high when there are precursors seen in the AE data. This is a crucial difference between the two materials.

\subsection{Microstructural differences glass beads and salt}
We interpret the differences between feature importance to be due to the micromechanics operating in both materials. Glass beads form force chains when shear stress is applied (figure \ref{forcechain}). When stress increases the beads change position and line up, many chains next to each other. When stress is applied even further the force chains deform, some beads may roll or a few chains may even break releasing some energy what we observe as a precursor in the AE data. The gradual increase in variance and kurtosis suggests that the system emits a small but progressively increasing amount of energy throughout the stress cycle (figure \ref{most_imp_glassbeads}), before abruptly releasing the accumulated energy when a slip event takes place and the sample displaces.
\\ \\
For salt however, we do not observe a gradual increase and these features (kurtosis and variance) are only high when there are precursors visible in the AE data, some time before the slip event (figure \ref{most_imp_salt620_wr6}). This is plausibly because the energy can be stored until the system reaches a critical stress state. We speculate that the precursors are the effect of very localized and very small scale slip. We do not observe precursors in the shear stress data, so this small scale slip did not cause a change in total stress state of the sample, but possibly did locally. This release of energy could also come from grainsize reduction \citep{scuderi2017}, however, the sample has been sheared for a long time before starting the continuous measurements so that is rather unlikely. Localized slip is therefore a more plausible explanation for the precursor activity.
\\ \\
Especially for the glass beads experiment it can be interpreted that the signal variance and kurtosis may be fingerprints of the friction and shear stress state, since it increases when stress increases. The features therefore carry quantitative frictional state information \citep{rouet17}. For both glass beads and salt, the model accurately predicts failure not only when failure is close, but also throughout the entire laboratory earthquake cycle, demonstrating that the system continuously progresses toward failure. This is somewhat unexpected, as impulsive precursors are only observed while the system is in a critical stress state. For salt the model is significantly better in predicting once precursors are observed. Furthermore, it seems that the model tends to use the resonance data for the first part of interseismic period. It would therefore be useful to filter out the resonance or reduce the effect of resonance by creating a longer seismic interval, to see if this energy increases throughout the stress cycle in a similar manner as glass beads or if energy is only released close to failure. With the experimental data on hand at this point, the latter one seems to be the case. Salt behaves differently to glass beads. With this reasoning, glass beads allow for better predictions than salt because the features evolve during the entire interseismic period.
\\ \\
From the most important features it seems that kurtosis \textsc{\char13}after\textsc{\char13} and variance \textsc{\char13}before\textsc{\char13} are favourable. We try to explain why the model would prefer the second part of a window for kurtosis and the first for variance. When \textsc{\char13}before\textsc{\char13} is favoured we observe that the feature varies most in the beginning of the interseismic period (figure \ref{most_imp_salt280} and \ref{most_imp_salt620}). While on the other hand, kurtosis increases steeply at the end of the interval, similar to the threshold and strain features (figure \ref{most_imp_salt280} and \ref{most_imp_salt620_wr6}). When a feature changes more in the beginning, the model chooses to take the first half of the data window (\textsc{\char13}before\textsc{\char13}) and if a feature varies more in the end of the interseismic period, it takes the second half of the window (\textsc{\char13}after\textsc{\char13}), since the first and last window of that interseismic period carry more useful information, respectively.
\\ \\
Another difference between the two materials is the fact that compaction and dilation mechanisms operate at different times. The glass beads sample dilate in the interseismic period when force chains form and compact during slip when one glass bead is squeezed out the force chain (figure \ref{forcechain}). This is opposite to salt, salt compacts during the interseismic and dilates during the stress drop. Compaction takes place due to deformation mechanisms like pressure solution \citep{niemeijer2010frictional} and dislocation glide and creep. The surfaces that are in contact weld together with an increase in stress and time. Once slip occurs, these contacts break and create a new surface with a different contact area, creating more gaps, resulting in a dilation of the material (figure \ref{frictionsalt}). 

\subsection{Microstructural observations}

\begin{wrapfigure}{R}{3in}
\centering 
\includegraphics[width=3in]{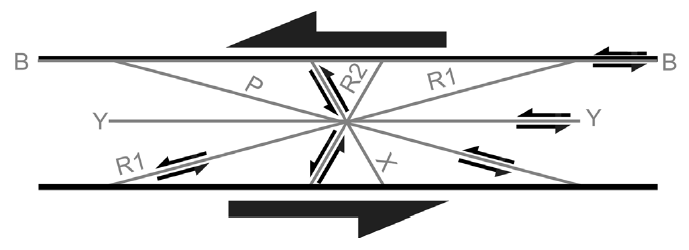}
\caption{Different shear orientations in a fault.}
\label{shear}
\end{wrapfigure}

\begin{figure}[ht!]
\subfigure[Cross section of a sheared salt sample in the stick-slip regime. The white arrows indicate the shear band formed in the upper boundary.]{
\label{imageshear}
\includegraphics[width=0.6\textwidth]{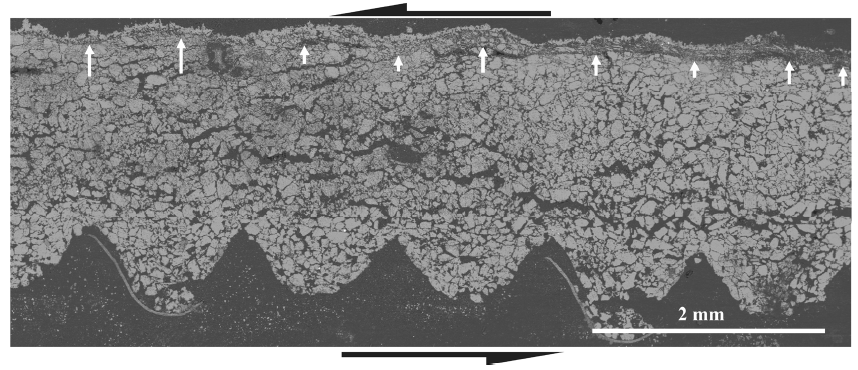}}
\qquad
\subfigure[Zoom in of \ref{imageshear} on the shear band. Fine grained material is sandwiched between larger grains. The white and black line indicate the boundary of the shear zone which is wavier at the upper part and more linear at the bottom part.]{
\label{zoomshear}
\includegraphics[width=0.3\textwidth]{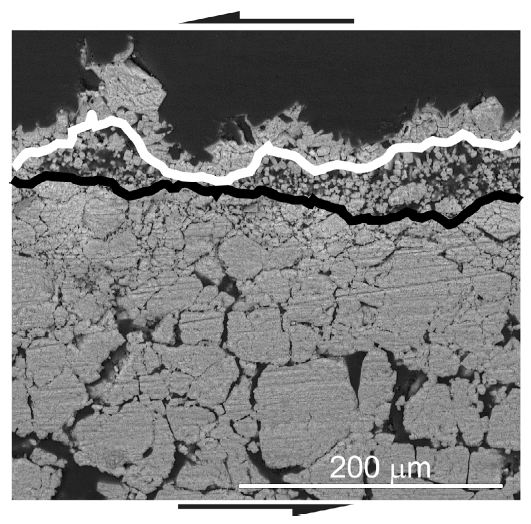}}
\caption{Microstructural observations of a shear experiment on salt. Figures from \citet{niemeijer2010frictional}}
\end{figure}

We can infer the microphysical processes operating in the salt experiment on microstructural observations from earlier research. \citet{niemeijer2010frictional} performed a shear experiment on salt in a double shear set up similar to the glass beads experiment. In their experiment they established stable sliding to stick slip behaviour with increasing sliding velocity and compared the microstructure in both regimes. We will look at what the microstructure looks like for stick slip behaviour. Riedel shears (R1 in figure \ref{shear}) were only present in the stable sliding phase and are not present anymore. While in the stick slip regime, they found a zone of fine grained material that formed boundary‐parallel Y or B shears (figure \ref{shear}). The shear band (indicated with the white arrows) is visible at the upper shear zone boundary (figure \ref{imageshear}). It is continuous along the length of the sample and is not present on the lower shear zone boundary, where the imprint of the teeth on the forcing block is preserved \citep{niemeijer2010frictional}. These teeth are also present in the upper boundary during the experiment but are not preserved while preparing this sample. Since the salt used in the experiment of \citet{korkolis} is sheared for a long time (1.45h) in the precompaction phase, it was subjected to much more shear so we expect that this fine grained shear zone extends a little in the vertical direction. Figure \ref{zoomshear} shows the shear band in more detail. 

\begin{wrapfigure}{R}{2.5in}
\centering 
\includegraphics[width=2.5in]{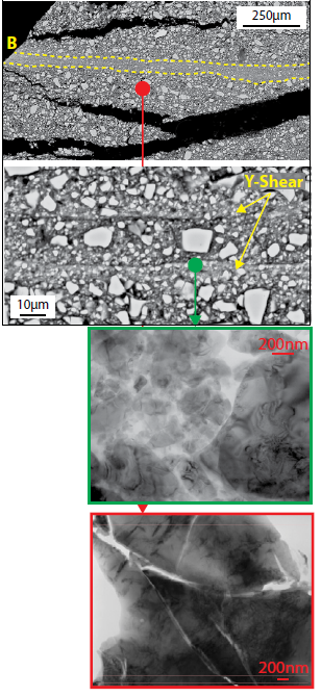}
\caption{Microstructure obtained for quartz in stick slips regime. Y-shears consist of nanoscale grains that contain intense dislocations and some fracturing (green inset). Outside the B-shear zone, grains are larger and contain fractures (red inset). Figure from \citet{scuderi2017}}
\label{quartz}
\end{wrapfigure}

\noindent Another research \citep{scuderi2017} investigated the evolution of shear fabric on quartz gouge, also from stable sliding to dynamic stick slip. Their microstructural observations show that with accumulated strain, deformation concentrates in shear zones containing sharp shear planes made of nanoscale grains, which favor the development of frictional instabilities (figure \ref{quartz}). Once this fabric (i.e., continuous B-shear zones containing interconnected Y-planes) is established, fault fabric does not change much with increasing slip velocity. 
\\ \\
Those observations are similar to the observations of salt gouge found by Niemeijer et al., 2010. Significant grain size reduction in the shear zone seems required for stick slip behavior, for both salt and quartz. Shear bands may form in the glass beads as well, even though it is not observed in experiments yet. The shear bands are likely to consist of force chains in this case. The glass beads do not break so grain size is not reduced. The formation of shear banding in glass beads is inferred from a theoretical model on soft glass beads [Jeannot Trampert, personal communication]. He found, in a numerical experiment, that when shear bands are present, regular stick-slip behavior is established, agreeing with the results of \citet{niemeijer2010frictional} and \citet{scuderi2017}. In their numerical experiment they enforced shear banding to observe regular stick slip behavior. Without the shear bands the timing of slip events was very irregular, to such an extent that as soon as the shear band was taken away, it was not possible in their model to predict time to failure. This irregularity, that is also found in earthquakes sequences on natural faults, should be simulated in the lab to test the model on more aperiodic cycles. The strength of this random forest model, however, is that it looks at what happens in between slip events, per window, so the evolution of features can still help in predicting time to failure even when the data set is highly irregular, on the condition that this information is also present in the training data set. 

\vspace{0.6cm}
\noindent We do have continuous data from a sheared quartz experiment as well (figure \ref{wendata}). However, in this experiment on quartz gouge there are no precursors observed, the data is therefore not useful for the ML model. We explain the lack of precursors by the fact that the sensors are located too far from the sample. It would be interesting to test quartz gouge in a similar set up as the salt experiment, to see if quartz releases energy in the critical stress state in the same way as either salt or glass beads.

\subsection{Effect of load point velocity}
We showed the model results for block 2 with relatively high and block 4 with relatively low load point velocity: both ML models perform very well (figure \ref{ch9_280} and \ref{ch9_620}). The resonance seems to help the prediction model in the early stages of the interseismic period so it is hard to be conclusive about the effect of load point velocity. However, we observe in the AE data that precursors are established earlier in the interseismic period with decreasing load point velocity. The absolute shear stress values do not show this correlation (figure \ref{precursors}). It seems that, when there is more time to build up stress, the sample enters the critical state of emitting bursts of energy earlier. What that critical state will be might be related to the conditions within that block. Further research is needed to be more conclusive about this observation. Keep in mind that we only evaluated three seconds of every block so clearer correlations might be found when all the data from a block are evaluated. 

\subsection{Asperity location}
Considering the material physics of salt, friction increases when there is an asperity located somewhere on the slip plane (figure \ref{frictionsalt}) and stress can build up locally. We evaluated the location of potential asperities by picking the first arrivals and inferring the location of energy release (figure \ref{circle}). If there would be one large asperity, the precursor could originate from that specific location throughout the experiment. In that case this asperity would rotate along with the displacement rate and figure \ref{circle} would show a movement of 70\degree through time. However, we do not observe that clearly. On the other hand, there are also no signals originated in the other half of the piston area [112.5\degree,292.5\degree]. We interpret this observation as misalignment of the sample in the apparatus. One half of the sample might consist of more material or the surfaces of the piston are not completely horizontally aligned. 

\subsection{Natural earthquakes}
Empirical friction laws are based on relative short term experiments which means their extrapolation to geological time scales is highly speculative. Long term ductile processes (e.g. slow relaxation, pressure-solution) are active within the upper crust \citep{voisin2007} and are difficult to simulate in a laboratory experiment. Besides, laboratory shear rates are orders of magnitude larger than Earth (mm/s versus mm/yr). Moreover, the laboratory temperature conditions do not resemble those in Earth, while the pressures could be representative of in situ pressures when fluid pressures are large \citep{rouet17}. However previous research has shown that some important fault frictional physics scale to Earth \citep{johnson13}. A laboratory experiment clearly cannot capture all of the physics of a complex rupture on Earth. Nevertheless, this machine learning approach can help revealing the microphysics operating and the next logical step would be to test this model on real Earth data at much larger scales.
\\ \\
The problem with testing real earthquakes is that it requires observing them in high detail, with data acquisition close to the frictional plane. At this point, earthquakes are largely unpredictable, so knowing where to put the instrumentation to make such observations is a challenge on its own \citep{bakun1985}. Besides, it requires a seismometer network rather than one seismometer, otherwise it is impossible to find out where the signal originated from.
\\ \\
Monitoring signals originating from one slip event seems to be the most obvious obstacle, but there is a need for precursors to be present in the data as well. Although the Parkfield earthquake was ideally located within a dense monitoring network designed to detect precursors, no significant signals were detected \citep{bakun1985}. The lack of precursors can be explained by a process called aseismic slip. A process when fault slips stably and no acoustic emissions are released. The Parkfield experiment showed that it is useful to maintain such specialized networks until a large earthquake occurs, and the detailed observations made at Parkfield demonstrate how valuable the information is for our understanding of earthquakes. Due to the lack of precursors, the question is whether the random forest model will be applicable to the data of that specific fault.

\subsection{Further research}
This research succeeded in producing a prediction model which works for both glass beads and rock salt. We observe that the differences in physical properties of glass beads and salt lead to a different feature evolution and therefore influence the model performance. The physical processes operating at microscale likely play a key role in the prediction models, but we have not validated specific microstructure processes for these experiments.
\\ \\
To prove the mechanisms operating on microscale in salt, localization of slip during precursory activity should be observed. One way of achieving that could be by placing markers on the sample to see if slips occurs locally. Different moments in time should be captured, right before and right after a precursor to see if something moved locally. Another suggestion would be to work with heat sensitive sensors. When at one spot slip occurs, it generates heat locally and the sensor will light up. Better waveform imaging could also help in localizing the source location related to the slip event.
\\ \\
Experiments on different materials can be evaluated to obtain a data base which shows the feature evolution for different materials. When the experiments are done in a similar set up, would other materials also exhibit precursor activity, gradually or only close before failure? Using AE data as a quantitative measure helps us understanding microscale processes for specific materials.
\\ \\
Besides focusing on the microphysics that can be explained by these type of laboratory experiments, a future step would be to apply the ML model to real earthquake data. As described earlier, it is hard to obtain useful data because of the importance of locating the seismometer accurately and the need of a long historical record. The Parkfield data is well monitored and even though there are almost no precursors observed, it could be useful to subject such data to the ML model to see if there is any feature evolution that would help in predicting the next earthquake. Finding another fault with repetitive earthquakes from which a long historical record is available would even be better. Testing that data in a machine learning model like this requires gigantic computational power since years of seismic data need to be evaluated. Obviously the window width will be chosen accordingly. 
\\ \\
The ultimate goal is to predict real earthquakes but to understand what factors influence the predictions, the microphysics need to be revealed by laboratory experiments. Developing the model to evaluate big real earthquake data, meanwhile reconstructing the physics that occur on microscales, contributes enormous to our understanding of earthquakes and their predictions.
\newpage

\section{Conclusion}
The machine learning model is able to predict laboratory earthquakes for both glass beads and salt. The statistics quantifying the signal amplitude distribution (variance and kurtosis) are highly effective at forecasting failure. For glass beads those features increases towards failure. For salt the presence of precursors seems to be more important in predicting the next slip event. The differences in feature evolution of both materials are explained by their difference in micromechanics. For glass beads the deformation of force chains releases energy gradually. Salt deforms elastically with increasing shear stress and when precursors are observed, that release in energy is interpreted to come from very localized microslip in fine grained shear zones. For lower loading velocities precursors occur earlier in the interseismic period. The next step would be to prove the microphysical interpretations made in this research by doing more laboratory experiments. Meanwhile the model should be extrapolated and prepared to predict earthquakes with data from nature.

\newpage

\bibliographystyle{unsrtnat}
\def\bibfont{\small}
\bibliography{references}

\newpage

\appendix
\section{Acoustic data block 1, 3 and 5}\label{appendix1}

\begin{figure*}[ht!]
    \centering
    \includegraphics[width=0.8\textwidth,scale=0.1]{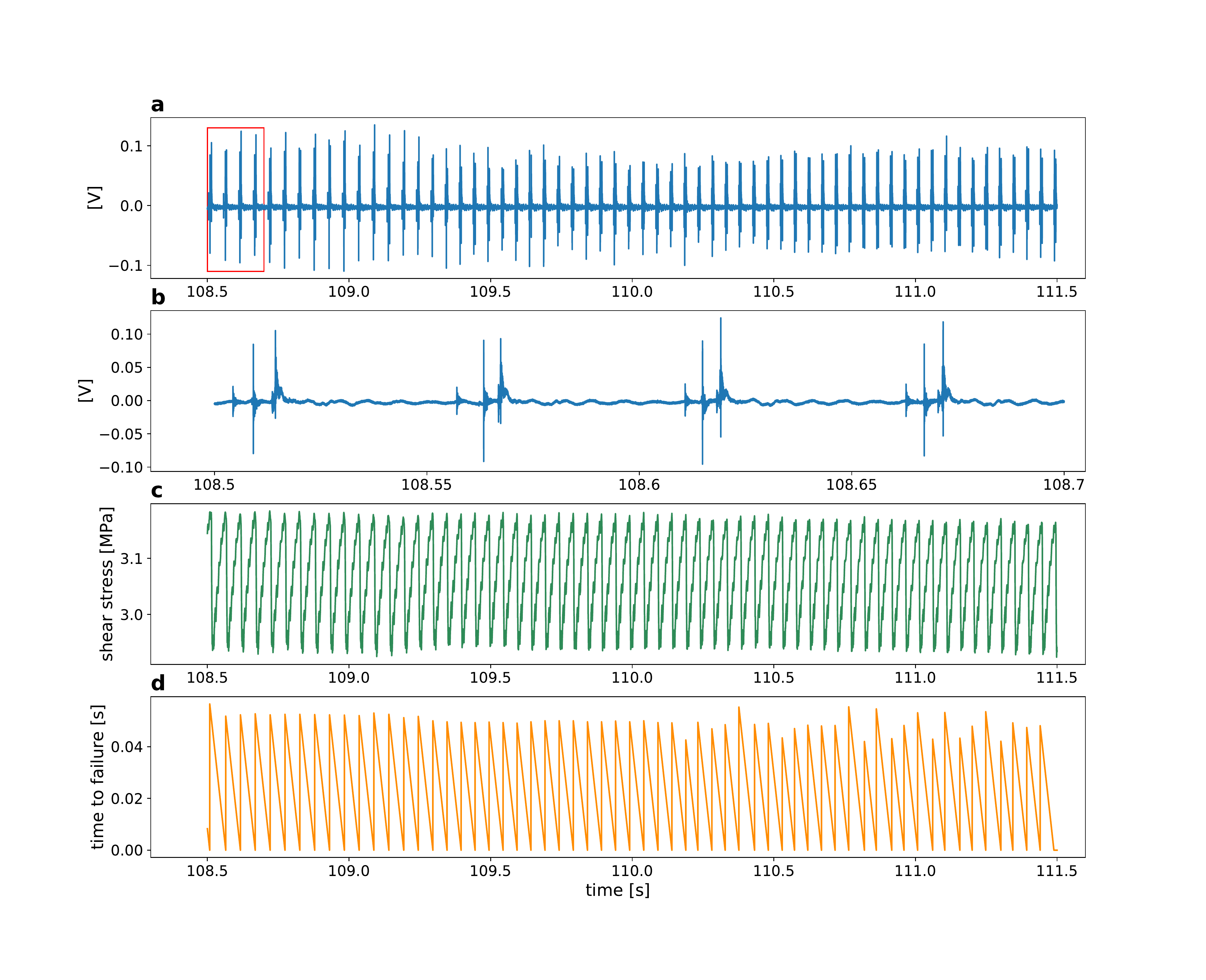}
    \caption{Continuously recorded data from the salt experiment. Sensor ch9, block 1, 110$\pm$1.5 seconds. a) Acoustic data. b) Zoom in acoustic data. c) Shear stress data. d) Time to failure curve. }
    \label{data_110}
\end{figure*}
\clearpage
\begin{figure*}[ht!]
    \centering
    \includegraphics[width=0.8\textwidth,scale=0.1]{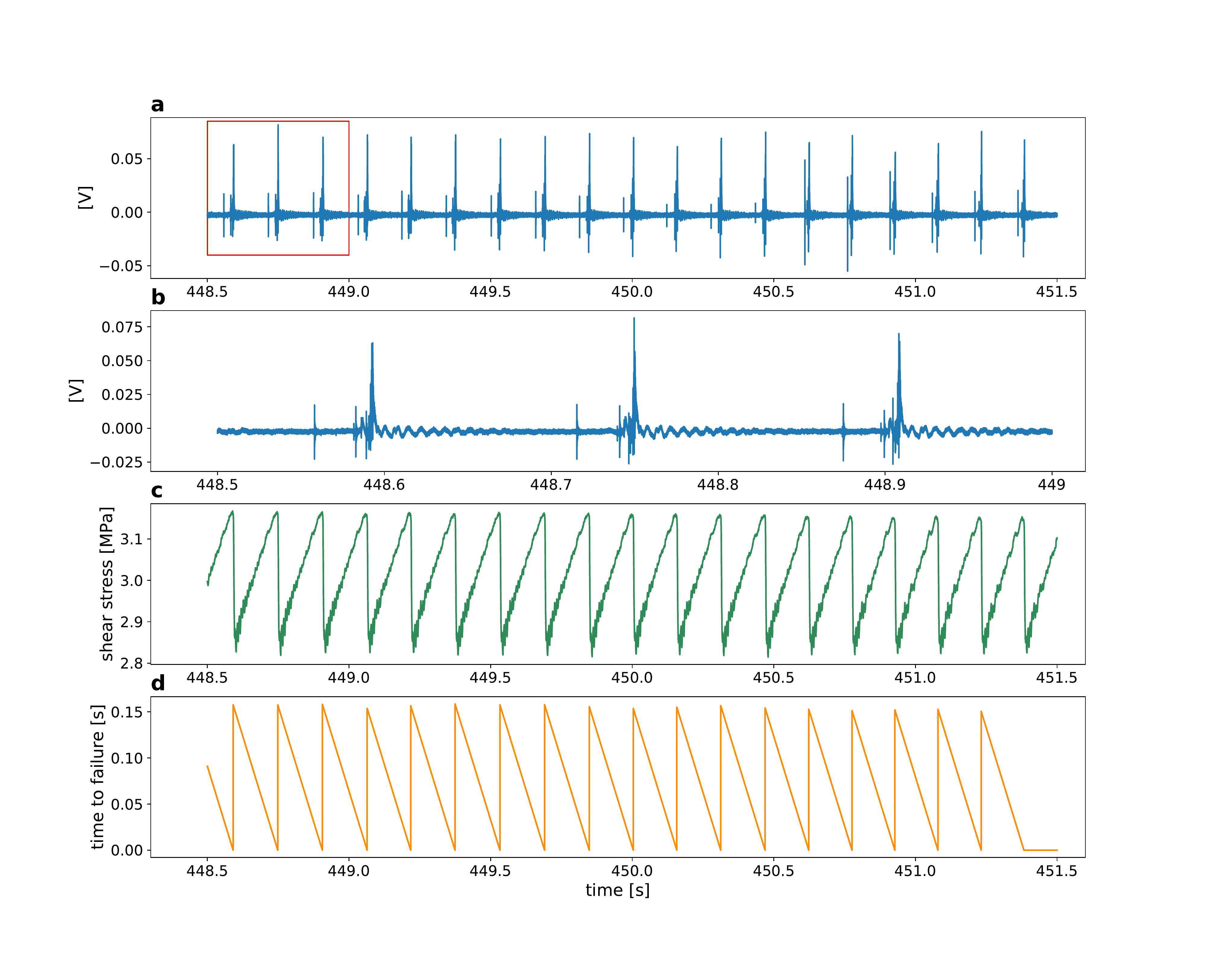}
    \caption{Continuously recorded data from the salt experiment. Sensor ch9, block 3, 450$\pm$1.5 seconds. a) Acoustic data. b) Zoom in acoustic data. c) Shear stress data. d) Time to failure curve. }
    \label{data_450}
\end{figure*}

\begin{figure*}[ht!]
    \centering
    \includegraphics[width=0.8\textwidth,scale=0.1]{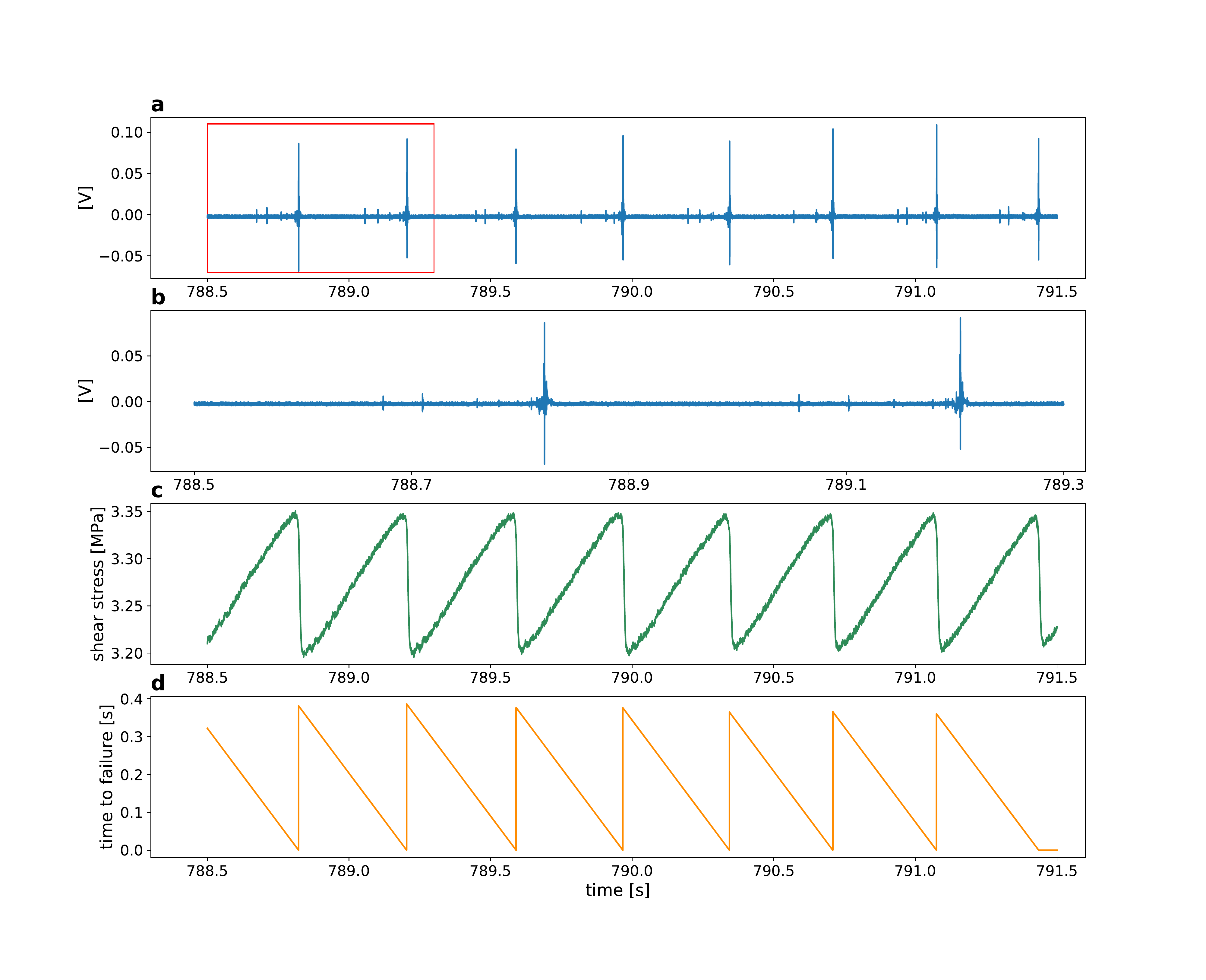}
    \caption{Continuously recorded data from the salt experiment. Sensor ch9, block 5, 790$\pm$1.5 seconds. a) Acoustic data. b) Zoom in acoustic data. c) Shear stress data. d) Time to failure curve. }
    \label{data_790}
\end{figure*}

\section{Model results block 1, 3 and 5}\label{appendix2}

\begin{figure}[ht]
    \centering
    \includegraphics[page=13,width=0.8\textwidth]{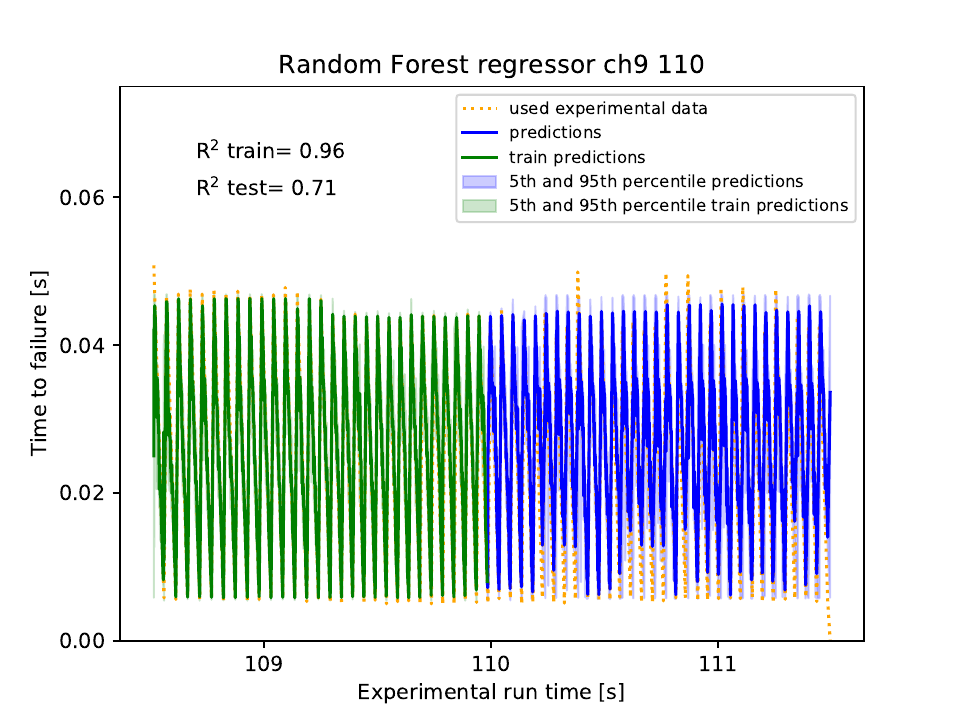}
    \caption{Random forest model for sensor 9, from 108.5 - 111.5 seconds. The performance of the training data set (green) is as good as the performance of the testing data set (blue) as indicated with the $r^{2}$ score. The shaded green and blue colors show the 5th and 95th percentile of the predictions corresponding to uncertainties in the predictions. The orange dashed line is the used time to failure data, inferred from the shear stress data, excluding the time windows that contain the failure event.}
    \label{ch9_110}
\end{figure}


\begin{figure}
    \centering
    \includegraphics[page=13,width=0.8\textwidth]{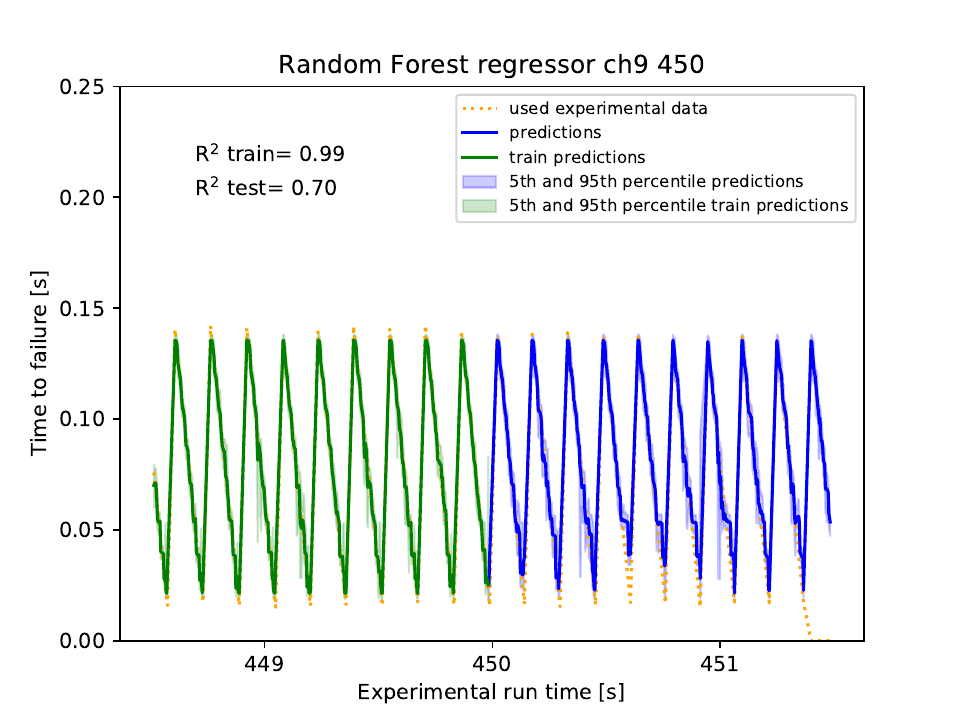}
    \caption{Random forest model for sensor 9, from 448.5 - 451.5 seconds. The performance of the training data set (green) is as good as the performance of the testing data set (blue) as indicated with the $r^{2}$ score. The shaded green and blue colors show the 5th and 95th percentile of the predictions corresponding to uncertainties in the predictions. The orange dashed line is the used time to failure data, inferred from the shear stress data, excluding the time windows that contain the failure event.}
    \label{ch9_450}
\end{figure}


\begin{figure}
    \centering
    \includegraphics[page=13,width=0.8\textwidth]{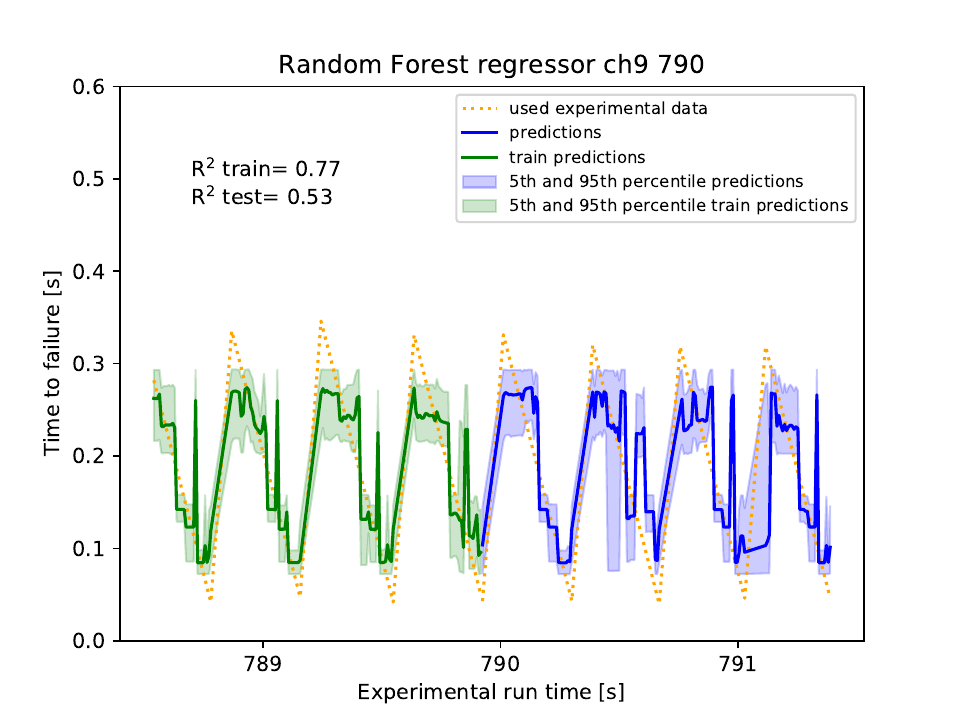}
    \caption{Random forest model for sensor 9, from 618.5 - 621.5 seconds. The performance of the training data set (green) is as good as the performance of the testing data set (blue) as indicated with the $r^{2}$ score. The shaded green and blue colors show the 5th and 95th percentile of the predictions corresponding to uncertainties in the predictions. The orange dashed line is the used time to failure data, inferred from the shear stress data, excluding the time windows that contain the failure event.}
    \label{ch9_790}
\end{figure}

\clearpage
\section{Precursor locations}\label{appendix3}

\begin{figure}[ht!]
    \centering
    \includegraphics[width=0.8\textwidth]{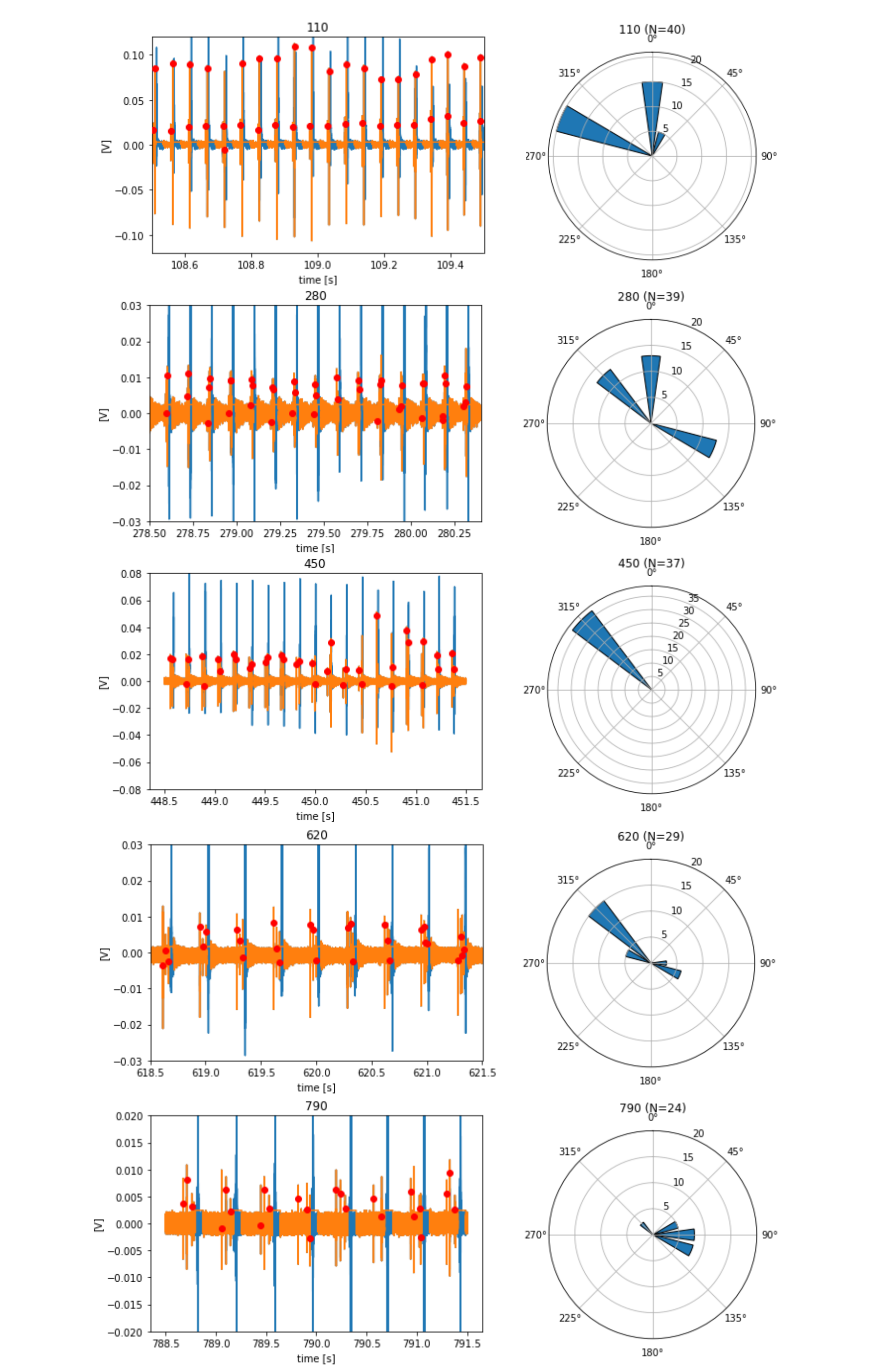}
    \caption{This figure shows from which precursors the first arrivals are evaluated. The orange color indicates the data to evaluate to make sure that failure events are not selected. The red dots indicate the precursors.}
    \label{precursors_5}
\end{figure}
\clearpage
\section{Precursor specifications}\label{appendix4}
\begin{table}[ht]
\small
\begin{tabular}{llllllllll}
110  &          &                                                        &                 &                                                                 &                                                                      &                                                                &                                                               &                                                                &                                                               \\
name & time (s) & \begin{tabular}[c]{@{}l@{}}first \\sensor\end{tabular} & class (\degree) & \begin{tabular}[c]{@{}l@{}}interseismic \\time (s)\end{tabular} & \begin{tabular}[c]{@{}l@{}}time   next \\slip event (s)\end{tabular} & \begin{tabular}[c]{@{}l@{}}relative \\timing (\%)\end{tabular} & \begin{tabular}[c]{@{}l@{}}min/max \\shear (MPa)\end{tabular} & \begin{tabular}[c]{@{}l@{}}absolute \\shear (MPa)\end{tabular} & \begin{tabular}[c]{@{}l@{}}relative \\shear (\%)\end{tabular} \\
1a   & 108.50   & 15                                                     & 292.5           &                                                                 & 108.51                                                               & 0.00                                                           &                                                               & 3.15                                                           & 0.00                                                          \\
1b   & 108.51   & 9                                                      & 0               &                                                                 &                                                                      & 0.00                                                           & 3.18                                                          & 3.18                                                           & 0.00                                                          \\
2a   & 108.56   & 15                                                     & 292.5           & 0.0565                                                          & 108.56                                                               & 0.86                                                           & 2.94                                                          & 3.16                                                           & 0.90                                                          \\
2b   & 108.56   & 9                                                      & 0               &                                                                 &                                                                      & 0.97                                                           & 3.18                                                          & 3.18                                                           & 1.00                                                          \\
3a   & 108.61   & 15                                                     & 292.5           & 0.0518                                                          & 108.62                                                               & 0.89                                                           & 2.93                                                          & 3.15                                                           & 0.89                                                          \\
3b   & 108.61   & 9                                                      & 0               &                                                                 &                                                                      & 0.97                                                           & 3.18                                                          & 3.18                                                           & 1.00                                                          \\
4a   & 108.66   & 15                                                     & 292.5           & 0.0523                                                          & 108.67                                                               & 0.88                                                           & 2.93                                                          & 3.16                                                           & 0.89                                                          \\
4b   & 108.67   & 9                                                      & 0               &                                                                 &                                                                      & 0.96                                                           & 3.18                                                          & 3.18                                                           & 0.99                                                          \\
5a   & 108.71   & 15                                                     & 292.5           & 0.0527                                                          & 108.72                                                               & 0.87                                                           & 2.93                                                          & 3.15                                                           & 0.88                                                          \\
5b   & 108.72   & 9                                                      & 0               &                                                                 &                                                                      & 0.97                                                           & 3.18                                                          & 3.18                                                           & 1.00                                                          \\
6a   & 108.77   & 15                                                     & 292.5           & 0.0523                                                          & 108.77                                                               & 0.87                                                           & 2.93                                                          & 3.15                                                           & 0.88                                                          \\
6b   & 108.77   & 9                                                      & 0               &                                                                 &                                                                      & 0.97                                                           & 3.18                                                          & 3.18                                                           & 1.00                                                          \\
7a   & 108.82   & 15                                                     & 292.5           & 0.0525                                                          & 108.83                                                               & 0.89                                                           & 2.94                                                          & 3.15                                                           & 0.88                                                          \\
7b   & 108.82   & 9                                                      & 22.5            &                                                                 &                                                                      & 0.96                                                           & 3.18                                                          & 3.18                                                           & 1.00                                                          \\
8a   & 108.87   & 15                                                     & 292.5           & 0.0525                                                          & 108.88                                                               & 0.88                                                           & 2.93                                                          & 3.15                                                           & 0.89                                                          \\
8b   & 108.88   & 9                                                      & 0               &                                                                 &                                                                      & 0.96                                                           & 3.18                                                          & 3.18                                                           & 1.00                                                          \\
9a   & 108.92   & 15                                                     & 292.5           & 0.0524                                                          & 108.93                                                               & 0.88                                                           & 2.93                                                          & 3.15                                                           & 0.89                                                          \\
9b   & 108.93   & 9                                                      & 0               &                                                                 &                                                                      & 0.97                                                           & 3.18                                                          & 3.18                                                           & 1.00                                                          \\
10a  & 108.98   & 15                                                     & 292.5           & 0.0523                                                          & 108.98                                                               & 0.88                                                           & 2.93                                                          & 3.16                                                           & 0.89                                                          \\
10b  & 108.98   & 9                                                      & 0               &                                                                 &                                                                      & 0.97                                                           & 3.18                                                          & 3.18                                                           & 0.99                                                          \\
11a  & 109.03   & 15                                                     & 292.5           & 0.0522                                                          & 109.04                                                               & 0.88                                                           & 2.93                                                          & 3.15                                                           & 0.89                                                          \\
11b  & 109.03   & 9                                                      & 0               &                                                                 &                                                                      & 0.97                                                           & 3.18                                                          & 3.18                                                           & 1.00                                                          \\
12a  & 109.08   & 15                                                     & 292.5           & 0.0520                                                          & 109.09                                                               & 0.89                                                           & 2.93                                                          & 3.15                                                           & 0.89                                                          \\
12b  & 109.09   & 9                                                      & 0               &                                                                 &                                                                      & 0.97                                                           & 3.18                                                          & 3.18                                                           & 0.99                                                          \\
13a  & 109.13   & 15                                                     & 292.5           & 0.0530                                                          & 109.14                                                               & 0.88                                                           & 2.92                                                          & 3.15                                                           & 0.88                                                          \\
13b  & 109.14   & 9                                                      & 22.5            &                                                                 &                                                                      & 0.96                                                           & 3.18                                                          & 3.18                                                           & 1.00                                                          \\
14a  & 109.19   & 15                                                     & 292.5           & 0.0525                                                          & 109.19                                                               & 0.88                                                           & 2.93                                                          & 3.14                                                           & 0.88                                                          \\
14b  & 109.19   & 9                                                      & 22.5            &                                                                 &                                                                      & 0.96                                                           & 3.17                                                          & 3.17                                                           & 1.00                                                          \\
15a  & 109.24   & 15                                                     & 292.5           & 0.0512                                                          & 109.24                                                               & 0.89                                                           & 2.93                                                          & 3.15                                                           & 0.90                                                          \\
15b  & 109.24   & 9                                                      & 22.5            &                                                                 &                                                                      & 0.97                                                           & 3.18                                                          & 3.17                                                           & 0.99                                                          \\
16a  & 109.29   & 15                                                     & 292.5           & 0.0517                                                          & 109.30                                                               & 0.88                                                           & 2.94                                                          & 3.16                                                           & 0.94                                                          \\
16b  & 109.29   & 9                                                      & 22.5            &                                                                 &                                                                      & 0.96                                                           & 3.18                                                          & 3.18                                                           & 1.00                                                          \\
17a  & 109.34   & 15                                                     & 292.5           & 0.0500                                                          & 109.35                                                               & 0.88                                                           & 2.94                                                          & 3.17                                                           & 0.94                                                          \\
17b  & 109.34   & 9                                                      & 0               &                                                                 &                                                                      & 0.95                                                           & 3.18                                                          & 3.17                                                           & 0.96                                                          \\
18a  & 109.39   & 15                                                     & 292.5           & 0.0496                                                          & 109.40                                                               & 0.88                                                           & 2.95                                                          & 3.17                                                           & 0.95                                                          \\
18b  & 109.39   & 9                                                      & 0               &                                                                 &                                                                      & 0.93                                                           & 3.18                                                          & 3.16                                                           & 0.90                                                          \\
19a  & 109.44   & 15                                                     & 292.5           & 0.0494                                                          & 109.45                                                               & 0.86                                                           & 2.94                                                          & 3.16                                                           & 0.93                                                          \\
19b  & 109.44   & 9                                                      & 0               &                                                                 &                                                                      & 0.91                                                           & 3.18                                                          & 3.16                                                           & 0.90                                                          \\
20a  & 109.49   & 15                                                     & 292.5           & 0.0493                                                          & 109.49                                                               & 0.87                                                           & 2.94                                                          & 3.16                                                           & 0.94                                                          \\
20b  & 109.49   & 9                                                      & 0               &                                                                 &                                                                      & 0.92                                                           & 3.18                                                          & 0.89                                                           &                                                              
\end{tabular}
\end{table}

\newpage

\begin{table}[ht]
\small
\begin{tabular}{llllllllll}
280  &          &                                                        &                 &                                                                 &                                                                      &                                                                &                                                               &                                                                &                                                               \\
name & time (s) & \begin{tabular}[c]{@{}l@{}}first \\sensor\end{tabular} & class (\degree) & \begin{tabular}[c]{@{}l@{}}interseismic \\time (s)\end{tabular} & \begin{tabular}[c]{@{}l@{}}time   next \\slip event (s)\end{tabular} & \begin{tabular}[c]{@{}l@{}}relative \\timing (\%)\end{tabular} & \begin{tabular}[c]{@{}l@{}}min/max \\shear (MPa)\end{tabular} & \begin{tabular}[c]{@{}l@{}}absolute \\shear (MPa)\end{tabular} & \begin{tabular}[c]{@{}l@{}}relative \\shear (\%)\end{tabular} \\
1a   & 278.60   & 15                                                     & 315             &                                                                 & 278.61                                                               & 0                                                              & 2.90                                                          & 3.17                                                           & 0.00                                                          \\
1b   & 278.60   & 11                                                     & 112.5           &                                                                 &                                                                      & 0                                                              & 3.20                                                          & 3.18                                                           & 0.00                                                          \\
2a   & 278.72   & 9                                                      & 0               & 0.1223                                                          & 278.73                                                               & 0.88                                                           & 2.88                                                          & 3.18                                                           & 0.93                                                          \\
2b   & 278.73   & 11                                                     & 112.5           &                                                                 &                                                                      & 0.93                                                           & 3.20                                                          & 3.18                                                           & 0.95                                                          \\
3a   & 278.83   & 15                                                     & 315             & 0.1223                                                          & 278.86                                                               & 0.81                                                           & 2.88                                                          & 3.16                                                           & 0.88                                                          \\
3b   & 278.84   & 9                                                      & 0               &                                                                 &                                                                      & 0.89                                                           & 3.20                                                          & 3.18                                                           & 0.94                                                          \\
3c   & 278.85   & 11                                                     & 112.5           &                                                                 &                                                                      & 0.93                                                           &                                                               & 3.18                                                           & 0.96                                                          \\
4a   & 278.96   & 15                                                     & 315             & 0.1230                                                          & 278.98                                                               & 0.81                                                           & 2.88                                                          & 3.16                                                           & 0.88                                                          \\
4b   & 278.96   & 9                                                      & 0               &                                                                 &                                                                      & 0.88                                                           & 3.20                                                          & 3.17                                                           & 0.91                                                          \\
4c   & 278.97   & 11                                                     & 112.5           &                                                                 &                                                                      & 0.92                                                           &                                                               & 3.18                                                           & 0.93                                                          \\
5a   & 279.08   & 15                                                     & 315             & 0.1218                                                          & 279.10                                                               & 0.82                                                           & 2.88                                                          & 3.16                                                           & 0.89                                                          \\
5b   & 279.09   & 9                                                      & 0               &                                                                 &                                                                      & 0.88                                                           & 3.20                                                          & 3.17                                                           & 0.91                                                          \\
5c   & 279.09   & 11                                                     & 112.5           &                                                                 &                                                                      & 0.92                                                           &                                                               & 3.18                                                           & 0.93                                                          \\
6a   & 279.20   & 15                                                     & 315             & 0.1220                                                          & 279.22                                                               & 0.79                                                           & 2.88                                                          & 3.15                                                           & 0.84                                                          \\
6b   & 279.21   & 9                                                      & 0               &                                                                 &                                                                      & 0.88                                                           & 3.20                                                          & 3.18                                                           & 0.93                                                          \\
6c   & 279.21   & 11                                                     & 112.5           &                                                                 &                                                                      & 0.91                                                           &                                                               & 3.18                                                           & 0.94                                                          \\
7a   & 279.32   & 15                                                     & 315             & 0.1211                                                          & 279.34                                                               & 0.81                                                           & 2.88                                                          & 3.16                                                           & 0.88                                                          \\
7b   & 279.33   & 9                                                      & 0               &                                                                 &                                                                      & 0.87                                                           & 3.20                                                          & 3.17                                                           & 0.92                                                          \\
7c   & 279.33   & 11                                                     & 112.5           &                                                                 &                                                                      & 0.91                                                           &                                                               & 3.18                                                           & 0.93                                                          \\
8a   & 279.44   & 15                                                     & 315             & 0.1226                                                          & 279.47                                                               & 0.81                                                           & 2.88                                                          & 3.16                                                           & 0.89                                                          \\
8b   & 279.45   & 9                                                      & 0               &                                                                 &                                                                      & 0.88                                                           & 3.20                                                          & 3.17                                                           & 0.93                                                          \\
8c   & 279.46   & 11                                                     & 112.5           &                                                                 &                                                                      & 0.92                                                           &                                                               & 3.18                                                           & 0.95                                                          \\
9a   & 279.57   & 9                                                      & 0               & 0.1225                                                          & 279.59                                                               & 0.88                                                           & 2.88                                                          & 3.17                                                           & 0.92                                                          \\
9b   & 279.58   & 11                                                     & 112.5           &                                                                 &                                                                      & 0.92                                                           & 3.20                                                          & 3.18                                                           & 0.95                                                          \\
10a  & 279.70   & 9                                                      & 0               & 0.1235                                                          & 279.71                                                               & 0.87                                                           & 2.88                                                          & 3.18                                                           & 0.92                                                          \\
10b  & 279.70   & 11                                                     & 112.5           &                                                                 &                                                                      & 0.91                                                           & 3.20                                                          & 3.18                                                           & 0.94                                                          \\
11a  & 279.81   & 15                                                     & 315             & 0.1231                                                          & 279.84                                                               & 0.78                                                           & 2.88                                                          & 3.15                                                           & 0.85                                                          \\
11b  & 279.82   & 9                                                      & 0               &                                                                 &                                                                      & 0.88                                                           & 3.20                                                          & 3.18                                                           & 0.93                                                          \\
11c  & 279.83   & 15                                                     & 315             &                                                                 &                                                                      & 0.94                                                           &                                                               & 3.19                                                           & 0.96                                                          \\
12a  & 279.93   & 15                                                     & 315             & 0.1244                                                          & 279.96                                                               & 0.78                                                           & 2.88                                                          & 3.15                                                           & 0.84                                                          \\
12b  & 279.94   & 9                                                      & 0               &                                                                 &                                                                      & 0.87                                                           & 3.20                                                          & 3.18                                                           & 0.93                                                          \\
12c  & 279.95   & 11                                                     & 112.5           &                                                                 &                                                                      & 0.91                                                           &                                                               & 3.18                                                           & 0.94                                                          \\
13a  & 280.06   & 15                                                     & 315             & 0.1233                                                          & 280.08                                                               & 0.81                                                           & 2.88                                                          & 3.17                                                           & 0.90                                                          \\
13b  & 280.07   & 9                                                      & 0               &                                                                 &                                                                      & 0.87                                                           & 3.20                                                          & 3.18                                                           & 0.93                                                          \\
13c  & 280.07   & 11                                                     & 112.5           &                                                                 &                                                                      & 0.92                                                           &                                                               & 3.18                                                           & 0.94                                                          \\
14a  & 280.18   & 15                                                     & 315             & 0.1213                                                          & 280.21                                                               & 0.80                                                           & 2.88                                                          & 3.15                                                           & 0.85                                                          \\
14b  & 280.18   & 15                                                     & 315             &                                                                 &                                                                      & 0.83                                                           & 3.20                                                          & 3.17                                                           & 0.90                                                          \\
14c  & 280.19   & 9                                                      & 0               &                                                                 &                                                                      & 0.89                                                           &                                                               & 3.18                                                           & 0.92                                                          \\
14d  & 280.20   & 11                                                     & 112.5           &                                                                 &                                                                      & 0.93                                                           &                                                               & 3.18                                                           & 0.93                                                         
\end{tabular}
\end{table}

\newpage

\begin{table}[ht]
\small
\begin{tabular}{llllllllll}
450  &          &                                                        &                 &                                                                 &                                                                      &                                                                &                                                               &                                                                &                                                               \\
name & time (s) & \begin{tabular}[c]{@{}l@{}}first \\sensor\end{tabular} & class (\degree) & \begin{tabular}[c]{@{}l@{}}interseismic \\time (s)\end{tabular} & \begin{tabular}[c]{@{}l@{}}time   next \\slip event (s)\end{tabular} & \begin{tabular}[c]{@{}l@{}}relative \\timing (\%)\end{tabular} & \begin{tabular}[c]{@{}l@{}}min/max \\shear (MPa)\end{tabular} & \begin{tabular}[c]{@{}l@{}}absolute \\shear (MPa)\end{tabular} & \begin{tabular}[c]{@{}l@{}}relative \\shear (\%)\end{tabular} \\
1a   & 448.56   & 15                                                     & 315             &                                                                 & 448.59                                                               &                                                                &                                                               & 3.12                                                           &                                                               \\
1b   & 448.58   & 15                                                     & 315             &                                                                 &                                                                      &                                                                &                                                               & 3.16                                                           &                                                               \\
2a   & 448.72   & 15                                                     & 315             & 0.1576                                                          & 448.75                                                               & 0.79                                                           & 2.83                                                          & 3.11                                                           & 0.85                                                          \\
2b   & 448.74   & 15                                                     & 315             &                                                                 &                                                                      & 0.95                                                           & 3.17                                                          & 3.16                                                           & 0.98                                                          \\
3a   & 448.87   & 15                                                     & 315             & 0.1575                                                          & 448.91                                                               & 0.80                                                           & 2.82                                                          & 3.11                                                           & 0.84                                                          \\
3b   & 448.90   & 15                                                     & 315             &                                                                 &                                                                      & 0.96                                                           & 3.17                                                          & 3.16                                                           & 0.98                                                          \\
4a   & 449.03   & 15                                                     & 315             & 0.1581                                                          & 449.06                                                               & 0.80                                                           & 2.83                                                          & 3.12                                                           & 0.87                                                          \\
4b   & 449.06   & 15                                                     & 315             &                                                                 &                                                                      & 0.95                                                           & 3.16                                                          & 3.16                                                           & 0.98                                                          \\
5a   & 449.19   & 15                                                     & 315             & 0.1537                                                          & 449.22                                                               & 0.80                                                           & 2.83                                                          & 3.12                                                           & 0.86                                                          \\
5b   & 449.21   & 15                                                     & 315             &                                                                 &                                                                      & 0.95                                                           & 3.16                                                          & 3.16                                                           & 0.99                                                          \\
6a   & 449.34   & 15                                                     & 315             & 0.1566                                                          & 449.37                                                               & 0.80                                                           & 2.83                                                          & 3.11                                                           & 0.86                                                          \\
6b   & 449.37   & 15                                                     & 315             &                                                                 &                                                                      & 0.95                                                           & 3.16                                                          & 3.15                                                           & 0.98                                                          \\
7a   & 449.50   & 15                                                     & 315             & 0.1585                                                          & 449.53                                                               & 0.81                                                           & 2.82                                                          & 3.11                                                           & 0.85                                                          \\
7b   & 449.53   & 15                                                     & 315             &                                                                 &                                                                      & 0.96                                                           & 3.16                                                          & 3.16                                                           & 0.98                                                          \\
8a   & 449.66   & 15                                                     & 315             & 0.1576                                                          & 449.69                                                               & 0.80                                                           & 2.82                                                          & 3.11                                                           & 0.85                                                          \\
8b   & 449.68   & 15                                                     & 315             &                                                                 &                                                                      & 0.96                                                           & 3.16                                                          & 3.16                                                           & 0.98                                                          \\
9a   & 449.81   & 15                                                     & 315             & 0.1577                                                          & 449.85                                                               & 0.78                                                           & 2.82                                                          & 3.11                                                           & 0.86                                                          \\
9b   & 449.84   & 15                                                     & 315             &                                                                 &                                                                      & 0.96                                                           & 3.16                                                          & 3.16                                                           & 0.99                                                          \\
10a  & 449.97   & 15                                                     & 315             & 0.1556                                                          & 450.00                                                               & 0.78                                                           & 2.82                                                          & 3.11                                                           & 0.87                                                          \\
10b  & 450.00   & 15                                                     & 315             &                                                                 &                                                                      & 0.95                                                           & 3.16                                                          & 3.16                                                           & 0.99                                                          \\
11a  & 450.12   & 15                                                     & 315             & 0.1536                                                          & 450.16                                                               & 0.77                                                           & 2.82                                                          & 3.11                                                           & 0.85                                                          \\
11b  & 450.16   & 9                                                      & 315             &                                                                 &                                                                      & 0.99                                                           & 3.16                                                          & 3.16                                                           & 0.98                                                          \\
12a  & 450.28   & 15                                                     & 315             & 0.155                                                           & 450.31                                                               & 0.78                                                           & 2.82                                                          & 3.11                                                           & 0.86                                                          \\
12b  & 450.30   & 15                                                     & 315             &                                                                 &                                                                      & 0.95                                                           & 3.16                                                          & 3.15                                                           & 0.98                                                          \\
13a  & 450.44   & 15                                                     & 315             & 0.1567                                                          & 450.47                                                               & 0.78                                                           & 2.82                                                          & 3.11                                                           & 0.85                                                          \\
13b  & 450.47   & 9                                                      & 315             &                                                                 &                                                                      & 0.99                                                           & 3.16                                                          & 3.15                                                           & 0.98                                                          \\
14a  & 450.61   & 15                                                     & 315             & 0.15                                                            & 450.62                                                               & 0.91                                                           & 2.81                                                          & 3.15                                                           & 0.97                                                          \\
14b  & 450.62   & 9                                                      & 0               &                                                                 &                                                                      & 0.96                                                           & 3.15                                                          & 3.15                                                           & 1.00                                                          \\
15a  & 450.76   & 15                                                     & 315             & 0.1528                                                          & 450.78                                                               & 0.90                                                           & 2.82                                                          & 3.13                                                           & 0.93                                                          \\
15b  & 450.77   & 15                                                     & 315             &                                                                 &                                                                      & 0.97                                                           & 3.16                                                          & 3.15                                                           & 0.99                                                          \\
16a  & 450.91   & 15                                                     & 315             & 0.1513                                                          & 450.93                                                               & 0.89                                                           & 2.82                                                          & 3.13                                                           & 0.94                                                          \\
16b  & 450.92   & 9                                                      & 315             &                                                                 &                                                                      & 0.98                                                           & 3.15                                                          & 3.15                                                           & 0.99                                                          \\
17a  & 451.06   & 15                                                     & 315             & 0.1521                                                          & 451.08                                                               & 0.87                                                           & 2.82                                                          & 3.12                                                           & 0.89                                                          \\
17b  & 451.08   & 9                                                      & 315             &                                                                 &                                                                      & 0.98                                                           & 3.16                                                          & 3.15                                                           & 0.99                                                          \\
18a  & 451.21   & 15                                                     & 315             & 0.1527                                                          & 451.23                                                               & 0.86                                                           & 2.82                                                          & 3.12                                                           & 0.89                                                          \\
18b  & 451.22   & 15                                                     & 315             &                                                                 &                                                                      & 0.95                                                           & 3.15                                                          & 3.15                                                           & 1.00                                                          \\
19a  & 451.36   & 15                                                     & 315             & 0.1505                                                          & 451.38                                                               & 0.86                                                           & 2.82                                                          & 3.11                                                           & 0.88                                                          \\
19b  & 451.38   & 15                                                     & 315             &                                                                 &                                                                      & 0.96                                                           & 3.15                                                          & 3.15                                                           & 1.00                                                         
\end{tabular}
\end{table}

\newpage

\begin{table}[ht]
\small
\begin{tabular}{llllllllll}
620  &          &                                                        &                 &                                                                 &                                                                      &                                                                &                                                               &                                                                &                                                               \\
name & time (s) & \begin{tabular}[c]{@{}l@{}}first \\sensor\end{tabular} & class (\degree) & \begin{tabular}[c]{@{}l@{}}interseismic \\time (s)\end{tabular} & \begin{tabular}[c]{@{}l@{}}time   next \\slip event (s)\end{tabular} & \begin{tabular}[c]{@{}l@{}}relative \\timing (\%)\end{tabular} & \begin{tabular}[c]{@{}l@{}}min/max \\shear (MPa)\end{tabular} & \begin{tabular}[c]{@{}l@{}}absolute \\shear (MPa)\end{tabular} & \begin{tabular}[c]{@{}l@{}}relative \\shear (\%)\end{tabular} \\
1a   & 618.61   & 11                                                     & 112.5           &                                                                 & 618.69                                                               &                                                                &                                                               & 3.19                                                           &                                                               \\
1b   & 618.64   & 9                                                      & 315             &                                                                 &                                                                      &                                                                &                                                               & 3.22                                                           &                                                               \\
1c   & 618.67   & 15                                                     & 315             &                                                                 &                                                                      &                                                                &                                                               & 3.24                                                           &                                                               \\
2a   & 618.95   & 11                                                     & 112.5           & 0.3339                                                          & 619.02                                                               & 0.77                                                           & 2.89                                                          & 3.20                                                           & 0.84                                                          \\
2b   & 618.97   & 9                                                      & 315             &                                                                 &                                                                      & 0.85                                                           & 3.26                                                          & 3.21                                                           & 0.88                                                          \\
2c   & 619.00   & 15                                                     & 292.5           &                                                                 &                                                                      & 0.94                                                           &                                                               & 3.24                                                           & 0.96                                                          \\
3a   & 619.28   & 11                                                     & 112.5           & 0.3326                                                          & 619.36                                                               & 0.78                                                           & 2.89                                                          & 3.19                                                           & 0.83                                                          \\
3b   & 619.31   & 9                                                      & 315             &                                                                 &                                                                      & 0.85                                                           & 3.25                                                          & 3.21                                                           & 0.89                                                          \\
3c   & 619.33   & 15                                                     & 292.5           &                                                                 &                                                                      & 0.93                                                           &                                                               & 3.23                                                           & 0.95                                                          \\
4a   & 619.62   & 11                                                     & 112.5           & 0.3323                                                          & 619.69                                                               & 0.78                                                           & 2.89                                                          & 3.19                                                           & 0.84                                                          \\
4b   & 619.64   & 9                                                      & 315             &                                                                 &                                                                      & 0.86                                                           & 3.25                                                          & 3.21                                                           & 0.90                                                          \\
4c   & 619.66   & 15                                                     & 292.5           &                                                                 &                                                                      & 0.93                                                           &                                                               & 3.23                                                           & 0.95                                                          \\
5a   & 619.95   & 11                                                     & 90              & 0.3322                                                          & 620.02                                                               & 0.78                                                           & 2.89                                                          & 3.19                                                           & 0.84                                                          \\
5b   & 619.97   & 9                                                      & 315             &                                                                 &                                                                      & 0.85                                                           & 3.25                                                          & 3.21                                                           & 0.89                                                          \\
5c   & 620.00   & 15                                                     & 292.5           &                                                                 &                                                                      & 0.93                                                           &                                                               & 3.23                                                           & 0.94                                                          \\
6a   & 620.28   & 11                                                     & 90              & 0.3337                                                          & 620.36                                                               & 0.78                                                           & 2.89                                                          & 3.19                                                           & 0.84                                                          \\
6b   & 620.31   & 9                                                      & 337.5           &                                                                 &                                                                      & 0.86                                                           & 3.25                                                          & 3.21                                                           & 0.91                                                          \\
6c   & 620.33   & 15                                                     & 315             &                                                                 &                                                                      & 0.92                                                           &                                                               & 3.23                                                           & 0.95                                                          \\
7a   & 620.62   & 11                                                     & 112.5           & 0.3320                                                          & 620.69                                                               & 0.78                                                           & 2.88                                                          & 3.19                                                           & 0.86                                                          \\
7b   & 620.64   & 9                                                      & 315             &                                                                 &                                                                      & 0.86                                                           & 3.25                                                          & 3.21                                                           & 0.91                                                          \\
7c   & 620.66   & 15                                                     & 315             &                                                                 &                                                                      & 0.91                                                           &                                                               & 3.22                                                           & 0.94                                                          \\
8a   & 620.95   & 11                                                     & 90              & 0.3302                                                          & 621.02                                                               & 0.79                                                           & 2.88                                                          & 3.20                                                           & 0.86                                                          \\
8b   & 620.97   & 9                                                      & 315             &                                                                 &                                                                      & 0.87                                                           & 3.25                                                          & 3.21                                                           & 0.91                                                          \\
8c   & 620.99   & 15                                                     & 292.5           &                                                                 &                                                                      & 0.91                                                           &                                                               & 3.23                                                           & 0.96                                                          \\
8d   & 621.00   & 9                                                      & 337.5           &                                                                 &                                                                      & 0.94                                                           &                                                               & 3.23                                                           & 0.97                                                          \\
9a   & 621.28   & 11                                                     & 112.5           & 0.3312                                                          & 621.35                                                               & 0.79                                                           & 2.89                                                          & 3.20                                                           & 0.85                                                          \\
9b   & 621.30   & 9                                                      & 337.5           &                                                                 &                                                                      & 0.87                                                           & 3.25                                                          & 3.22                                                           & 0.91                                                          \\
9c   & 621.32   & 15                                                     & 315             &                                                                 &                                                                      & 0.91                                                           &                                                               & 3.23                                                           & 0.94                                                          \\
9d   & 621.33   & 9                                                      & 337.5           &                                                                 &                                                                      & 0.95                                                           &                                                               & 3.24                                                           & 0.97                                                         
\end{tabular}
\end{table}

\newpage

\begin{table}[ht]
\small
\begin{tabular}{llllllllll}
790  &          &                                                        &                 &                                                                 &                                                                      &                                                                &                                                               &                                                                &                                                               \\
name & time (s) & \begin{tabular}[c]{@{}l@{}}first \\sensor\end{tabular} & class (\degree) & \begin{tabular}[c]{@{}l@{}}interseismic \\time (s)\end{tabular} & \begin{tabular}[c]{@{}l@{}}time   next \\slip event (s)\end{tabular} & \begin{tabular}[c]{@{}l@{}}relative \\timing (\%)\end{tabular} & \begin{tabular}[c]{@{}l@{}}min/max \\shear (MPa)\end{tabular} & \begin{tabular}[c]{@{}l@{}}absolute \\shear (MPa)\end{tabular} & \begin{tabular}[c]{@{}l@{}}relative \\shear (\%)\end{tabular} \\
1a   & 788.67   & 11                                                     & 90              &                                                                 & 788.82                                                               &                                                                &                                                               & 3.29                                                           &                                                               \\
1b   & 788.71   & 11                                                     & 112.5           &                                                                 &                                                                      &                                                                &                                                               & 3.31                                                           &                                                               \\
1c   & 788.76   & 11                                                     & 67.5            &                                                                 &                                                                      &                                                                &                                                               & 3.33                                                           &                                                               \\
2a   & 789.06   & 11                                                     & 90              & 0.3815                                                          & 789.20                                                               & 0.61                                                           & 3.20                                                          & 3.29                                                           & 0.64                                                          \\
2b   & 789.10   & 11                                                     & 112.5           &                                                                 &                                                                      & 0.73                                                           & 3.35                                                          & 3.31                                                           & 0.78                                                          \\
2c   & 789.14   & 11                                                     & 67.5            &                                                                 &                                                                      & 0.84                                                           &                                                               & 3.33                                                           & 0.90                                                          \\
3a   & 789.45   & 11                                                     & 90              & 0.3865                                                          & 789.59                                                               & 0.63                                                           & 3.20                                                          & 3.30                                                           & 0.66                                                          \\
3b   & 789.48   & 11                                                     & 112.5           &                                                                 &                                                                      & 0.72                                                           & 3.35                                                          & 3.31                                                           & 0.76                                                          \\
3c   & 789.53   & 11                                                     & 67.5            &                                                                 &                                                                      & 0.84                                                           &                                                               & 3.33                                                           & 0.88                                                          \\
4a   & 789.82   & 11                                                     & 90              & 0.3769                                                          & 789.97                                                               & 0.61                                                           & 3.20                                                          & 3.29                                                           & 0.63                                                          \\
4b   & 789.91   & 11                                                     & 112.5           &                                                                 &                                                                      & 0.84                                                           & 3.35                                                          & 3.33                                                           & 0.91                                                          \\
4c   & 789.94   & 9                                                      & 337.5           &                                                                 &                                                                      & 0.92                                                           &                                                               & 3.34                                                           & 0.97                                                          \\
5a   & 790.20   & 11                                                     & 90              & 0.3763                                                          & 790.34                                                               & 0.61                                                           & 3.20                                                          & 3.29                                                           & 0.63                                                          \\
5b   & 790.24   & 11                                                     & 112.5           &                                                                 &                                                                      & 0.72                                                           & 3.35                                                          & 3.31                                                           & 0.77                                                          \\
5c   & 790.29   & 11                                                     & 67.5            &                                                                 &                                                                      & 0.85                                                           &                                                               & 3.33                                                           & 0.89                                                          \\
6a   & 790.57   & 11                                                     & 90              & 0.3649                                                          & 790.71                                                               & 0.62                                                           & 3.20                                                          & 3.30                                                           & 0.65                                                          \\
6b   & 790.65   & 11                                                     & 112.5           &                                                                 &                                                                      & 0.84                                                           & 3.35                                                          & 3.33                                                           & 0.91                                                          \\
7a   & 790.94   & 11                                                     & 90              & 0.3660                                                          & 791.07                                                               & 0.63                                                           & 3.20                                                          & 3.30                                                           & 0.64                                                          \\
7b   & 790.97   & 11                                                     & 112.5           &                                                                 &                                                                      & 0.72                                                           & 3.35                                                          & 3.31                                                           & 0.71                                                          \\
7c   & 791.03   & 11                                                     & 67.5            &                                                                 &                                                                      & 0.87                                                           &                                                               & 3.33                                                           & 0.90                                                          \\
7d   & 791.04   & 9                                                      & 337.5           &                                                                 &                                                                      & 0.90                                                           &                                                               & 3.34                                                           & 0.92                                                          \\
8a   & 791.30   & 11                                                     & 90              & 0.3604                                                          & 791.43                                                               & 0.62                                                           & 3.20                                                          & 3.29                                                           & 0.63                                                          \\
8b   & 791.33   & 11                                                     & 112.5           &                                                                 &                                                                      & 0.71                                                           & 3.35                                                          & 3.30                                                           & 0.71                                                          \\
8c   & 791.38   & 9                                                      & 337.5           &                                                                 &                                                                      & 0.85                                                           &                                                               & 3.33                                                           & 0.88                                                         
\end{tabular}
\end{table}

\end{document}